\documentclass[11pt]{article}
\usepackage[utf8]{inputenc}
\usepackage{geometry}
\usepackage{amsmath}
\usepackage{amssymb}
\usepackage{graphicx}
\usepackage{bm}
\usepackage{mathtools}
\usepackage{hyperref}
\usepackage{tikz}
\usepackage{setspace}
\usepackage[round]{natbib}
\usepackage{subcaption}
\usepackage{bbm}
\usepackage{mathbbol}
\usepackage{array}
\usepackage{enumitem}

\usetikzlibrary{shapes.geometric, arrows}

\usepackage{fullpage}
\newtheorem{assumption}{\small\sc Assumption}

\newcommand{\indep}{\rotatebox[origin=c]{90}{$\models$}}

\newcommand{\E}{\mathbbm{E}}

\DeclareMathOperator{\var}{var}
\DeclareMathOperator{\sd}{sd}
\DeclareMathOperator{\se}{se}
\DeclareMathOperator{\cov}{cov}
\DeclareMathOperator{\cor}{cor}

\tikzstyle{stepstyle} = [rectangle, rounded corners, minimum width=3cm, minimum height=1cm, draw=black, text width = 5cm]
\tikzstyle{decisionstyle} = [diamond, rounded corners, minimum width=3cm, minimum height=1cm, draw=black, text width = 5cm]

\tikzstyle{arrow} = [thick,->,>=stealth]

\title{Sensitivity of weighted least squares estimators to omitted variables\thanks{We thank Erin Hartman, Onyebuchi Arah, Mark Handcock, Melody Huang, Angelica Anabel Remache Lopez, the UCLA Practical Causal Inference Lab, and anonymous reviewers for their valuable comments and suggestions. We also thank Wolfgang Brightenburg for his extensive contributions to the \texttt{R} Package \href{https://github.com/lwainstein/sensewls}{\texttt{sensewls}}, which implements the methods proposed here.}} 
\date{\today}

\footnotesize
 \author{
 	Leonard Wainstein\thanks{
		Assistant Professor, Mathematics and Statistics Department, Reed College. 
		\newline
		Mailing Address: 3203 SE Woodstock Blvd, Portland, OR 97202
		\newline
		Email: \href{mailto:lwainstein@reed.edu}{lwainstein@reed.edu}
	}
	\ \ \ \ \ \ \ \ \ \ \ \ \ 
	Chad Hazlett\thanks{
		Professor, Departments of Statistics and Political Science, University of California Los Angeles.
		\newline
		Mailing Address: 8125 Math Sciences Bldg, Box 951554, Los Angeles, CA 90095
            \newline
        Email: \href{mailto:chazlett@ucla.edu}{chazlett@ucla.edu}
		\newline
		URL: \href{http://www.chadhazlett.com}{http://www.chadhazlett.com}
	}
} 

\begin{document}

\setcounter{page}{0}
\maketitle

\thispagestyle{empty}

\begin{abstract}
We introduce tools for assessing the sensitivity, to unobserved confounding, of a common estimator of causal effects  that employs weights: the weighted linear regression of the outcome on the treatment and observed covariates. This estimator's bias is a function of two intuitive weighted partial $\hat{R}^2$ values: (i) the proportion of weighted variance in the treatment that unobserved confounding explains given the covariates and (ii) the proportion of weighted variance in the outcome that unobserved confounding explains given the covariates and the treatment. Following previous work, we define sensitivity statistics for routine reporting,  derive formal bounds on the strength of unobserved confounding with (multiples of) the strength of certain covariates, and propose adjusted inference procedures. A key choice we make is to examine only how the outcome model is influenced by unobserved confounding, instead of how the weights have been affected. One benefit of this choice is that our tools apply with any (non-negative) weights (e.g., inverse propensity score, matching, or covariate balancing). Another benefit is that we can rely on omitted variable bias approaches that impose no distributional assumptions on the data or unobserved confounding, and can address misspecification bias. The tools are available in the \href{https://github.com/lwainstein/sensewls}{\texttt{sensewls}}  package for \texttt{R}.
\end{abstract}

\pagebreak
\clearpage

\onehalfspacing

\newpage
\setcounter{page}{1}

\setcounter{page}{1}

\section{Introduction}\label{sec:intro}

Researchers often seek the causal effect of a
treatment, $D$, on an outcome of interest, $Y$. In observational settings, estimating this unbiasedly requires accounting for all confounders in the relationship between $D$ and $Y$. 
In many traditions, this has come in the form of a linear regression of $Y$ on $D$ and a host of observed covariates, $X$.
However, when $D$ is binary, it is commonplace to 
utilize weights that leave the treated and control (i.e., untreated) groups more similar on $X$. 
For example, weights based on the probability of being treated given $X$, or 
the propensity score (\citealp{rosenbaum1983central}),
can be motivated as the weights that would, in expectation, equate the distribution of $X$ in the control group to that in the treated group. 
Other examples include 
``balancing" weights, which directly aim to equate the means (or other functions) of $X$ in both groups exactly (e.g., \citealp{hainmueller2012entropy}; \citealp{chan2016globally}) or approximately (e.g.,  
\citealp{wang2020minimal}; \citealp{kallus2020generalized}; \citealp{hazlett2018kernel}). Most ``matching'' methods (e.g., \citealp{rosenbaum1983central, iacus2012causal, sekhon2009opiates}) are also forms of weighting. So too are estimators that employ stratification or sub-classification to produce adjusted differences in means, or equivalently 
compute treatment effects conditionally on strata and marginalize over them.   

After weights are chosen, estimates of the effect of $D$ on $Y$ are often produced either by taking a weighted difference in means in $Y$ or by some form of \textit{weighted} linear regression of $Y$ on $D$ and $X$, where  
the weighting is meant to reduce dependence on the estimated linear model  (\citealp{ho2007matching}). The latter approach is preferable in many or most cases (see e.g. \citealp{residualizedSEs}) because (i) with perfect mean balancing weights this has no impact on the point estimate but allows the resulting standard errors to ``take credit for'' the reduced variance in the estimate achieved by the weighting procedure, while (ii) with approximate or in-expectation balancing (e.g., inverse propensity score weights), it additionally provides a model-based tool to address residual imbalances, and has the interpretation of an augmented estimator (e.g., \citealp{robins1994estimation};  \citealp{hirshberg2017augmented}), as described below. Post-weighting regression has accordingly become a standard approach, including by default in software packages such \texttt{WeightIT} (\citealp{greifer2019package}).

However, the first-order concern in observational studies is typically the risk of unobserved confounding that leads to persistent biases in the estimate regardless of the conditioning technology used. Specifically, the claim that all variables that must be accounted for to achieve identification  is unlikely to hold in many real-world cases. The resulting bias in the estimate will be driven by the extent to which unobserved confounders, $Z$, are related to $D$ and $Y$ conditionally on the observables, $X$. Transparency thus requires that we assess the sensitivity of one's conclusions to unobserved confounding. 

Many sensitivity analyses have been proposed, both for outcome-oriented models such as regression (e.g., \citealp{cinelli2020making}) and for weight-based estimators such as inverse propensity score weighting or matching (e.g., \citealp{shen2011sensitivity, hong2020did, rosenbaum2002sensitivity}). We propose a simple strategy for investigators to employ with weighting:  choose (non-negative) weights by any means, and conduct the sensitivity analysis with respect to the weighted regression. This approach specifically differs from those that begin with the question of ``how the weights would change'' had an omitted confounder been present. For example, if the weights are intended to represent inverse propensity score weights, then the user would need to consider how the estimated propensity score model using observed variables differs from the propensity score model that uses the observed variables \textit{and} the omitted confounders.

Our alternative makes its own assumption, but is designed to provide two major benefits. First, it enables us to adapt the powerful yet simple tools proposed for sensitivity analysis of regressions by \cite{cinelli2020making}, henceforth ``C\&H''. Consequently, this approach shares the conveniences of omitted-variable bias approaches---for example, no assumption needs to be made on the number or distribution of the unobserved confounders.
Second, this approach is agnostic to the origin of the weights, other than the requirement that the weights be non-negative. It applies in any setting where (non-negative) weights are used, whether they are assumed to be inverse propensity score weights (in actuality or by an equivalence argument), calibration/balancing weights, or the result of matching or stratification procedures that can be represented by weights.

Concretely, we show that regardless of the logic motivating the choice of weights, the bias of the weighted regression is a function of two intuitive sensitivity parameters: (i) the proportion of weighted variance in $D$ that $Z$ explains given $X$ and (ii) the proportion of weighted variance in $Y$ that $Z$ explains given $X$ and $D$, i.e., two weighted partial $\hat{R}^2$ values that quantify the strengths of the relationships between $Z$ and $D$, and $Z$ and $Y$, respectively. Following C\&H, we define sensitivity statistics that lend themselves well to routine reporting, and a benchmarking procedure to formally bound the strength of $Z$ 
with (a multiple of) the strength of select dimensions of $X$, helping determine if unobserved confounding that would alter one's conclusions is plausible. We propose and find good performance with three inference procedures: two that involve the bootstrap and are more agnostic, and one closed-form alternative that requires stronger assumptions. 
Additionally, we note that when the weights exactly balance $X$, 
our proposed tools also apply to the point estimate from the simple weighted difference in means in $Y$. 

To outline, Section~\ref{sec:background} details notation and other preliminaries. 
Section~\ref{sec:wsa_results} develops the proposed sensitivity tools and Section~\ref{sec:app1} demonstrates them in an applied setting: estimating the effect of exposure to violence in Darfur on attitudes toward peace (\citealp{hazlett2020angry}), employing inverse propensity score weights, matching, and balancing weights. Section~\ref{sec:discussion} provides further discussion and concludes. We have also developed the \href{https://github.com/lwainstein/sensewls}{\texttt{sensewls}} package to implement our proposed tools in the \texttt{R} programming language.

\section{Background}\label{sec:background}

Let $i \in \{1, \dots, n\}$ index the units of observation and let $p(\cdot)$ be the density function of an arbitrary random variable. Then, let $D$ be the treatment, with $\mathbf{D} = [D_1 \ \dots \ D_n]^{\top}$ being the vector of treatment statuses for the sample, and let $X$ be an observed $P$-dimensional vector of covariates, with $\mathbf{X}$ being the matrix of $X_i$ for the sample,
	\begin{align}
    		X &= \begin{bmatrix} X^{(1)} \\ \vdots \\ X^{(P)} \end{bmatrix} \in \mathbbm{R}^{P}  \ , \   \mathbf{X} = \begin{bmatrix} X_{1}^{\top} \\ \vdots \\ X_{n}^{\top} \end{bmatrix}  \in \mathbbm{R}^{n \times P}
	\end{align}
Note that $X$ here may include, or be exchanged, with its nonlinear transformations (e.g., polynomial terms, or basis functions). We also allow $X$ to potentially include functions of the covariates \textit{and} $D$, as we describe in Section~\ref{subsec:estimators}. However, we continue to use $X$ to encompass these possibilities in the interest of simplifying notation. Next, let $Y$ be the outcome of interest, with $\mathbf{Y} = [Y_1 \ \dots \ Y_n]^{\top}$. In accordance with the potential outcomes framework (\citealp{splawa1990application}; \citealp{rubin1974estimating}), let $Y (d)$ be the potential outcome under treatment status $d$, so $Y = Y (D)$ is observed (maintaining the consistency assumption). Implicit in achieving this consistency is the stable unit treatment value assumption (SUTVA), i.e., the potential outcomes for unit $i$ are not functions of the treatment statuses of other units, and that each treatment status $d$ is administered the same across the units. Additionally, the tuples $( X_i, D_i,  Y_i (d) )$ are assumed independent and identically distributed (iid) unless otherwise noted.

We consider binary treatments, $D \in \{0,1\}$ where $\sum_{i=1}^n D_i = n_1$ is the number of treated units and $n_0 = n - n_1$ is the number of control units. 
We consider estimating the average treatment effect (ATE),
	\begin{align}
		\mathrm{ATE} = \E [ Y (1) - Y (0) ]
	\end{align}
where $\E (\cdot)$ is the expectation over $p(\cdot)$, i.e., the super-population. We also consider 
the average treatment effect on the treated (ATT) and the average treatment effect on the controls (ATC), 
	\begin{align}
		\mathrm{ATT} = \E[Y (1) - Y (0) \ | \ D = 1]  \ \ \ \text{and} \ \ \ \mathrm{ATC} = \E[Y (1) - Y (0) \ | \ D = 0]
	\end{align} 

Finally, let $w_i$ be a non-negative (i.e., $w_i \geq 0)$ weight for unit $i$. Without loss of generality, additionally let the $w_i$ sum to $n$ (i.e., $\sum_{i=1}^n w_i = n$). We primarily focus on ``honest" weights, or weights that are a function of \textit{only} $\mathbf{X}$ and $\mathbf{D}$, and are thus independent of $\mathbf{Y}$ (conditional on $\mathbf{X}$ and $\mathbf{D}$). However, our sensitivity tools below do apply with (non-negative) weights that depend on the outcomes (e.g., \citealp{ning2020robust, wainstein2022targeted}).

\subsection{Primary estimator of interest: weighted least squares}\label{subsec:estimators}

We primarily develop tools to assess the sensitivity to unobserved confounding of the estimator that results from a weighted regression with covariates. However, before formally defining this estimator, we first consider the traditional, \textit{unweighted} regression,
	\begin{align}\label{eq:ols_ydx}
		& ( \hat{\mu}_{\mathrm{ols}}, \hat{\tau}_{\mathrm{ols}}, \hat{\beta}_{\mathrm{ols}} ) = \underset{\mu, \tau, \beta}{\mathrm{argmin}} \ \frac{1}{n} \sum_{i=1}^{n} \biggr( Y_i - (\mu + \tau D_i + X_i^{\top} \beta) \biggr)^2
	\end{align}
in which $\hat{\tau}_{\mathrm{ols}}$ is of interest. Without transformations to $X$, this linear model is misspecified, even absent unobserved confounding, when treatment effects are heterogeneous in $X$ and treatment probability changes in $X$, which leads to the apparent upweighting of strata in which the probability of treatment is nearer to 50\% (\citealp{hazlett2024understanding, chattopadhyay2023implied, angrist1995estimating}). However, this is resolved by replacing $X$ in Expression~\ref{eq:ols_ydx} with $(X - m(X), D*(X - m(X))$ where $m(X)$ is the appropriate sample mean of $X$ for the desired estimand. For example, $m (X) = \frac{1}{n} \sum_{i=1}^n X_i$ targets the ATE, and the resulting $\hat{\tau}_{\mathrm{ols}}$ takes the form of the estimator studied by \cite{lin_agnostic_2013}. Further, $m (X) = \frac{1}{n_1} \sum_{i:D_i=1} X_i$ targets the ATT, and $m (X) = \frac{1}{n_0} \sum_{i:D_i=0} X_i$ targets the ATC. 
This is advisable and we recommend it in many cases, though for simplicity of notation, we write regression expressions below without adding the interaction term or centered covariates.


Our sensitivity tools focus on a generalization of Expression~\ref{eq:ols_ydx} that weights each unit's squared error by a non-negative $w_i$, i.e., the weighted least squares regression
	\begin{align}\label{eq:wls_ydx}
		& ( \hat{\mu}_{\mathrm{wls}}, \hat{\tau}_{\mathrm{wls}}, \hat{\beta}_{\mathrm{wls}}  ) = \underset{\mu, \tau, \beta}{\mathrm{argmin}} \ \frac{1}{n} \sum_{i=1}^n w_i \biggr( Y_i - ( \mu + \tau D_i + X_i^{\top} \beta) \biggr)^2
	\end{align}
where $\hat{\tau}_{\mathrm{wls}}$ is the estimated treatment effect.\footnote{Note that even if some starting weights $\tilde{w}_i$ do not sum to $n$, they can be replaced with normalized weights $w_i = n \times \frac{\tilde{w}_i}{\sum_{i=1}^n \tilde{w}_i}$  that do sum to $n$ without changing $\hat{\tau}_{\mathrm{wls}}$. Thus, our requirement that the weights sum to $n$ does not restrict at all the (non-negative) weights one can employ.} While Expression~\ref{eq:ols_ydx} treats each unit's squared error equally, the weighted regression in Expression~\ref{eq:wls_ydx} prioritizes minimizing unit $i$'s squared error over that of unit $j$ if $w_i > w_j$. When $D$ is binary, the goal of the weights is to make the distributions of $X$ more similar between the treated and control groups, making  $\hat{\tau}_{\mathrm{wls}}$ more robust to violations to the estimated linear model (e.g., \citealp{ho2007matching}). Further, weights can be chosen to 
target the ATE, ATT, or ATC. For example, Entropy Balancing (\citealp{hainmueller2012entropy}) may select for control units the $w_i$ of maximum entropy, $-\sum_{i: D_i=0} \frac{w_i}{n_0} \mathrm{log}(\frac{w_i}{n_0})$, that equate 
the means of $X$ in the treated and control groups:\footnote{
Weights that equate the means of $X$ in both groups are often referred to 
as ``balancing" weights, or more precisely, mean balancing weights. Alternatives may target only approximate balance, or may achieve balance on moments/functions of $X$ instead of (or in addition to) the untransformed covariates in an effort to enforce broader distributional balance (e.g., 
\citealp{chan2016globally}, 
\citealp{wang2020minimal}; \citealp{kallus2020generalized}; \citealp{hazlett2018kernel}).}
	\begin{align}\label{eq:ebal}
		& \underset{w} {\mathrm{argmax}}  \biggr[ -\sum_{i: D_i=0} \frac{w_i}{n_0} \mathrm{log}(\frac{w_i}{n_0}) \biggr] \ \ \text{where} \ \ \frac{1}{n_0} \sum_{i: D_i=0} w_i X_i = \frac{1}{n_1} \sum_{i: D_i=1} X_i \ \ \text{and} \ \ \sum_{i: D_i = 0} w_i = n_0
	\end{align}
Using $w_i$ from Expression~\ref{eq:ebal} for control units and $w_i = 1$ for treated units then yields an estimate of 
the ATT. Another example of weighting is the inverse propensity score weight, which estimates a model 
for the probability of treatment given the covariates, $\pi (X) = p(D = 1 \ | \ X)$ 
or the propensity score (\citealp{rosenbaum1983central}). These weights can be understood as equating the distributions of $X$ in the control and treated groups in expectation when $\pi(X)$ has been consistently estimated. When estimating the ATE, inverse propensity score weights choose
    \begin{align}\label{eq:psweights}
        w_i \propto \begin{cases} 
            \frac{1}{1 - \hat{\pi} (X_i)} & \ \text{if} \ D_i = 0 \\
            \ \frac{1}{\hat{\pi} (X_i)} & \ \text{if} \ D_i = 1 \\   
        \end{cases}
    \end{align}
where $\hat{\pi} (X) $ is an estimate of $\pi (X)$, and units are weighted inversely proportional to their (estimated) probability of receiving the treatment status they were ultimately given.\footnote{
To estimate the ATC and the ATT, respectively, inverse propensity score weights choose
	\begin{align}
		w_i \propto
		\begin{cases} 
			\ \ \ \ 1 & \text{if} \ D_i = 0 \\
			\frac{1 - \hat{\pi} (X_i)}{\hat{\pi} (X_i)} & \text{if} \ D_i = 1
		\end{cases} \ \ \ \text{and} \ \ \ 
		w_i \propto
		\begin{cases} 
			\frac{\hat{\pi} (X_i)}{1 - \hat{\pi} (X_i)} & \text{if} \ D_i = 0 \\
			\ \ \ \ 1 & \text{if} \ D_i = 1
		\end{cases}\nonumber
	\end{align}
} 
Another commonly used family of weights are matching weights. For example, one-to-one propensity score matching for the ATT matches each treated unit with a control unit that has the closest $\hat{\pi} (X)$. There is also ``exact" matching, in which units are only matched if they have the exact same $X$. When  ATT matching is done \textit{with} replacement, the same control unit can be matched to multiple treated units. This results in weights where $w_i \propto 1$ for treated units, and for control units, $w_i \propto $ the number of times matched. When ATT matching is done \textit{without} replacement, this results in weights where $w_i \propto 1$ for treated units (unless the unit is dropped for lack of a match), and $w_i \propto I(\text{unit $i$ matched})$ for control units.

\subsection{Weighted distributions}\label{subsec:w_distr}

The analyses below rely on an understanding of how weighted regression can be viewed as regression in a sample where the distribution of $X$, $D$, $Y$ and unobserved confounders ($Z$) have been altered by applying (non-uniform) weights. Accordingly, we define here sample statistics for the weighted distribution that are analogous to the usual sample mean (i.e., $\frac{1}{n} \sum_{i=1}^n X_i$), covariance, and others, and will be used to parametrize the bias, adjusted inference, and other proposed sensitivity tools.

\paragraph{Intuition.} 
While our sensitivity tools apply with any non-negative weights, we first consider a simplified setting where $w_i = w (X_i, D_i)$ for some weight function $w ( \cdot )$ to motivate the weighted sample statistics we utilize below. The OLS regression in Expression~\ref{eq:ols_ydx} finds coefficients that have probability limit
	\begin{align}
		( \hat{\mu}_{\mathrm{ols}}, \hat{\tau}_{\mathrm{ols}}, \hat{\beta}_{\mathrm{ols}} ) \overset{p}{\rightarrow} \underset{\mu, \tau, \beta}{\mathrm{argmin}} \  \E \biggr[ \biggr( Y - (\mu + \tau D + X^{\top} \beta ) \biggr)^2  \biggr] 
	\end{align}
In other words, these coefficients minimize the mean squared error over $p(X, D, Y)$ in expectation. In the weighted regression in Expression~\ref{eq:wls_ydx}, however, the coefficients instead minimize the expected squared error over the \textit{weighted} distribution, $p_w (X, D, Y) = w(X, D) p(X, D, Y)$. 
To see this, note that the coefficients in Expression~\ref{eq:wls_ydx} have the probability limit
	\begin{align}
		( \hat{\mu}_{\mathrm{wls}}, \hat{\tau}_{\mathrm{wls}}, \hat{\beta}_{\mathrm{wls}} ) \overset{p}{\rightarrow}  \underset{\mu, \tau, \beta}{\mathrm{argmin}} \  \E_{w} \biggr[ \biggr( Y - (\mu + \tau D + X^{\top} \beta ) \biggr)^2  \biggr] 
	\end{align}
where $\E_w (\cdot) $ is the expectation assuming that $(X_i, D_i, Y_i) \overset{iid}{\sim}  p_w (X, D, Y) $. Thus, the $w_i$ shift the distribution under which the coefficients minimize the model's mean squared error. 
Accordingly, within the sample, these weights shift the empirical distribution, yielding $\hat{p} (X_i, D_i, Y_i) = \frac{1}{n}$, 
to the weighted empirical distribution, yielding $\hat{p}_w (X_i, D_i, Y_i) = \frac{w_i}{n}$. 

\paragraph{Sample statistics and weighted $\hat{R}^2$.}

Sample statistics for the weighted empirical distribution are thus required. Let $A$ and $B$ be random vectors. Define weighted sample means, covariances, and variances, respectively, as
	\begin{align}
		\widehat{\E}_w (A) &= \frac{1}{n} \sum_{i=1}^n w_i A_i \\
		 \widehat{\cov}_w (A, B) &= \frac{1}{n} \sum_{i=1}^n w_i \biggr( A_i - \widehat{\E}_w (A) \biggr) \biggr( B_i - \widehat{\E}_w (B) \biggr)^{\top} \ \text{and} \ \ \widehat{\var}_w (A) = \widehat{\cov}_w (A, A)
	\end{align}
Then, if $A$ and $B$ are scalar random variables, define weighted standard deviations and correlations, respectively, as
	\begin{align}
		\widehat{\sd}_w (A) =  \sqrt{\widehat{\var}_w (A)} \ \ \ \text{and} \ \ \ \hat{R}_w (B \sim A) &= \frac{\widehat{\cov}_w (A, B)}{\widehat{\sd}_w (A) \widehat{\sd}_w (B) }
	\end{align}

These sample statistics give meaning to the coefficients from weighted regressions such as Expression~\ref{eq:wls_ydx} and  Expression~\ref{eq:wls_ydxz} to come, 
and to an analogous $\hat{R}^2$, or percent of variation explained. Let the $A_i$ and $B_i$ 
be centered by their weighted sample means (i.e., $\widehat{\E}_w (\cdot)$), and let $B$ be one-dimensional. Then,
	\begin{align}\label{eq:wls_as_samplestats}
		\underset{\nu}{\mathrm{argmin}} \ \frac{1}{n} \sum_{i=1}^{n} w_i ( B_i - A_i^{\top} \nu )^2 = [\widehat{\var}_w (A)]^{-1}  \widehat{\cov}_w (A, B) 
	\end{align}
This allows the ``partialing out" of $A$ from $B$ in the weighted distribution, or residualizing $B$ after the regression in Expression~\ref{eq:wls_as_samplestats} above, to be defined as
	\begin{align}
		B_i^{\perp_w A} = B_i - A_i^{\top} \biggr( [\widehat{\var}_w (A)]^{-1}  \widehat{\cov}_w (A, B) \biggr) 
	\end{align}
The $B_i^{\perp_w A}$ can additionally be thought of the portion of the $B_i$ that is uncorrelated, or orthogonal, to the $A_i$ in the weighted distribution. This also allows definitions for a weighted $\hat{R}^2$ and a weighted partial $\hat{R}^2$,  respectively:\footnote{Note that the definition for partial $\hat{R}_w^2$ here is a slight abuse of notation. The conditioning on $X$ in $\hat{R}_w^2 (B \sim A | X)$ does not mean this value is the $\hat{R}_w^2$ for a set value of $X$. Instead, it means the value is the $\hat{R}_w^2$ after partialing out $X$ from $A$ and $B$.}
	\begin{equation}\label{eq:weightedR2}
		\hat{R}_w^2 (B \sim A) = \frac{\widehat{\var}_w (B) - \widehat{\var}_w (B^{\perp_w A})}{\widehat{\var}_w (B) } \ \ \ \text{and} \ \ \ \hat{R}_w^2 (B \sim A | X) = \hat{R}_w^2 (B^{\perp_w X}  \sim A^{\perp_w X} ) 
	\end{equation}
Weighted $\hat{R}^2$ thus has a similar intuition as it does with uniform weights: the proportion of variance explained, which is bounded between 0 and 1. The key difference is that $\hat{R}_w^2$ is the proportion of variance explained \textit{in the weighted empirical distribution}. 
Furthermore, in the case of two scalar random variables, $\hat{R}_w^2$ is the square of their weighted correlation, i.e., $\hat{R}_w^2 (B \sim A ) = [ \hat{R}_w (B \sim A) ] ^2 $. Finally, note that the traditional sample statistics follow from those above (up to a degrees of freedom adjustment) when all $w_i = 1$ (e.g., $\frac{1}{n} \sum_{i=1}^n X_i = \widehat{\E}_w (X)$ when all $w_i=1$). We therefore omit the $w$-subscript to refer to them (e.g., $\widehat{\E} (X) = \frac{1}{n} \sum_{i=1}^n X_i$).


\section{Sensitivity analysis tools}\label{sec:wsa_results}

We now develop sensitivity tools for $\hat{\tau}_{\mathrm{wls}}$. 
These tools are mostly weighted generalizations of those from C\&H, who assess the sensitivity of $\hat{\tau}_{\mathrm{ols}}$ (from Expression~\ref{eq:ols_ydx}) from an omitted variable bias (OVB) perspective. 
However, two novel contributions are required.  
First, we offer three options for adjusted inference that allow for heteroscedasticity, two of which involve bootstrap procedures. Second, because $D$ and $X$ are often left uncorrelated in the weighted distribution, a strict generalization of C\&H's method for benchmarking 
the strength of unobserved confounding with (a multiple of) that of $X$ is infeasible. We thus develop a novel benchmarking 
approach that is robust to the case where $D$ and $X$ have zero weighted correlation by appealing to a \textit{semi}-weighted distribution, where ``semi-weights" leave correlation between $D$ and one (or several) of the covariates.\footnote{Because of these novel contributions, our sensitivity tools \textit{cannot} be replicated in practice with C\&H's \href{https://cran.r-project.org/web/packages/sensemakr/index.html}{\texttt{sensemakr}} package by simply transforming the data. For example, it is well-known that multiplying each $Y_i$, $D_i$, and $X_i$ (and an intercept term) by the corresponding $\sqrt{w_i}$ and performing a standard \textit{un}weighted least squares would recover the point estimate, $\hat{\tau}_{\mathrm{wls}}$. However, feeding the resulting model into the \texttt{sensemakr} package would \textit{not} yield the same benchmarking bounds or inference-related results. Thus, our \href{https://github.com/lwainstein/sensewls}{\texttt{sensewls}} package is required to implement our tools.} 

\subsection{Identification and specification bias}\label{subsec:defining_bias}

In order to conduct sensitivity analyses for $\hat{\tau}_{\mathrm{wls}}$,
an expression for its 
bias is required. However, we first consider the conditions under which it 
may show bias. Identification of the causal effect of $D$ on $Y$ hinges on the assumption that conditioning on $X$ is sufficient to eliminate all confounding in the relationship between $D$ and $Y$, often referred to as ``conditional ignorability", or the ``no unobserved confounding" assumption (e.g., \citealp{rosenbaum1983central}),
	\begin{assumption}[No Unobserved Confounding]\label{asm:ci}
        		$Y (d) \ \indep \ D \ | \ X$
	\end{assumption}
Informally, Assumption~\ref{asm:ci} states that ``accounting" for $X$ is sufficient to unbiasedly estimate the desired causal effect. $\hat{\tau}_{\mathrm{wls}}$ attempts to do this with its weights and by modeling $\E[Y (d) \ | \ X]$.  There are two sources of bias to consider: specification and identification. First, even if  Assumption~\ref{asm:ci} holds, we might mispecify the relationships between $X$, $D$, and $Y$. This involves producing incorrect weights (e.g., using inverse propensity score weights that misspecify $\pi (X)$), or mis-modeling $Y$ given $X$ and $D$ (e.g., using a linear model, when $Y$ is nonlinear in $X$ and $D$). The second source of bias is an identification concern: Assumption~\ref{asm:ci} may not hold, and thus accounting for $X$---even if done correctly---is insufficient for unbiased or consistent estimation. 

One way to attack both biases is through the omitted variable bias approach: consider the existence of an unobserved variable, $Z$ with $\mathbf{Z} = [Z_1 \ \dots \ Z_n]^{\top}$, such that accounting for $Z$ would eliminate bias by correcting the identification or specification error. Had $Z$ been observed, it could in principle prompt two alterations to $\hat{\tau}_{\mathrm{wls}}$: (i) choosing weights that involve $Z$ in addition to $X$, and (ii) estimating $\E[Y (d) \ | \ X, Z]$ instead of $\E[Y (d) \ | \ X]$. Several existing methods for sensitivity analyses with weights have focused on how causal estimates change after the first of these (e.g., \citealp{shen2011sensitivity}; \citealp{hong2020did}), but we focus exclusively on the latter. In other words, \textit{we leave the weights unchanged even though they are not expected to properly account for both $X$ and $Z$}, and instead consider how estimates would change were one to estimate  $\E[Y (d) \ | \ X, Z]$. Concretely, we consider a generalization of the weighted regression that yields $\hat{\tau}_{\mathrm{wls}}$ in Expression~\ref{eq:wls_ydx}:
	\begin{align}\label{eq:wls_ydxz}
		( \hat{\mu}_{\mathrm{target}}, \hat{\tau}_{\mathrm{target}}, \hat{\beta}_{\mathrm{target}}, \hat{\gamma}_{\mathrm{target}}  ) = \underset{\mu, \tau, \beta, \gamma}{\mathrm{argmin}} \ \frac{1}{n} \sum_{i=1}^n w_i \biggr( Y_i - ( \mu + \tau D_i + X_i^{\top} \beta + \gamma Z_i) \biggr)^2
	\end{align}
where $Z$ has been added as a regressor, and $\hat{\tau}_{\mathrm{target}}$ is the adjusted estimate of the causal effect of $D$ on $Y$. We then define the bias of $\hat{\tau}_{\mathrm{wls}}$ as: 
    \begin{align}\label{eq:bias_function}
        \widehat{\mathrm{bias}} ( \hat{\tau}_{\mathrm{wls}} )= \hat{\tau}_{\mathrm{wls}} - \hat{\tau}_{\mathrm{target}} 
    \end{align}
As shown in Section~\ref{subsubsec:wsa_bias}, $\hat{\tau}_{\mathrm{target}}$ is defined by $Z$'s in-sample relationships with $D$ and $Y$. Thus, with $Z$ unknown, our tools vary these two relationships to assess the sensitivity of 
conclusions from $\hat{\tau}_{\mathrm{wls}}$. 

We recognize that our choice of focusing only on the regression's sensitivity is counterintuitive in the sense that, were $Z$ observed, one would certainly use it to adjust the weights, but $\hat{\tau}_{\mathrm{target}}$ employs the original weights that we suspect have yielded a biased $\hat{\tau}_{\mathrm{wls}}$. 
However, we emphasize the generality that this choice allows. First and foremost, the weights are left arbitrary in $\hat{\tau}_{\mathrm{target}}$, implying that \textit{our tools apply for any choice of (non-negative) weights.} It thus applies to inverse propensity score weighting, balancing weights of any kind, matching, or sub-classification/stratification estimators. This is a key advantage over most sensitivity analysis procedures in the literature that involve weights (e.g., \citealp{shen2011sensitivity, hong2020did, mccaffrey2004propensity, soriano2021interpretable, zhao2017sensitivity, huang2024sensitivity, huang2024variance, hartman2024sensitivity, 10.1162/REST.a.1705}), and the motivation for developing this approach. Second, beyond the requirement that $Z$ does not render $\hat{\tau}_{\mathrm{target}}$ non-unique or undefined (e.g., by collinearity with $X$), \textit{our tools do not require any distributional assumptions on $Z$}.\footnote{See examples of such assumptions in \cite{vanderweele2011unmeasured},  \cite{ichino2008temporary}, \cite{carnegie2016assessing}, or \cite{huang2020sensitivity}.} Relatedly, this brings us to the ability of this approach to address misspecification bias: $Z$ need not only be an unobserved confounder, but could instead be a function of $X$ that addresses misspecification of the relationship between $Y$ and $X$ (and $D$) inherent in the linear model, or misspecification of the relationship between $D$ and $X$ inherent in the weights that induces bias (i.e., also relates to $Y$).  In fact, a $Z$ can be defined such that $\hat{\tau}_{\mathrm{target}}$ simultaneously corrects for bias from both unobserved confounding (identification bias) and influential omitted non-linear functions (specification bias). We show this concretely in Section~\ref{subsec:multiple_bias}.
\subsection{Sensitivity of the point estimate, and the sensitivity parameters}\label{subsubsec:wsa_bias}
Through the omitted variable bias framework, the bias of $\hat{\tau}_{\mathrm{wls}}$ decomposes as
	\begin{align}\label{eq:wsa_bias}
		\widehat{\mathrm{bias}} (\hat{\tau}_{\mathrm{wls}}) = \frac{ \hat{R}_w (Y \sim Z | D, X) \times \hat{R}_w (D \sim Z | X)}{\sqrt{1 - {} \hat{R}^2_w (D \sim Z | X) }} \times \frac{\widehat{\sd}_w (Y^{\perp_w X, D})}{\widehat{\sd}_w (D^{\perp_w X})}
	\end{align} 
where $\hat{R}_w (Y \sim Z | D, X)$ and $\hat{R}_w (D \sim Z | X)$ are unknown, but are freely varying in both magnitude and sign. See Appendix~\ref{app:proofs_wsa_bias} for proof. $\hat{R}_w^2 (Y \sim Z | D, X)$ and  $\hat{R}_w^2 (D \sim Z | X)$ therefore determine the bias of $\hat{\tau}_{\mathrm{wls}}$ --- a favorable result, as these values are 
intuitive (i.e., the proportion of leftover weighted variance in $Y$ and $D$ that $Z$ explains after controlling for $D$ and $X$), and are bounded between 0 and 1. Therefore, $\hat{R}^2_w (Y \sim Z | D, X)$ and  $\hat{R}^2_w (D \sim Z | X)$ will henceforth be 
referred to 
as the ``sensitivity parameters".\footnote{Note that because both of these sensitivity parameters are partial $\hat{R}^2_w$ conditional on (at least) $X$, one may assume without loss of generality that $\hat{R}^2_w (Z \sim X) = 0$. In practice, this means that when postulating values for the sensitivity parameters, one should consider the parts of $Z$ that are \textit{not} already explained by $X$.}

The sensitivity parameters in Expression~\ref{eq:wsa_bias} are exact, in-sample quantities: if known, plugging them in would recover $\hat{\tau}_{\text{target}}$ exactly. In practice they are unknown, and investigators instead reason about plausible values for them---reasoning that is naturally framed in population rather than sample terms. This leaves a gap between the finite-sample values the bias formula actually uses and the population values the investigator is contemplating. We regard this gap as immaterial. It is small at any reasonable sample size; and more importantly, the plausibility ranges entertained in sensitivity analysis are necessarily coarse and conservative, with ample slack to absorb such finite-sample deviations.


\subsection{Sensitivity of the standard error}\label{subsubsec:wsa_se}
 
Variance estimators in this setting vary along two conceptual axes. The first is the \textit{resampling design}. Under a \textit{fixed design}, or model-based approach, $(\mathbf{X}, \mathbf{D}, \mathbf{Z})$ (and thus any honest weights $w$) are treated as fixed and variation arises only from $\mathbf{Y} \ | \ \mathbf{X}, \mathbf{D}, \mathbf{Z}$. Under a \textit{random design}, or super-population framework, the units themselves are hypothetically redrawn from a larger super-population, so that $(\mathbf{X}, \mathbf{D}, \mathbf{Z})$ (and the resulting weights) also vary. 

The second axis  regards the \textit{target variance estimand} relevant for sensitivity analysis: the variance of the \textit{adjustment procedure}---how $\hat{\tau}_{\mathrm{target}}$, at fixed sensitivity parameter values, would vary under resampling, versus the variance of the \textit{implied coefficient}, i.e., the standard error the regression would have produced had a $Z$ characterized by the postulated $\hat{R}^2_w$ values actually been included. The latter is more in keeping with the OVB analysis built on the question of how the estimate would look if we were able to include the omitted $Z$. It is the approach taken by C\&H, and it behaves qualitatively differently from the adjustment procedure target: the estimated variance of the implied coefficient can shrink when $\hat{R}^2_w (Y \sim Z \mid D, X)$ is large because $Z$ would have absorbed residual variation in $Y$, and can also inflate when $\hat{R}^2_w (D \sim Z \mid X)$ is large because including $Z$ leaves less variation in $D$ after partialing out $(X, Z)$. Unfortunately, the choice between these variance estimands is not entirely a free one: complications arise due to approximations required to identify these variances. While the implied-coefficient target may be conceptually preferred, its estimation engages further assumptions, which can be problematically violated.\footnote{While C\&H provide a closed-form standard error for the variance of the implied coefficient in an unweighted regression under spherical errors, this approach becomes problematic with weights because the variance of $\hat{\tau}_{\mathrm{target}}$ involves squared weights that are not compatible with the sensitivity parameters. It is easily shown that with a spherical error assumption that $\var(Y \ | \ X, D, Z) = \sigma^2$,
 \begin{align*}
     \var( \hat{\tau}_{\mathrm{target}} \ | \ X, D, Z) = \frac{\sigma^2}{n} \biggr( \widehat{\var} (D^{\perp_w X, Z} ) \biggr)^{-2} \biggr( \frac{1}{n} \sum_{i=1}^n w_i^2 (D_i^{\perp_w X, Z})^2 \biggr)
 \end{align*}
The right-most term in the above expression is (proportional to) a sample variance in the $w^2$-weighted distribution, which would require additional claims beyond the $\hat{R}^2_w$ sensitivity parameters. A solution is possible with a $\sqrt w$-weighted transformation of the data, but this introduces heteroskedasticity. Addressing this requires assumptions, described in our inference Option 3.  
} 

Considering these varied goals and tradeoffs, we propose three adjusted inference options for investigators, which occupy three points along the axes above. 
We outline them briefly here, saving extensive derivation and discussion to Appendix~\ref{app:variance_choices}. Appendix~\ref{subsec:se_extension} also describes extensions to these inference options that account for clustering, which we utilize in our application in Section~\ref{sec:app1} due to the nature of the data.

\paragraph*{Option 1: Pairs bootstrap with weight re-estimation.} This option targets the adjustment procedure's variance under the random design/superpopulation framework. Draw $B$ bootstrap samples, each consisting of $n$ tuples $(D_i, X_i, Y_i)$ drawn with replacement. For each iterate, re-estimate the weights and use them to construct the values required for adjustment: $\hat{\tau}_{\mathrm{wls}}$, $\widehat{\sd}_w (Y^{\perp_w X, D})$, and $\widehat{\sd}_w (D^{\perp_w X})$. For fixed postulated values of the sensitivity parameters, use Expression~\ref{eq:wsa_bias} to calculate $\hat{\tau}_{\mathrm{target}}$. The standard error $\widehat{\se}_{\mathrm{boot}} (\hat{\tau}_{\mathrm{target}}) = \widehat{\sd} ( \{ \hat{\tau}_{\mathrm{target}}^{(1)}, \dots, \hat{\tau}_{\mathrm{target}}^{(B)} \} )$ can be reported or used to construct normal-approximation confidence intervals; alternatively, confidence intervals can be formed directly from the desired quantiles of the bootstrap distribution.

\paragraph{Option 2: Fixed weights with bootstrap.} This is a hybrid approach offering a conservative estimate for the variance of the adjustment procedure under a fixed design and honest weights. Taking interest in the fixed design result, we can regard the honest $w$ as fixed since they are a function of only $\textbf{X}$ and $\textbf{D}$. We then conduct a pairs bootstrap over the sample as in Option 1, but now drawing the corresponding $w_i$ along with each tuple $(D_i, X_i, Y_i)$, and then applying those fixed, resampled weights instead of re-estimating them for each iterate.

\paragraph{Option 3: Closed-form implied-coefficient standard error for a fixed design.} If we wish to see the variance of the coefficient as if estimated in the presence of $Z$, we can apply a closed-form estimator akin to that of C\&H. However, this must be done on the $\sqrt{w}$-transformed data, which forces us to consider heteroskedasticity-consistent (HC) standard errors. This creates a new challenge: unlike the spherical result, HC0 standard errors (\citealp{white1980heteroskedasticity}) cannot be computed without additional assumptions on how the inclusion of $Z$ would have changed the pointwise residuals. We thus invoke the following:
\begin{assumption}[Uniform Shrinkage]\label{asm:uniform_shrinkage}
For all $i$, the following statements hold:
\begin{enumerate}
    \item $(Y_i^{\perp_w X, D, Z})^2 \approx (Y_i^{\perp_w X, D})^2 * \big( 1 - \hat{R}^2_w (Y \sim Z | X, D) \big)$
    \item $(D_i^{\perp_w X, Z})^2 \approx (D_i^{\perp_w X})^2 * \big( 1 - \hat{R}^2_w (D \sim Z | X) \big)$
\end{enumerate}
\end{assumption}
\noindent That is, each unit's squared residual in the weighted outcome (respectively, treatment) regression shrinks by a uniform proportion---determined by the relevant sensitivity parameter---when $Z$ is introduced. These statements hold \textit{on average} in the weighted distribution by algebra (see Appendix~\ref{subsec:se_closed}); Assumption~\ref{asm:uniform_shrinkage} strengthens this to hold approximately at the \textit{unit} level. Under it, a closed-form HC0 standard error for the sensitivity-adjusted coefficient can be computed via the $\sqrt{w}$-transformed regression:
\begin{align}\label{eq:adjusted_hc0}
    \widehat{\se}_{\text{adj-}\mathrm{HC0}} ( \hat{\tau}_{\mathrm{target}}) \approx \sqrt{ \frac{1 - \hat{R}^2_w (Y \sim Z | X, D)}{1 - \hat{R}^2_w (D \sim Z | X )} } \times \widehat{\se}_{\mathrm{HC0}} ( \hat{\tau}_{\mathrm{wls}})
\end{align}
Proof is given in Appendix~\ref{app:proofs_wsa_se}, where we also extend this idea to other HC standard errors (e.g., HC1). 

\paragraph{Discussion.} The choice among these three options involves a number of tradeoffs. Option 1 corresponds neatly to the random design or super-population view and requires few additional assumptions. Accordingly, Appendix~\ref{app:bootstrap}  demonstrates Option 1's validity empirically in a random design scenario with smooth weights.  However, this option can be computationally costly due to re-estimation of the weights on each bootstrap iterate. Option 2 targets the fixed design analog of the same adjustment procedure quantity as does Option 1, with all randomness arising from $\mathbf{Y} \ | \ \mathbf{X}, \mathbf{D}, \mathbf{Z}$ rather than the sampling of different units, and is computationally simpler because the weights only need to be estimated once. However, combining the fixed design logic with bootstrap resampling makes Option 2 conservative for the true standard error under a fixed design: the bootstrap introduces variation in $(\mathbf{X}, \mathbf{D})$ that is absent in the fixed design truth, and the overstatement grows with the postulated $\hat{R}^2_w (Y \sim Z \mid X, D)$ because the unobserved $Z$ is necessarily folded into noise. We demonstrate this through simulation in Appendix~\ref{app:fixed_bootstrap}.

Option 3 is the only option that targets the uncertainty in the implied coefficient had $Z$ been included in the regression, and can therefore be smaller or larger than the adjustment procedure-level options depending on whether residual reduction in $Y$ (through $\hat{R}^2_w (Y\sim Z | X, D)$) or $D$ (through $\hat{R}^2_w (D\sim Z | X)$) 
dominates. However, its accuracy hinges on Assumption~\ref{asm:uniform_shrinkage}. In Appendix~\ref{subsec:se_closed} we explain mathematically when this assumption fails.
Then, in Appendix~\ref{app:fixed_bootstrap}, we present settings in which Option 3 meaningfully overestimates or underestimates the standard error of $\hat{\tau}_{\mathrm{target}}$ 
because the per-unit shrinkage approximation in Assumption~\ref{asm:uniform_shrinkage} breaks down.

Finally, when weights are chosen by matching, Option 1 is worrying as bootstrap resampling can generate non-smoothness in the weights and invalid uncertainty intervals \citep{abadie2008failure}. Thus, Options 2 and 3 are recommended with matching weights.
\subsection{Sensitivity statistics: robustness values and extreme scenarios}\label{subsubsec:wsa_stats}
A contour-plot with the two sensitivity statistics as axes can plot 
$\hat{\tau}_{\mathrm{target}}$ on the contours (or its standard error, boundaries of the confidence interval, or p-values), fully characterizing how one's results would change according to the strength of hypothetical unobserved confounding.  But for ease-of-use and standardized reporting, C\&H also define summary statistics that more succinctly characterize the types of $Z$ that would alter one's conclusions: ``robustness values" and $\hat{R}^2 (Y \sim D | X)$ as an extreme scenario. Both are easily translated to the weighted setting.

\paragraph{Robustness values.}
Were $Z$ to explain equal leftover weighted variance in $Y$ and $D$ (i.e., were the sensitivity parameters equal), robustness values (RV) quantify how strong $Z$ would need to be to (i) reduce the estimated effect by $(100 \times q)\%$, for some $q$, or (ii) render $\hat{\tau}_{\mathrm{target}}$ insignificant at the $\alpha$ level, for some $\alpha$. Starting with the former, a $Z$ that were to explain
	\begin{align}\label{eq:wsa_rvq}
		\mathrm{RV}_q = \frac{1}{2} \biggr( \sqrt{ \omega^4_q + 4 \omega^2_q} - \omega_q^2 \biggr) \ \ \ \text{where} \ \ \ \omega_q = q \times \biggr| \frac{\hat{R}_w(Y \sim D | X)}{\sqrt{1 - \hat{R}_w^2(Y \sim D | X)}} \biggr|
	\end{align}
of the remaining weighted variation in $Y$ and $D$ would reduce $\hat{\tau}_{\mathrm{wls}}$ 
by $(100 \times q)\%$. Then, we define $\mathrm{RV}_\alpha$ to be the minimum value of the sensitivity parameters that renders $\hat{\tau}_{\mathrm{target}}$ insignificant at the $\alpha$ level, which we define as the corresponding $100\times (1 - \alpha)$\% confidence interval including 0. Note that $\mathrm{RV}_\alpha$ changes depending on the inference procedure chosen from Section~\ref{subsubsec:wsa_se}. Naturally, the RVs that reduce the estimate to 0 (i.e., $\mathrm{RV}_{q=1} $) or render $\hat{\tau}_{\mathrm{target}}$ insignificant at the 0.05 level (i.e., $\mathrm{RV}_{\alpha=0.05} $) are useful statistics. See Appendix~\ref{app:proofs_wsa_rv} for the derivation of $\mathrm{RV}_q$.

\paragraph{Extreme scenarios.}
Second, in the extreme scenario where $Z$ explains the remaining weighted variation in $Y$ (i.e., $\hat{R}_w^2 (Y \sim Z| D, X) = 1$), a $Z$ would be strong enough to bring $\hat{\tau}_{\mathrm{target}}$ to 0 if $\hat{R}^2_w(D \sim Z | X) = \hat{R}^2_w(Y \sim D | X)$. See Appendix~\ref{app:proofs_wsa_extreme} for proof. Therefore, $\hat{R}^2_w (Y \sim D | X)$ is another useful diagnostic, analogous to the result of one-parameter sensitivity analyses that make such a worst-case assumption on the relationship of confounding with the outcome (e.g. \citealp{rosenbaum1987sensitivity}).

\subsection{Benchmarking 
$\hat{R}_w^2 (Y \sim Z | D, X)$ and $\hat{R}_w^2 (D \sim Z | X)$ using observed covariates}\label{subsubsec:wsa_benchmark}

We 
demonstrate here how to benchmark a $Z$'s strength by comparing it to that of a chosen covariate, $X^{(j)}$, extending the benchmarking tools from C\&H. 
Informally, our benchmarking tools allow one to entertain a $Z$ that is ``as strong" or ``multiple times as strong" as is $X^{(j)}$ in its relationships with $D$ and $Y$. The researcher may then use these benchmarks to determine if such a $Z$ would change one's conclusions. For example, if one hypothesizes that $X^{(j)}$ is stronger than any potential unobserved confounding, and a $Z$ as strong as $X^{(j)}$ fails to switch the sign of $\hat{\tau}_{\mathrm{wls}}$ or render the estimate insignificant at the 0.05 level, then the conclusions are robust to unobserved confounding under those assumptions. 

Such an exercise requires a formal definition of the relative strength of $Z$ in relation to that of $X^{(j)}$. First, let $X^{(-j)}$ be the remainder of the covariates after removing $X^{(j)}$ from $X$. Then, define $w_i^{(-j)}$ to be ``semi-weights", formed by the same process as are $w_i$, but using only $X^{(-j)}$.\footnote{What is included in the ``process" of forming the weights is ultimately at the researcher's discretion. However, note that, as will be discussed below, the main goal of the semi-weights is to leave imbalance in the held-out covariate, $X^{(j)}$. Thus, it is often sufficient to only treat the ``process" of forming weights as the final function that inputs $\mathbf{X}$ and $\mathbf{D}$ (and potentially $\mathbf{Y}$) and outputs a vector of weights. For example, in a matching setting where balance on all covariates is infeasible, the researcher may have gone through multiple rounds of matching and balance checks (that are potentially automated) until deciding on a subset of covariates to match on, of which $X^{(j)}$ is one of them. Here, to form semi-weights we would simply match on that same subset of covariates \textit{after removing} $X^{(j)}$. We would \textit{not} repeat the iterative matching and balance checks holding out $X^{(j)}$ from the very start, even though that could potentially arrive at a different subset of covariates to match on. }
Further, when $X$ is one-dimensional (i.e., $X = X^{(j)}$), semi-weights are simply uniform weights (i.e., all $w_i^{(-j)} = 1$). Then, let the relative strength of $Z$ be defined by
	\begin{align}\label{eq:wsa_kappa}
		\kappa_{w/w^{(-j)}} (D) := \frac{ \hat{R}_w^2 (D \sim Z | X^{(-j)})}{ \hat{R}_{w^{(-j)}}^2 (D \sim X^{(j)} | X^{(-j)})} \ \ \ \text{and} \ \ \ \kappa_w (Y) := \frac{ \hat{R}_w^2 (Y \sim Z | D, X^{(-j)})}{ \hat{R}_w^2 (Y \sim X^{(j)} | D, X^{(-j)})}
	\end{align}
In words, $\kappa_w (Y)$ is the weighted variance in $Y$ that $Z$ explains (given $D$ and $X^{(-j)}$), compared to what $X^{(j)}$ explains (also given $D$ and $X^{(-j)}$). This  quantifies how much better (or worse) $Z$ is than is $X^{(j)}$ at predicting $Y$. For example, if $\kappa_w (Y)=1$, then $Z$ may be thought of as being ``as strong" as $X^{(j)}$ in its relationship with $Y$. 

Similarly, the term $\kappa_{w/w^{(-j)}} (D)$ in Expression~\ref{eq:wsa_kappa} describes how much stronger (or weaker) $Z$ is than $X^{(j)}$ in terms of its relationship with $D$. However, this term is complicated by the switch between the full weights ($w$) and the semi-weights ($w^{-j}$) in the denominator. To further investigate this, consider first the alternative choice akin to $\kappa_w (Y)$, \begin{align}\label{eq:wsa_kappad_maybe}
		\kappa_{w} (D) = \frac{ \hat{R}_w^2 (D \sim Z | X^{(-j)})}{ \hat{R}_{w}^2 (D \sim X^{(j)} | X^{(-j)})}
	\end{align}
The problem with this, however, is that the weighting procedure will make $\hat{R}_w^2 (D \sim X) \approx 0$ (and thus $ \hat{R}_{w}^2 (D \sim X^{(j)} | X^{(-j)}) \approx 0$ in the denominator of Expression~\ref{eq:wsa_kappad_maybe}) when the weights effectively render the treated and control groups similar on $X$. For example, $\hat{R}_w^2 (D \sim X) = 0$ when the weighted means of $X$ are equal in the treated and control groups (as achieved by the balancing weights in Expression~\ref{eq:ebal}). This quantity is then not useful when reasoning about how strong unobserved confounding relates to treatment compared to observables, since the quantity the user must instead reason about---reflecting the influence of that $X^{(j)}$ on $D$---refers to relationships in the unweighted data, in which that relationship has not been destroyed by weighting. 
Thus we employ $\kappa_{w/w^{(-j)}} (D)$ in Expression~\ref{eq:wsa_kappa}, which exchanges the denominator of $\kappa_{w} (D)$ in Expression~\ref{eq:wsa_kappad_maybe} with its analog in the semi-weighted distribution, because while $D$ and $X^{(-j)}$ may be uncorrelated in the semi-weighted distribution, $D$ and $X^{(j)}$ are likely still correlated. 

Next, rewriting $\kappa_{w/w^{(-j)}} (D)$ as the product of two terms illuminates its meaning:
    \begin{align}\label{eq:wsa_kappa_rewrite}
       \kappa_{w/w^{(-j)}} (D) =  \underbrace{ \biggr[ \frac{ \hat{R}_{w^{(-j)}}^2 (D \sim Z | X^{(-j)})}{ \hat{R}_{w^{(-j)}}^2 (D \sim X^{(j)} | X^{(-j)})} \biggr]}_{\text{``semi-strength"}} \times \underbrace{ \biggr[ \frac{ \hat{R}_w^2 (D \sim Z | X^{(-j)})}{ \hat{R}_{w^{(-j)}}^2 (D \sim Z | X^{(-j)})} \biggr] }_{\text{``translator"}}
    \end{align}
The first term in Expression~\ref{eq:wsa_kappa_rewrite}, or the ``semi-strength", describes the predictive strength of $Z$ in relation to that of $X^{(j)}$ in the semi-weighted distribution. The second term in Expression~\ref{eq:wsa_kappa_rewrite}, or the ``translator", then translates $Z$'s predictive power in the semi-weighted distribution to that in the weighted distribution. Thus, by thinking of the multiplication in Expression~\ref{eq:wsa_kappa_rewrite} as the translator converting the relative strength of $Z$ in the semi-weighted distribution to that in the weighted distribution, $\kappa_{w/w^{(-j)}} (D)$ captures the strength of $Z$ relative to that of $X^{(j)}$.\footnote{We have also considered maximizing $\hat{R}_w (D \sim Z |X)$ over $Z$ given a constraint on the semi-strength in Expression~\ref{eq:wsa_kappa_rewrite} (e.g., $\text{semi-strength} \leq 2$). This obviates the need to consider the translator term from Expression~\ref{eq:wsa_kappa_rewrite}. However, these bounds quickly become too large to be useful.}

It is tempting to assume the translator is 1, and only consider the semi-strength in Expression~\ref{eq:wsa_kappa_rewrite}. However, the translator can be large when there are large differences between the weighted and semi-weighted distributions. For example, Appendix~\ref{app:translator} demonstrates a setting where the translator is over 7. While the data generating process required here is extreme, it is still instructive: in settings where the weighted and semi-weighted distributions are very different, one should entertain larger $\kappa_{w/w^{(-j)}} (D)$ than they might otherwise. In particular, the researcher should look out for instances where the weighted distribution is putting much more emphasis on certain values of $X$ than is the semi-weighted distribution. This can happen easily when multiple units' $w_i$ have become extreme (i.e., large or near zero) when the corresponding $w_i^{(-j)}$ were not previously, or vice-versa. Another sign is when the Effective Sample Size (ESS) of the weights,
	\begin{align}\label{eq:ess}
		\mathrm{ESS} (w) = \frac{(\sum_{i=1}^n w_i)^2}{\sum_{i=1}^n w_i^2}
	\end{align}
is significantly smaller (or larger) than the ESS of the semi-weights.\footnote{While weighted regressions consider all $n$ units when the $w_i > 0$, units are nearly discarded when their $w_i$ are close to 0. ESS is one way to estimate how many units a weighted regression meaningfully incorporates. We discuss ESS more thoroughly in Appendix~\ref{subsec:w_distr_normalize}, where we also recommend normalizing all weights so that the overall ESS reflects the ESS in the treatment and control groups.} We provide guiding examples of this in Section~\ref{sec:app1}.

Finally, the purpose in postulating values for $\kappa_{w/w^{(-j)}} (D)$ and $\kappa_w (Y)$ is that they imply bounds on the sensitivity parameters,
	\begin{align}\label{eq:wsa_benchmarkd}
		\ \ \ \ \ \hat{R}_w^2 (D \sim Z | X) &= \kappa_{w/w^{(-j)}} (D) \times \frac{\hat{R}_{w^{(-j)}}^2 (D \sim X^{(j)} | X^{(-j)})}{1 - \hat{R}_{w}^2 (D \sim X^{(j)} | X^{(-j)})}
	\end{align}
	\vspace{-0.3in}
	\begin{align}\label{eq:wsa_benchmarky}
		\hat{R}_w^2 (Y \sim Z | D, X) &\leq \eta_{w/w^{(-j)}}^2 \times \frac{\hat{R}_w^2 (Y \sim X^{(j)} | D, X^{(-j)})}{1 - \hat{R}_w^2 (Y \sim X^{(j)} | D, X^{(-j)})}
	\end{align}
where $\eta_{w/w^{(-j)}}^2$ is a function of $\kappa_{w/w^{(-j)}} (D)$ and $\kappa_w (Y)$. Proof is given in Appendix~\ref{app:proofs_wsa_benchmark}, where we also extend these bounds to allow researchers to benchmark the strength of $Z$ using \textit{multiple} covariates. We also note that were the weights and semi-weights set to uniform weights, these results are equivalent to the bounds on the sensitivity parameters in C\&H. 

Researchers may set the sensitivity parameters equal to the bounds in Expressions~\ref{eq:wsa_benchmarkd} and~\ref{eq:wsa_benchmarky} to directly translate their statements of the relative strength of $Z$ (with $\kappa_{w/w^{(-j)}} (D)$ and $\kappa_{w} (Y)$) into adjusted estimates and inference. Note that while the result for $\hat{R}_w^2 (D \sim Z | X)$ is already an equality, we further suggest setting $\hat{R}_w^2 (Y \sim Z | D, X)$ to be equal to the upper bound in Expression~\ref{eq:wsa_benchmarky} for two reasons. First, the inequality becomes an equality when $\hat{R}_w^2 (D \sim X)=0$, i.e., the weights equate the means of $X$ in the treated and control groups. Second, even if $\hat{R}_w^2 (D \sim X) \neq0$, the bound is sharp in that in any sample, a $Z$ vector can always be chosen such that the inequality becomes an equality. 

%

\subsection{Allowing $Z$ to encompass multiple sources of bias}\label{subsec:multiple_bias}

Although we have treated $Z$ as univariate to this point, $Z$ can encompass more than just a single unobserved confounder, and can even adjust for misspecification of the relationships between $X$, $D$, and $Y$. To show this, we adapt an analogous result from C\&H (see Section 4.5) to the weighted setting. Let $\tilde{Z}$ be a \textit{vector} of omitted variables that we wished we had included as regressors in the weighted regression on $Y$. This could also include functions of $X$ that we mistakenly omitted from the initial weighted regression, or interactions of (centered) omitted variables with $D$ as in the estimator studied by \cite{lin_agnostic_2013}. Were all of $\tilde{Z}$ observed, we would have ideally estimated the following model:
    \begin{align}\label{eq:multivariateZ_wls}
		( \hat{\mu}, \hat{\tau}, \hat{\beta}, \hat{\phi}  ) = \underset{\mu, \tau, \beta, \phi}{\mathrm{argmin}} \ \frac{1}{n} \sum_{i=1}^n w_i \biggr( Y_i - ( \mu + \tau D_i + X_i^{\top} \beta + \tilde{Z}_i^{\top} \phi) \biggr)^2
    \end{align}
where $\hat{\tau}$ is the adjusted estimate, and $\hat{\phi}$ is the estimated vector of corresponding coefficients for $\tilde{Z}$. Letting $Z \equiv \tilde{Z}^{\top} \hat{\phi}$ in Expression~\ref{eq:wls_ydxz}, the $\hat{\tau}$ in Expression~\ref{eq:multivariateZ_wls} would  exactly equal $\hat{\tau}_{\mathrm{target}}$. 

At first glance, the clever choice of $Z\equiv \tilde{Z}^{\top} \hat{\phi}$ makes the sensitivity parameters less intuitive, and one might wish they could instead reason about $\hat{R}^2_w (Y \sim \tilde{Z} | D, X)$ and $\hat{R}^2_w (D \sim \tilde{Z} | X)$. However, we propose that one simply does reason about these $\hat{R}_w^2$ values with $\tilde{Z}$, and then treats them as being equal to the sensitivity parameters. As shown in C\&H, the resulting sensitivity analysis is guaranteed to be conservative, so long as the investigator is reasoning about how much of $D$ can be explained by $\tilde{Z}$ (given $X$ and $D$) in any linear combination. This is because, first, $\hat{R}^2_w (Y \sim \tilde{Z} | D, X) = \hat{R}^2_w (Y \sim Z | D, X)$ when $Z \equiv \tilde{Z}^{\top} \hat{\phi}$. Second, $\hat{R}^2_w (D \sim \tilde{Z} | X) \geq \hat{R}^2_w (D \sim Z | X)$ when $Z \equiv \tilde{Z}^{\top} \hat{\phi}$, because the $\tilde{Z}^{\top} \hat{\phi}$ generating bias can explain no more of $D$ than the maximum over the linear span of $\tilde{Z}$. The consequent bias created by the omission of $\tilde{Z}$ must therefore be no larger than the bias that would be generated by confounding of the strength postulated.

This fact also strengthens the justification for our general approach of neglecting how the weights might change were $Z$ observed. One could imagine that $\tilde{Z}$ includes a set of variables sufficient to allow $\hat{\tau}$ in Expression~\ref{eq:multivariateZ_wls} above to be unbiased for the estimand of interest.\footnote{This is certainly possible because a univariate $Z$ can always be defined such that $\hat{\tau}_{\mathrm{target}}$ is unbiased for the target estimand when $D$ is binary. Consider a setting in which Assumption~\ref{asm:ci} may not hold, but conditioning on $( X, \tilde{Z})$ for some vector of unobserved confounders $\tilde{Z}$ is sufficient to achieve the desired conditional independence: $Y (d) \ \indep \ D \ | \ X, \tilde{Z}$. Then, if the ATE is the target estimand, for example, defining
	\begin{align*}
		Z = 
		\begin{cases} 
			\E[ Y (0) \ | \ X, \tilde{Z} ] & \text{if} \ D = 0 \\
			\E[ Y (1) \ | \ X, \tilde{Z} ] - \mathrm{ATE} & \text{if} \ D = 1
		\end{cases}
	\end{align*}
yields $\E( \hat{\tau}_{\mathrm{target}} ) = \mathrm{ATE}$, assuming the $w_i$ are entirely defined by $\mathbf{D}$ and $\mathbf{X}$. This follows because, 
	\begin{align*}
		\mu + \tau D + X^{\top} \beta + \gamma Z = 
		\begin{cases} 
			\gamma \E[ Y (0) \ | \ X, \tilde{Z} ] + (\mu + X^{\top} \beta) & \text{if} \ D = 0 \\
			\gamma \E[ Y (1) \ | \ X, \tilde{Z} ] + (\mu + X^{\top} \beta) + (\tau - \gamma \mathrm{ATE})    & \text{if} \ D = 1
		\end{cases}
	\end{align*}
and since $Y (d) \ \indep \ D \ | \ X, \tilde{Z}$,
	\begin{align*}
		\E[ Y \ | \ D, X, Z] = 
		\begin{cases} 
			\E[ Y (0) \ | \ X, \tilde{Z} ] & \text{if} \ D = 0 \\
			\E[ Y (1) \ | \ X, \tilde{Z} ] & \text{if} \ D = 1
		\end{cases}
	\end{align*}
Meaning that 
    \begin{align*}
		Y = \mu + \tau D + X^{\top} \beta + \gamma Z  + \epsilon_i \  \ \text{where} \ \ \E(\epsilon \ | \ D, X, Z) = 0
	\end{align*}
holds with $\mu = \beta = 0$, $\gamma = 1$, and $\tau = \mathrm{ATE}$. Thus, $\E( \hat{\tau}_{\mathrm{target}} ) = \mathrm{ATE}$ if the $w_i$ are entirely defined by $\mathbf{D}$ and $\mathbf{X}$.} Because $\hat{\tau}_{\mathrm{target}} = \hat{\tau}$ for a proper choice of univariate $Z$, that means $\hat{\tau}_{\mathrm{target}}$ would also be unbiased. So, replacing $\hat{\tau}_{\mathrm{target}}$ as the reference point in Expression~\ref{eq:bias_function} with an unbiased estimator that instead, or additionally, uses $Z$ to adjust its weights would yield the same bias in expectation. Thus, if using $Z$ to re-estimate the model for $Y$ recovers an unbiased estimate with or without adjusted weights, we argue that leaving the weights unchanged (and obtaining $\hat{\tau}_{\mathrm{target}}$) is reasonable.

\subsection{Extension to the weighted difference in means}\label{subsec:dwim_maintext}

Although we focus principally on $\hat{\tau}_{\mathrm{wls}}$ here, we also note a direct extension of our sensitivity tools to the weighted difference in means when $D$ is binary,
	\begin{align}\label{eq:tau_wdim}
		\hat{\tau}_{\mathrm{wdim}} = \frac{\sum_{i: D_i = 1} w_i Y_i}{\sum_{i: D_i = 1} w_i} -  \frac{ \sum_{i: D_i = 0} w_i Y_i }{\sum_{i: D_i = 0} w_i} 
	\end{align} 
Note that the $\hat{\tau}_{\mathrm{wdim}}$ above is a H\'{a}jek style estimator, as it normalizes the weights within the treated and control groups. When the weights exactly equate the means of $X$ in the treatment and control groups (e.g., balancing weights from Expression~\ref{eq:ebal}),
\begin{align}\label{eq:balance}
    \frac{\sum_{i:D_i=0} w_i X_i}{\sum_{i:D_i=0} w_i} =  \frac{\sum_{i:D_i=1} w_i X_i}{\sum_{i:D_i=1} w_i}
\end{align}
it follows that $\hat{\tau}_{\mathrm{wdim}} = \hat{\tau}_{\mathrm{wls}}$. Thus, when Expression~\ref{eq:balance} above holds, the proposed tools entirely apply to $\hat{\tau}_{\mathrm{wdim}}$.

\section{Application: Exposure to violence in Darfur}\label{sec:app1}

Section~\ref{sec:app1} demonstrates the sensitivity tools from Section~\ref{sec:wsa_results} in a real-data example.  \cite{hazlett2020angry} applies the sensitivity tools from C\&H when estimating the effect of exposure to violence in Darfur on attitudes toward peace. 
The same setting is considered here, where we use inverse propensity score weights (Section~\ref{subsec:app1.psdr}), matching (Section~\ref{subsec:app1.psmatch}), and balancing weights (Section~\ref{subsec:app1.ebal}) to estimate the effect.

\subsection{Setting and initial results}\label{subsec:app1.setting}

In Darfur, a western region of Sudan, government forces and the ``Janjaweed", a pro-government militia, committed a campaign of violence against its citizens, with peak intensity in 2003-2004, killing an estimated 200,000 (\citealp{flint2008darfur}). \cite{hazlett2020angry} investigates the effect of direct harm by such violence ($D$) on  attitudes toward peace ($Y$) among Darfurian refugees in eastern Chad. 

\cite{hazlett2020angry} describes the main determinants of whether or not an individual would eventually experience direct harm. It is possible that certain villages experienced higher rates of violence than others, whether by the government's intention or due to features such as size, proximity to armed group bases, etc. Within villages, there is little basis for targeting some individuals rather than others:  Any bombs or debris dropped from aircrafts were not precisely guided, and the aim of the Janjaweed militia was primarily to depopulate the village, not to kill or interrogate specific individuals or types of individuals. However, the Janjaweed did target women for sexual assault and rape. These observations support the argument for conditioning on village and gender in an effort to address confounding. Among other estimation methods, \cite{hazlett2020angry} does so by including village and gender fixed effects as covariates ($X$) in a linear regression of individuals' attitudes toward peace and exposure to violence. The covariates also include several other characteristics, such as age, whether or not the individual was a farmer, herder, or merchant/trader, household size, and whether or not the individual had voted before. As argued in \cite{hazlett2020angry}, because gender is expected to be especially likely to relate to harm, and is observed to be a strong influence on attitudes in this context, it is also a useful benchmark variable to consider in sensitivity analyses.

Throughout, we use a subset of the original dataset from \cite{hazlett2020angry} that only retains villages in which there were treated and untreated individuals. This subset of the data describes 807 individuals, of which 339 (42\%) were exposed to violence. Further, this data is clustered: individuals are clustered within villages. When data is clustered, it may be necessary to account for intracluster dependence in the outcome in uncertainty estimates, or else risk downwardly biased standard error estimates for estimated regression coefficients (e.g., \citealp{hazlett2022understanding}). Because it is plausible that attitudes toward peace could be more similar for individuals within the same village than for individuals from different villages, we cluster standard errors by village, or perform a cluster bootstrap, for all estimators (see Appendix~\ref{subsec:se_extension}).

Table~\ref{tab:app1.lm} presents the results of a linear regression of $Y$ on $D$ and $X$, yielding $\hat{\tau}_{\mathrm{ols}}$ from Expression~\ref{eq:ols_ydx}. We find $\hat{\tau}_{\mathrm{ols}}$ is positive and significant at the 0.05 level, implying that direct harm positively influenced attitudes toward peace. 
	\begin{table}[!h]
	\vspace{0.15in}
	\begin{center}
	\caption{Sensitivity results for $\hat{\tau}_{\mathrm{ols}}$ from C\&H}\label{tab:app1.lm}
     \begin{tabular}{ m{2cm} m{3cm} | m{2cm} m{2cm} m{3.5cm} }
     \hline
     \hline
     Estimate & 95\% CI & $\mathrm{RV}_{q=1}$ & $\mathrm{RV}_{\alpha=0.05}$ & $\hat{R}^2 (Y \sim D | X)$ \\
     \hline
     0.096* & (0.047, 0.146) & 0.142 & 0.077 & 0.023 \\
     \hline
     \multicolumn{5}{ l }{Bound ($Z$ as strong as \textit{Female}): $\hat{R}^2 (Y \sim Z | D, X)$=0.121,  $\hat{R}^2 (D \sim Z | X)$=0.010} \\
     \multicolumn{5}{ l }{Adjusted Estimate ($Z$ as strong as \textit{Female}): $\hat{\tau}=$0.074*} \\
     \multicolumn{5}{ l }{Adjusted 95\% CI ($Z$ as strong as \textit{Female}): (0.031, 0.117)} \\
     \end{tabular}
     \vspace{0.20in}
	\subcaption*{\textit{Note:}  95\% confidence interval (CI) employs standard errors clustered by village (\citealp{white1984asymptotic}). Starred (*) values indicate significance at the 0.05 level. When bounding the sensitivity parameters with a ``$Z$ as strong as \textit{Female}", $X^{(j)}$ consists of the dummy variable for female, and, using C\&H's notation,  
	$k_D=k_Y=1$. C\&H's $k_D$ and $k_Y$ are analogous to our $\kappa_{w / w^{(-j)}} (D)$ and $\kappa_{w} (Y)$, respectively, and are in fact equivalent when the weights and semi-weights are uniform (i.e., all $w_i = w^{(-j)}_i = 1$).}
	\end{center}
	\vspace{-0.25in}
	\end{table}
Table~\ref{tab:app1.lm} also includes sensitivity statistics from C\&H, which show this conclusion is robust to omitted confounding ($Z)$ as influential (on treatment and outcome) as gender. 
Confounding as strong as gender would not bring the estimate to 0, as the bounded values for the sensitivity parameters are both smaller than $\mathrm{RV}_{q=1}$. The adjusted estimate is also still statistically significant at the 0.05 level. Additionally, if confounding explained all remaining variation in $Y$ and as much of $D$ as gender, it still would not bring the estimate to 0, as the bounded value of $\hat{R}^2 (D \sim Z | X) = 0.010$ is below the extreme scenario value of $\hat{R}^2 (Y \sim D | X) = 0.023$.

\subsection{Inverse propensity score weighted regression for the ATE}\label{subsec:app1.psdr}


We now demonstrate our sensitivity tools for $\hat{\tau}_{\mathrm{wls}}$ with inverse propensity score weights for the ATE. We estimate the propensity score with logistic regression, with log-odds linear in $X$. The weights take the form of those in Expression~\ref{eq:psweights}. Additionally, we normalize the weights within the treatment and control groups as we recommend in Appendix~\ref{subsec:w_distr_normalize}. 

Table~\ref{tab:app1.ate} reports the estimate and corresponding sensitivity tools from Section~\ref{sec:wsa_results}.
	\begin{table}[!h]
	\vspace{0.15in}
	\begin{center}
	\caption{Estimating the ATE with inverse propensity score weighted $\hat{\tau}_{\mathrm{wls}}$ 
	}\label{tab:app1.ate}
     \begin{tabular}{ m{2cm} m{3cm} | m{2cm} m{2cm} m{3.5cm} }
     \hline
     \hline
     Estimate & 95\% CI & $\mathrm{RV}_{q=1}$ & $\mathrm{RV}_{\alpha=0.05}$ & $\hat{R}_w^2 (Y \sim D | X)$ \\
     \hline
     0.089* & (0.036, 0.138) & 0.139 & 0.058 & 0.022 \\
     \hline
     \multicolumn{5}{ l }{Bound ($Z$ as strong as \textit{Female}): $\hat{R}_w^2 (Y \sim Z | X, D)$=0.108,  $\hat{R}_w^2 (D \sim Z | X)$=0.011} \\
     \multicolumn{5}{ l }{Adjusted Estimate ($Z$ as strong as \textit{Female}): $\hat{\tau}_{\mathrm{target}}$=0.069*} \\
     \multicolumn{5}{ l }{Adjusted 95\% CI ($Z$ as strong as \textit{Female}): (0.015, 0.117) } \\
     \end{tabular}
     \vspace{0.20in}
	\subcaption*{\textit{Note:}  Starred (*) values indicate significance at the 0.05 level. 
    All inference-related results come from the adjusted inference Option 1, using 1000 bootstrapped samples, and modified with a cluster-bootstrap by village (see Appendix~\ref{subsec:se_extension}). 95\% confidence intervals are found through the percentile method.
    When bounding the sensitivity parameters with a ``$Z$ as strong as \textit{Female}", $X^{(j)}$ consists of the dummy variable for female, and $\kappa_{w/w^{(-j)}} (D) = \kappa_w (Y) = 1$.} 
	\end{center}
	\vspace{-0.25in}
	\end{table}
$\hat{\tau}_{\mathrm{wls}}$ 
is  similar to $\hat{\tau}_{\mathrm{ols}}$ in Table~\ref{tab:app1.lm} in terms of magnitude and significance. Like the point estimates, the robustness values and extreme scenario values in Tables~\ref{tab:app1.lm} and~\ref{tab:app1.ate} are very similar. Both estimates (and their statistical significance) are robust to confounding as strong as gender, with adjusted estimates that are very close. 

\begin{figure}[!h]
	\vspace{0.15in}
	\begin{center}
	\caption{Comparison of inverse propensity score weights and semi-weights for estimating the ATE}\label{fig.app1.weights.ate}
    \vspace{-.25in}
    \begin{subfigure}{.45\textwidth}
    \begin{center}
    \includegraphics[scale=0.50]{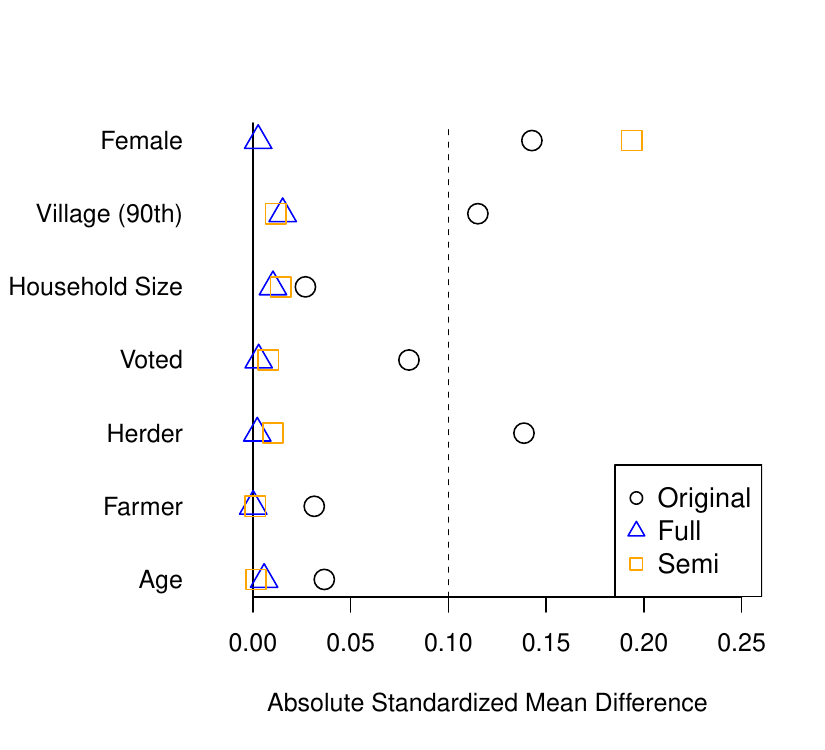} 
    \end{center}
    \vspace{-.25in}
    \subcaption{Balance}\label{fig.app1.weights.ate.balance}
    \end{subfigure}
    \begin{subfigure}{.45\textwidth}
    \begin{center}
    \includegraphics[scale=0.50]{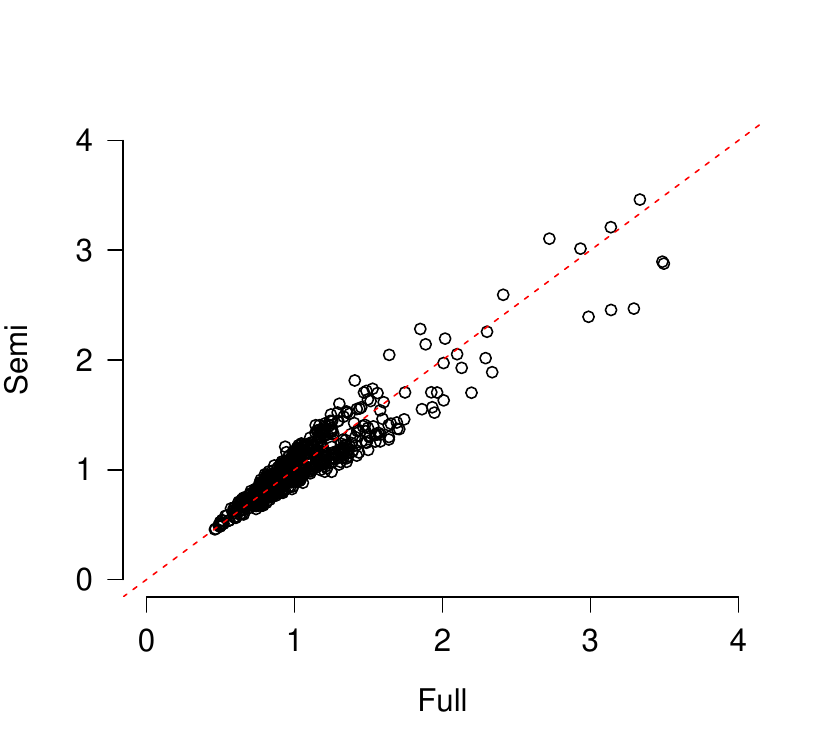} 
    \end{center}
    \vspace{-.25in}
    \subcaption{Weights comparison}\label{fig.app1.weights.ate.comparison}
    \end{subfigure}
    \subcaption*{\textit{Note:} Comparison of the weights $w_i$ (Full) and semi-weights $w_i^{(-j)}$ (Semi).   The weights have $\mathrm{ESS}(w) = 708.5$ (87.8\% of original sample) and the semi-weights have $\mathrm{ESS}(w^{(-j}) = 722.3$ (89.5\%). The $w_i$ and $ w_i^{(-j)}$ are correlated at 0.940. \textit{(a)} ASMD in the covariates before weighting (Original), with the weights, and with the semi-weights. We report the 90th percentile value of ASMDs among the village indicator variables (Village (90th)). The dashed line indicates the target benchmark of 0.10 (\citealp{austin2009balance, normand2001validating}). \textit{(b)} Plot of weights and semi-weights. Points have the coordinates $(w_i, w_i^{(-j)})$ across $i$, and the dashed line indicates equality.}
	\end{center}
	\vspace{-0.25in}
	\end{figure}

We do not entertain a $\kappa_{w/w^{(-j)}} (D)$ larger than 1 here. This is because the weighted and semi-weighted distributions are 
very similar, as can be seen in Figure~\ref{fig.app1.weights.ate}.
Figure~\ref{fig.app1.weights.ate.balance} reports the Absolute Standardized Mean Difference (ASMD) for the covariates before weighting, with the weights, and with the semi-weights.\footnote{The ASMD in a covariate is the absolute value of the difference in the (weighted) means of that covariate between the treatment and control groups, standardized by dividing this difference by a combination of the (weighted) standard deviations in the treatment and control groups.} Both sets of weights achieve good balance on all of the covariates, except for the semi-weights for the female indicator variable, where the ASMD is well over the target benchmark of 0.10 (\citealp{austin2009balance, normand2001validating}). Though this is expected, as the semi-weights here omit gender as a regressor in the model for the propensity score. Moving to Figure~\ref{fig.app1.weights.ate.comparison}, both sets of weights show modest variation, with most falling between 0.5 and 2, a few over 3, and none near 0. The $w_i$ and $w_i^{(-j)}$ are also highly correlated (0.940) with few substantial differences. The weights and semi-weights also have very similar ESS, reflecting 87.8\% and 89.5\% of the original sample size, respectively.  

Finally, we apply our adjusted inference Option 1 
for the adjusted 95\% confidence interval in Table~\ref{tab:app1.ate}. We choose Option 1 because the weights here are smooth, and the dataset is relatively small ($n=807$), so Option 1 is not too computationally intensive. However, for completeness, we provide in Table~\ref{tab:app1.ate.allinference} below the adjusted inference results from all three inference options described in Section~\ref{subsubsec:wsa_se}. Note that the adjusted standard errors, 95\% confidence intervals, and $\mathrm{RV}_{\alpha=0.05}$ from the three options are very similar. All three adjusted confidence intervals indicate that the statistical significance of $\hat{\tau}_{\mathrm{wls}}$ is robust to omitted confounding as strong as gender.
	\begin{table}[!h]
	\begin{center}
	\caption{All adjusted inference options for a $Z$ as strong as \textit{Female} for inverse propensity score weighted $\hat{\tau}_{\mathrm{wls}}$ for the ATE}\label{tab:app1.ate.allinference}
     \vspace{-0.05in}
     \begin{tabular}{ m{6cm} | m{2.5cm} m{3cm} m{2cm}  }
     \hline
     \hline
     Inference Option & Adjusted SE & Adjusted 95\% CI & $\mathrm{RV}_{\alpha=0.05}$  \\
     \hline
     Option 1: Pairs bootstrap & 0.0260 & (0.0154, 0.1174) & 0.0582 \\
     Option 2: Fixed weights bootstrap & 0.0267 & (0.0169, 0.1201) & 0.0616 \\
     Option 3: Closed-form & 0.0263 & (0.0172, 0.1201) & 0.0572 \\
     \hline
     \end{tabular}

     \vspace{0.05in}
	\subcaption*{\textit{Note:} Inference Options 1 
    and 2 
    use 1000 bootstrapped samples, and are modified with a cluster-bootstrap (see Appendix~\ref{subsec:se_extension}) by village. 95\% confidence intervals for Options 1 and 2 are found through the percentile method. Inference Option 3 
    applies adjusted cluster-robust standard errors (see Appendix~\ref{app:proofs_wsa_se_crse}), clustered by village. When bounding the sensitivity parameters with a ``$Z$ as strong as \textit{Female}", $X^{(j)}$ consists of the dummy variable for female, and $\kappa_{w/w^{(-j)}} (D) = \kappa_w (Y) = 1$.} 
	\end{center}
	\vspace{-0.25in}
	\end{table}

\subsection{Matching estimators}\label{subsec:app1.psmatch}

We now demonstrate how our weighted sensitivity analysis applies when the weights are derived from a matching procedure, using propensity score matching (Section~\ref{subsec:app1.psmatch.background}) and exact matching (Section~\ref{subsec:app1.psmatch.exactmatching}).

\subsubsection{Propensity score matching for the ATT}\label{subsec:app1.psmatch.background}

First, we estimate the ATT with one-to-one nearest neighbor matching with replacement on the estimated propensity score, $\hat{\pi}(X)$. 
As described in Section~\ref{subsec:estimators}, this results in weights where $w_i \propto 1$ for treated units, and for control units, $w_i \propto $ the number of times matched. We again normalize the weights per Appendix~\ref{subsec:w_distr_normalize}. 
We use the same model for $\hat{\pi} (X_i)$ described in Section~\ref{subsec:app1.psdr}.\footnote{All matching was done with the \texttt{MatchIt} package in \texttt{R}.}. Table~\ref{tab:app1.att} displays the results and sensitivity tools for a propensity score matched $\hat{\tau}_{\mathrm{wls}}$ for the ATT.
	\begin{table}[!h]
	\vspace{0.15in}
	\begin{center}
	\caption{Estimating the ATT with propensity score matched $\hat{\tau}_{\mathrm{wls}}$}\label{tab:app1.att}
     \begin{tabular}{ m{2.25cm} m{3.25cm} | m{2.25cm} m{2.25cm} m{3.75cm} }
     \hline
     \hline
     Estimate & 95\% CI & $\mathrm{RV}_{q=1}$ & $\mathrm{RV}_{\alpha=0.05}$ & $\hat{R}_w^2 (Y \sim D | X)$ \\
     \hline
     0.078* & (0.008, 0.153) & 0.109 & 0.022 & 0.013 \\
     \hline
     \multicolumn{5}{ l }{Bound ($Z$ is 2 times as strong as \textit{Female} and \textit{Age} for $D$, and 1 times for $Y$):} \\
     \multicolumn{5}{ l }{\ \ \ \ \ \ \ $\hat{R}_w^2 (Y \sim Z | X, D)$=0.126,  $\hat{R}_w^2 (D \sim Z | X)$=0.016} \\
     \multicolumn{5}{ l }{Adjusted Estimate ($Z$ is 2 times as strong as \textit{Female} and \textit{Age} and for $D$, and 1 times for $Y$):} \\
     \multicolumn{5}{ l }{\ \ \ \ \ \ \  $\hat{\tau}_{\mathrm{target}}$=0.048} \\
     \multicolumn{5}{ l }{Adjusted 95\% CI ($Z$ is 2 times as strong as \textit{Female} and \textit{Age} for $D$, and 1 times for $Y$):} \\
     \multicolumn{5}{ l }{\ \ \ \ \ \ \  (-0.016, 0.120) } \\
     \end{tabular}

	\vspace{0.05in}
	\subcaption*{\textit{Note:} Starred (*) values indicate significance at the 0.05 level. All inference-related results come from the adjusted inference Option 2, 
    using 1000 bootstrapped samples, and modified with a cluster-bootstrap (see Appendix~\ref{subsec:se_extension}) by village. 95\% confidence intervals are found through the percentile method. When bounding the sensitivity parameters where ``$Z$ is 2 times as strong as \textit{Female} and \textit{Age} for $D$, and 1 times for $Y$", $X^{(j)}$ consists of age and the dummy variable for female, $\kappa_{w/w^{(-j)}} (D) = 2$, and $\kappa_w (Y) = 1$.}
	\end{center}
	\vspace{-0.25in}
	\end{table}
Like that for the ATE in the previous sections, the estimate for the ATT is positive and significant at the 0.05 level, implying that direct harm increased attitudes for peace for those who were harmed.

Shifting focus to the sensitivity results, the bounds in Table~\ref{tab:app1.att} imply that these conclusions are sensitive to omitted confounding twice as strong as are gender \textit{and age} in their relationship with $D$, and (one times) as strong in their relationship with $Y$. Here, semi-weights are found by matching on estimated propensity scores from a logistic regression that omits gender and age as regressors. A $Z$ this strong would yield an adjusted estimate that is still positive, but with a 95\% confidence interval of (-0.016, 0.120), which contains 0. Additionally, the bounded value of $\hat{R}_w^2 (D \sim Z | X) = 0.016$ is just over the extreme scenario value of $\hat{R}_w^2 (Y \sim D | X) = 0.013$, meaning that a $Z$ this strong would bring the estimate to 0 if it were to explain the remaining weighted variation in $Y$. 

\begin{figure}[!h]
	\vspace{0.15in}
	\begin{center}
	\caption{Comparison of propensity score matching weights and semi-weights for estimating the ATT}\label{fig.app1.weights.att}
    \vspace{-.25in}
    \begin{subfigure}{.45\textwidth}
    \begin{center}
    \includegraphics[scale=0.50]{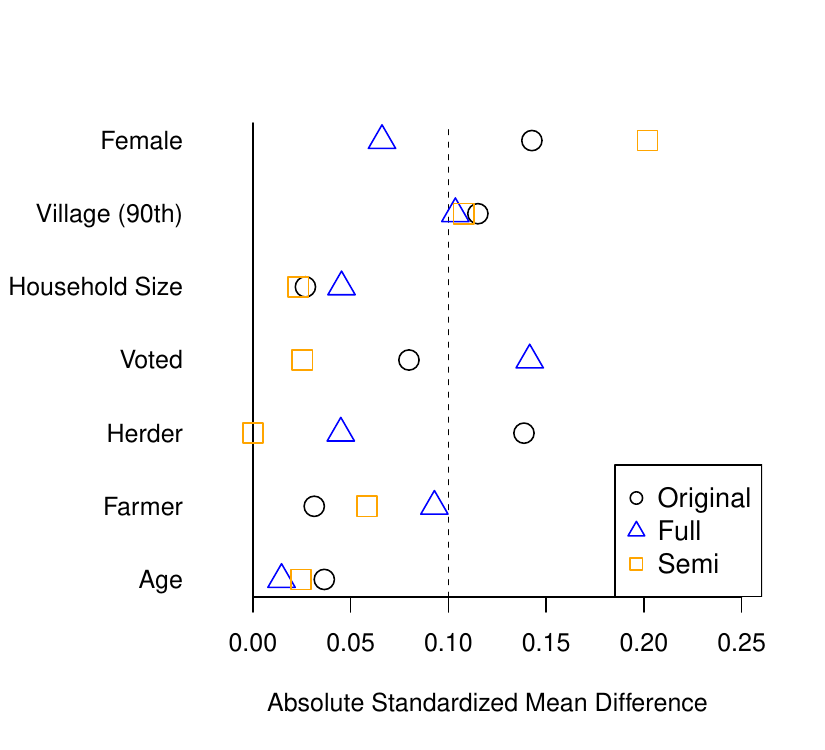} 
    \end{center}
    \vspace{-.25in}
    \subcaption{Balance}\label{fig.app1.weights.att.balance}
    \end{subfigure}
    \begin{subfigure}{.45\textwidth}
    \begin{center}
    \includegraphics[scale=0.50]{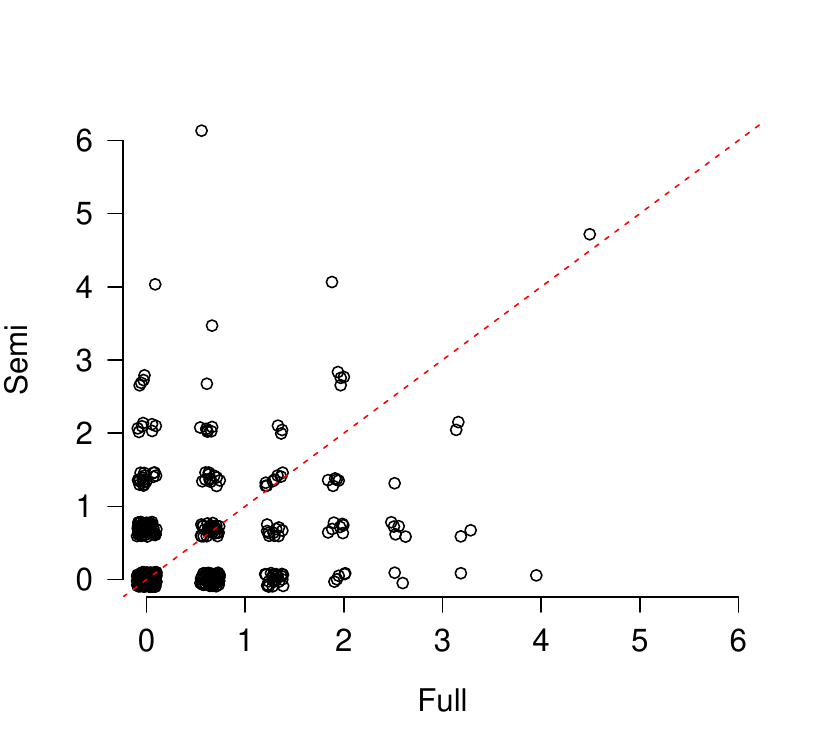} 
    \end{center}
    \vspace{-.25in}
    \subcaption{Weights comparison}\label{fig.app1.weights.att.comparison}
    \end{subfigure}
    \subcaption*{\textit{Note:} Comparison of the weights $w_i$ (Full) and semi-weights $w_i^{(-j)}$ (Semi). Overall, the weights have $\mathrm{ESS}(w) = 464.3$ (57.5\% of original sample) and the semi-weights have $\mathrm{ESS}(w^{(-j}) = 476$ (59.0\%). In the control group, the weights have $\mathrm{ESS}_0(w) = 125.3$ (26.8\% of control group) and the semi-weights have $\mathrm{ESS}_0 (w^{(-j}) = 137.0$ (29.3\%). The $w_i$ and $ w_i^{(-j)}$ are correlated at 0.633 overall, and at only 0.238 in the control group. \textit{(a)} ASMD in the covariates before weighting (Original), with the weights, and with the semi-weights. We report the 90th percentile value of ASMDs among the village indicator variables (Village (90th)). The dashed line indicates the target benchmark of 0.10 (\citealp{austin2009balance, normand2001validating}). \textit{(b)} Plot of weights and semi-weights in the control group. Points have the coordinates $(w_i, w_i^{(-j)})$ across $i$, and the dashed line indicates equality. Coordinates have been slightly jittered on both axes because of overlapping points.}
	\end{center}
	\vspace{-0.25in}
	\end{figure}

Note that we consider a $Z$ twice as strong as the chosen covariates in their relationship with $D$ here, rather than just one times as strong as was done in Section
~\ref{subsec:app1.psdr}. This is because the weighted and semi-weighted distributions show more differences here than they did with inverse propensity score weights, as can be seen in Figure~\ref{fig.app1.weights.att}. 
In Figure~\ref{fig.app1.weights.att.balance}, the weights and semi-weights achieve similar levels of balance on village, household size, and age, but show more differences for the other covariates. Again, the semi-weights achieve poor balance on the female indicator because of its omittance in the semi-weighting process. In Figure~\ref{fig.app1.weights.att.comparison}, the matching weights and semi-weights often deviate significantly, with many instances of the $w_i$ being 0 but the $w^{(-j)}_i$ being non-zero, and vice versa, indicating that the matching from both processes is utilizing different sets of control units. Further, the weights and semi-weights are correlated at 0.633 overall, and at only 0.238 in the control group.

Finally, we apply inference Option 2 
here because the weights are matching weights, which Option 1 
does not allow. Additionally, the assumed $\hat{R}^2_w (Y \sim Z | X, D) = 0.126$ when benchmarking for a $Z$ as strong as gender and age is low, so we are unconcerned with Option 2's potential over-conservativeness. Nevertheless, our conclusions here are robust to using Option 3 
for adjusted inference (results available upon request). 

\subsubsection{Exact matching on gender and village for the ATT}\label{subsec:app1.psmatch.exactmatching}


The identification strategy in this setting requires only conditioning on village and gender. This would suggest a sub-classification or stratification estimator using village and gender, or equivalently, a weighted difference in means ($\hat{\tau}_{\mathrm{wdim}}$) after exact matching on village and gender, where each treated unit is matched (with replacement) to all control units who share the same village and gender. We again normalize the weights within the treatment and control group as we recommend in Appendix~\ref{subsec:w_distr_normalize}, though this does not ultimately influence the H\'{a}jek style $\hat{\tau}_{\mathrm{wdim}}$.  Because the exact matching induces exact mean balance on the gender and village dummy variables, the resulting $\hat{\tau}_{\mathrm{wdim}}$ is exactly equal to $\hat{\tau}_{\mathrm{wls}}$ with the same weights where $X$ only includes the gender and village dummy variables (per Section~\ref{subsec:dwim_maintext}). Thus, our proposed sensitivity tools apply directly to this estimator.

We demonstrate such an analysis here (in Table~\ref{tab:app1.att.exact}). While this is a non-parametric option for achieving our conditioning requirements, this generality comes at a cost that some treated units cannot be matched. Here, 35 treated individuals are dropped because there are no control individuals in the data who share the same gender \text{and} village. This changes the estimand, making direct comparisons difficult.
Nevertheless, the estimated effect of $\hat{\tau}_{\mathrm{wls}} = 0.071$ is only slightly lower than the estimated treatment effects from the methods tried earlier (see Tables~\ref{tab:app1.lm},~\ref{tab:app1.ate}, and~\ref{tab:app1.att}). Additionally, the estimate is still statistically significant at the 0.05 level. 
	\begin{table}[!h]
	\vspace{0.15in}
	\begin{center}
	\caption{Estimating the ATT with exact matching on $\textit{Female}$ and $\textit{Village}$}\label{tab:app1.att.exact}
     \begin{tabular}{ m{2.25cm} m{3.25cm} | m{2.25cm} m{2.25cm} m{3.75cm} }
     \hline
     \hline
     Estimate & 95\% CI & $\mathrm{RV}_{q=1}$ & $\mathrm{RV}_{\alpha=0.05}$ & $\hat{R}_w^2 (Y \sim D | X)$ \\
     \hline
     0.071* & (0.026, 0.120) & 0.110 & 0.042 & 0.014 \\
     \hline
     \multicolumn{5}{ l }{Bound ($Z$ is 2 times as strong as \textit{Female} for $D$, and 1 times for $Y$):} \\
     \multicolumn{5}{ l }{\ \ \ \ \ \ \ $\hat{R}_w^2 (Y \sim Z | X, D)$=0.064,  $\hat{R}_w^2 (D \sim Z | X)$=0.017} \\
     \multicolumn{5}{ l }{Adjusted Estimate ($Z$ is 2 times as strong as \textit{Female} for $D$, and 1 times for $Y$):} \\
     \multicolumn{5}{ l }{\ \ \ \ \ \ \  $\hat{\tau}_{\mathrm{target}}$=0.051*} \\
     \multicolumn{5}{ l }{Adjusted 95\% CI ($Z$ is 2 times as strong as \textit{Female} for $D$, and 1 times for $Y$):} \\
     \multicolumn{5}{ l }{\ \ \ \ \ \ \  (0.005, 0.100) } \\
     \end{tabular}
	\vspace{0.05in}
	\subcaption*{\textit{Note:}  Starred (*) values indicate significance at the 0.05 level. All inference-related results come from the adjusted inference Option 2, using 1000 bootstrapped samples, and modified with a cluster-bootstrap (see Appendix~\ref{subsec:se_extension}) by village. 95\% confidence intervals are found through the percentile method. When bounding the sensitivity parameters where ``$Z$ is 2 times as strong as \textit{Female} for $D$, and 1 times for $Y$", $X^{(j)}$ consists of the dummy variable for female, $\kappa_{w/w^{(-j)}} (D) = 2$, and $\kappa_w (Y) = 1$.} 
	\end{center}
	\vspace{-0.25in}
	\end{table}
We again use the proposed tools to benchmark the strength of unobserved confounding using gender, applying adjusted inference Option 2 
because the weights are matching weights. Semi-weights are found by exact matching only on village. We find the sign of the estimate is robust to omitted confounding that is twice as strong as is gender in its relationship with $D$ ($\kappa_{w/w^{(-j)}} (D) = 2$) and as strong in its relationship with $Y$ ($\kappa_w (Y) = 1$), as the adjusted estimate is still positive ($\hat{\tau}_{\mathrm{target}} = 0.051$). The adjusted 95\% confidence interval (0.005, 0.100) also does not contain 0, meaning the statistical significance (at the 0.05 level) of the estimate is robust to omitted confounding this strong. 

\begin{figure}[!h]
	\vspace{0.15in}
	\begin{center}
	\caption{Comparison of exact matching weights and semi-weights for estimating the ATT}\label{fig.app1.weights.ate.exact}
    \vspace{-.25in}
    \begin{subfigure}{.45\textwidth}
    \begin{center}
    \includegraphics[scale=0.50]{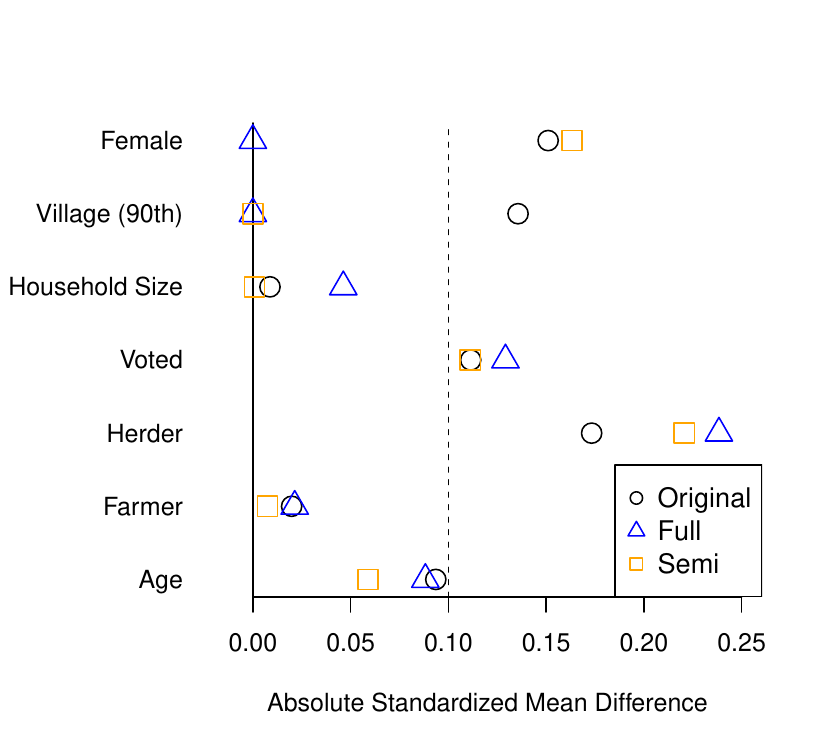} 
    \end{center}
    \vspace{-.25in}
    \subcaption{Balance}\label{fig.app1.weights.ate.balance.exact}
    \end{subfigure}
    \begin{subfigure}{.45\textwidth}
    \begin{center}
    \includegraphics[scale=0.50]{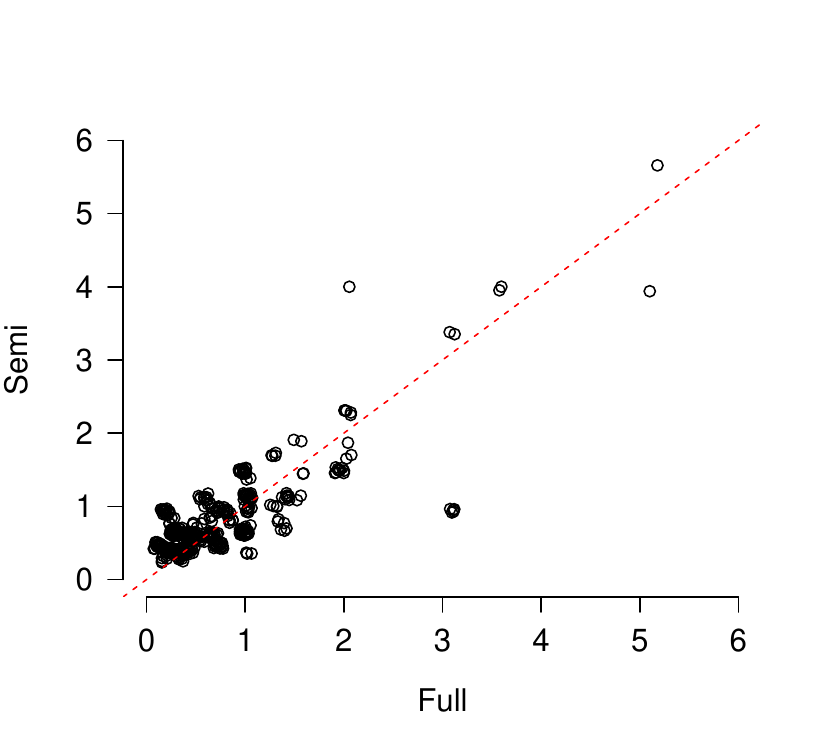} 
    \end{center}
    \vspace{-.25in}
    \subcaption{Weights comparison}\label{fig.app1.weights.ate.comparison.exact}
    \end{subfigure}
    \subcaption*{\textit{Note:} Comparison of the weights $w_i$ (Full) and semi-weights $w_i^{(-j)}$ (Semi). Overall, the weights have $\mathrm{ESS}(w) = 538.0$ (74.9\% of original sample) and the semi-weights have $\mathrm{ESS}(w^{(-j}) = 583.0$ (81.2\%). In the control group, the weights have $\mathrm{ESS}_0(w) = 234.0$ (56.5\% of control group) and the semi-weights have $\mathrm{ESS}_0 (w^{(-j}) = 279.0$ (67.4\%). The $w_i$ and $ w_i^{(-j)}$ are correlated at 0.832 overall, and at 0.793 in the control group. \textit{(a)} ASMD in the covariates before weighting (Original), with the weights, and with the semi-weights. We report the 90th percentile value of ASMDs among the village indicator variables (Village (90th)). The dashed line indicates the target benchmark of 0.10 (\citealp{austin2009balance, normand2001validating}). \textit{(b)} Plot of weights and semi-weights in the control group. Points have the coordinates $(w_i, w_i^{(-j)})$ across $i$, and the dashed line indicates equality. Coordinates have been slightly jittered on both axes because of overlapping points.}
	\end{center}
	\vspace{-0.25in}
	\end{figure}

Here, we entertain omitted confounding twice as strong as gender in its relationship with the treatment because while the weighted and semi-weighted distributions are largely similar (they are correlated at 0.832 overall, and 0.793 in the control group), they do show some clear differences in Figure~\ref{fig.app1.weights.ate.exact}. For example, the effective sample size of the weights in the control group (234.0) is noticeably smaller than that of the semi-weights (279.0). Figure~\ref{fig.app1.weights.ate.comparison.exact} also depicts some clusters of observations where the weights and semi-weights differ greatly (e.g., the cluster of points where $w_i \approx 3$ and $w_i^{(-j)} \approx 1$), and some observations where the $w_i$ are near 0, but the $w_i^{(-j)}$ are well over 0. Note also that in Figure~\ref{fig.app1.weights.ate.balance.exact} the weights and semi-weights, as expected, differ greatly in their balance on the female indicator, with the weights achieving perfect balance, and the semi-weights showing an ASMD similar to the original imbalance before weighting.

\newpage
\subsection{Mean balancing for the ATT}\label{subsec:app1.ebal}

Finally, we demonstrate that our methods also apply to weights chosen to optimize covariate balance. Here, we use maximum entropy weights (\citealp{hainmueller2012entropy}), as in Expression~\ref{eq:ebal}, to achieve exact mean balance on village and gender, weighting the controls units to match the treated and thus targeting the ATT.\footnote{Entropy balancing weights are found with the \texttt{ebalance} package in \texttt{R}.} We then again normalize the weights as we recommend in Appendix~\ref{subsec:w_distr_normalize}. The results for this estimation approach are in Table~\ref{tab:app1.att.ebal.femvillage}, where we apply adjusted inference Option 1 
because the weights here are smooth.\footnote{Although it is not required, we restrict $X$ in the weighted regression to be the gender and village dummy variables as in Section~\ref{subsec:app1.psmatch.exactmatching}. This also makes it so the resulting $\hat{\tau}_{\mathrm{wls}}$ is exactly equal to a $\hat{\tau}_{\mathrm{wdim}}$ with the same weights, due to the exact mean balance on the gender and village dummy variables.}
	\begin{table}[!h]
	\vspace{0.15in}
	\begin{center}
	\caption{Estimating the ATT with mean balancing on $\textit{Female}$ and $\textit{Village}$}\label{tab:app1.att.ebal.femvillage}
     \begin{tabular}{ m{2cm} m{3cm} | m{2cm} m{2cm} m{3.5cm} }
     \hline
     \hline
     Estimate & 95\% CI & $\mathrm{RV}_{q=1}$ & $\mathrm{RV}_{\alpha=0.05}$ & $\hat{R}_w^2 (Y \sim D | X)$ \\
     \hline
     0.096* & (0.049, 0.140) & 0.150 & 0.082 & 0.026 \\
     \hline
     \multicolumn{5}{ l }{Bound ($Z$ as strong as \textit{Female}): $\hat{R}_w^2 (Y \sim Z | X, D)$=0.101,  $\hat{R}_w^2 (D \sim Z | X)$=0.006} \\
     \multicolumn{5}{ l }{Adjusted Estimate ($Z$ as strong as \textit{Female}): $\hat{\tau}_{\mathrm{target}}$=0.082*} \\
     \multicolumn{5}{ l }{Adjusted 95\% CI ($Z$ as strong as \textit{Female}): (0.034, 0.126) } \\
     \end{tabular}

	\vspace{0.05in}
	\subcaption*{\textit{Note:}  Starred (*) values indicate significance at the 0.05 level. All inference-related results come from the adjusted inference Option 1 
    using 1000 bootstrapped samples, and modified with a cluster-bootstrap (see Appendix~\ref{subsec:se_extension}) by village. 95\% confidence intervals are found through the percentile method. When bounding the sensitivity parameters with a ``$Z$ as strong as \textit{Female}", $X^{(j)}$ consists of the dummy variable for female, and $\kappa_{w/w^{(-j)}} (D) = \kappa_w (Y) = 1$.} 
	\end{center}
	\vspace{-0.25in}
	\end{table}
As with the other estimators tried, the resulting estimated effect ($\hat{\tau}_{\mathrm{wls}} = 0.096$) is positive, and statistically significant at the 0.05 level. In fact, this estimate is about the same as the unweighted least squares estimate (see Table~\ref{tab:app1.lm}).

\begin{figure}[!h]
	\vspace{0.15in}
	\begin{center}
	\caption{Comparison of mean balancing weights and semi-weights for estimating the ATT}\label{fig.app1.weights.ate.ebal}
    \vspace{-.25in}
    \begin{subfigure}{.45\textwidth}
    \begin{center}
    \includegraphics[scale=0.50]{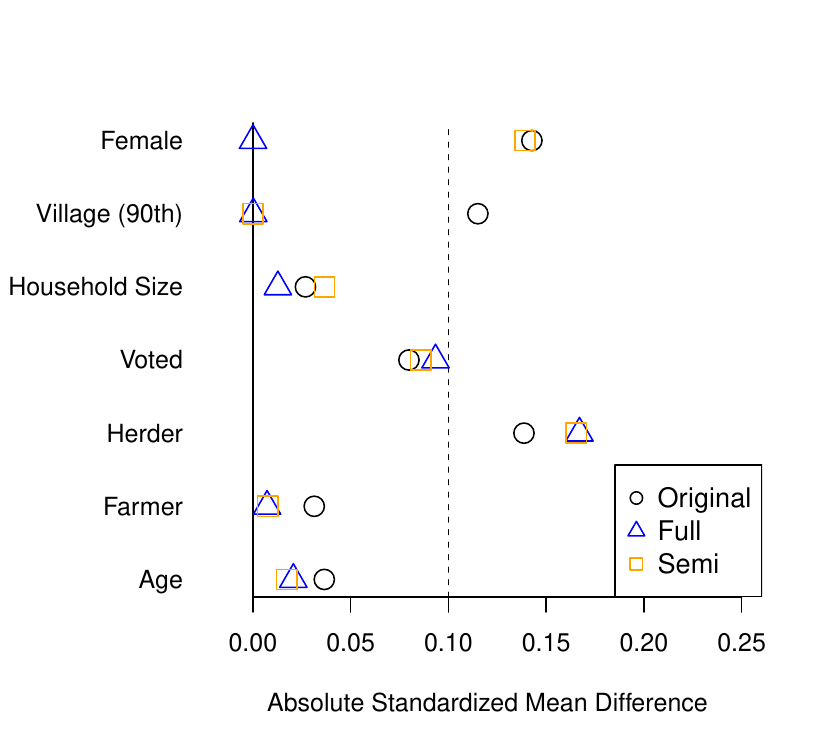} 
    \end{center}
    \vspace{-.25in}
    \subcaption{Balance}\label{fig.app1.weights.ate.balance.ebal}
    \end{subfigure}
    \begin{subfigure}{.45\textwidth}
    \begin{center}
    \includegraphics[scale=0.50]{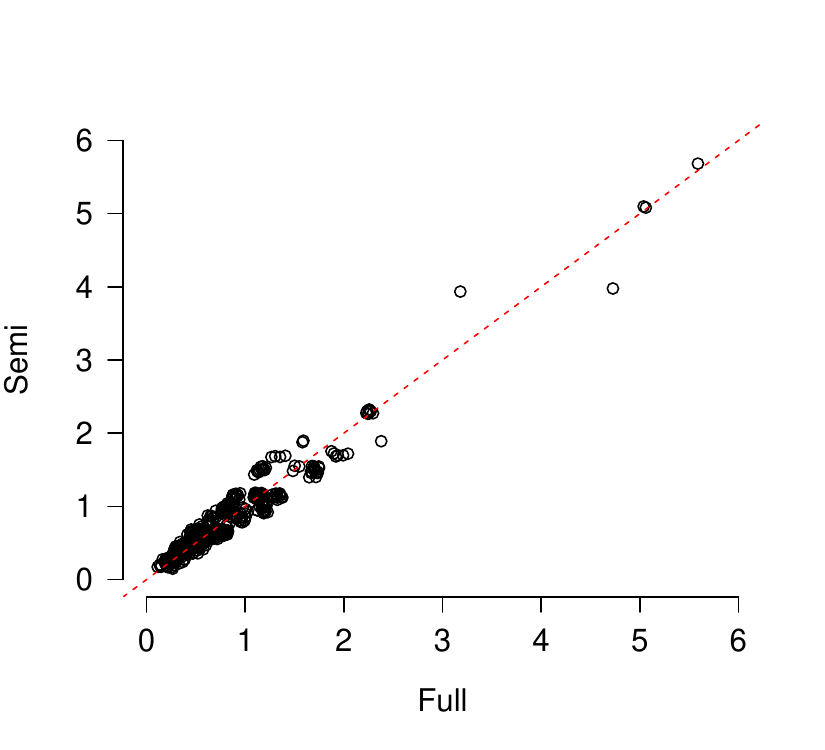} 
    \end{center}
    \vspace{-.25in}
    \subcaption{Comparison of weights}\label{fig.app1.weights.ate.comparison.ebal}
    \end{subfigure}
    \subcaption*{\textit{Note:} Comparison of the weights $w_i$ (Full) and semi-weights $w_i^{(-j)}$ (Semi). Overall, the weights have $\mathrm{ESS}(w) = 643.5$ (79.7\% of original sample) and the semi-weights have $\mathrm{ESS}(w^{(-j}) = 649.8$ (80.5\%). In the control group, the weights have $\mathrm{ESS}_0(w) = 304.5$ (65.1\% of control group) and the semi-weights have $\mathrm{ESS}_0 (w^{(-j}) = 310.8$ (66.4\%). The $w_i$ and $ w_i^{(-j)}$ are correlated at 0.975 overall, and at 0.970 in the control group. \textit{(a)} ASMD in the covariates before weighting (Original), with the weights, and with the semi-weights. We report the 90th percentile value of ASMDs among the village indicator variables (Village (90th)). The dashed line indicates the target benchmark of 0.10 (\citealp{austin2009balance, normand2001validating}). \textit{(b)} Plot of weights and semi-weights in the control group. Points have the coordinates $(w_i, w_i^{(-j)})$ across $i$, and the dashed line indicates equality. Coordinates have been slightly jittered on both axes because of overlapping points.}
	\end{center}
	\vspace{-0.25in}
	\end{figure}

Table~\ref{tab:app1.att.ebal.femvillage} also shows that the sign and statistical significance of this weighted estimate are robust to omitted confounding that is as strong as gender: the adjusted estimate is $\hat{\tau}_{\mathrm{target}} = 0.082$, with an adjusted 95\% confidence interval (0.034, 0.126) that does not contain 0. Here, semi-weights only equate the means of the village dummy variables.  Further, note that even though the original estimate is about the same as the unweighted least squares estimate, the adjusted estimate here is noticeably higher than the adjusted estimate for the unweighted least squares in Table~\ref{tab:app1.lm} ($\hat{\tau} = 0.074$).  Finally, we do not entertain a $\kappa_{w/w^{(-j)}} (D)$ larger than 1 here, as Figure~\ref{fig.app1.weights.ate.comparison.ebal} depicts weighted and semi-weighted distributions that are very similar, and very highly correlated (at 0.975 overall, and at 0.970 in the control group). Figure~\ref{fig.app1.weights.ate.balance.ebal} also shows that the weights and semi-weights have similar ASMD for all of the covariates, other than the female indicator because the semi-weighting process omits it.

\section{Discussion}\label{sec:discussion}

In summary, we employ an omitted variable bias perspective to develop tools for assessing the sensitivity of a wide variety of weighting-based estimators to unobserved confounding. This includes the sensitivity of the point estimate as well as that of  inference. Our overall approach focuses on the sensitivity of a weighted regression step, asking how omitted variables in that regression affect the conclusions, rather than asking how omitted variables affect the weights themselves. The impact of unobserved confounding then relies on only two intuitive sensitivity parameters: (i) the proportion of weighted variance in the treatment that unobserved confounding explains given the covariates, and (ii) the proportion of weighted variance in the outcome that unobserved confounding explains given the covariates and the treatment. This focus on omitted variable bias in the weighted regression allows our approach to apply without reference to the origin of the weights (e.g., inverse propensity score, matching, or covariate mean balancing). It also avoids the need for assumptions on the dimension or distribution of unobserved confounding, and can address bias due to misspecification. We also extended the ``robustness value" and extreme scenario sensitivity statistics from C\&H to the weighted setting, which lend themselves well to routine reporting. Finally, we developed and explored the current challenges of a benchmarking procedure, related to that from C\&H, to formally bound the sensitivity parameters using (a multiple of) the strength of select dimensions of the observed covariates. We make these tools available in the \href{https://github.com/lwainstein/sensewls}{\texttt{sensewls}} package for the \texttt{R} statistical computing language.

\subsection{Comparison to other methods}\label{subsec:comparison}

As demonstrated above, the key benefit of our tools is their generality: by not considering how the weights would change were $Z$ observed, our tools apply to \textit{any} choice of (non-negative) weights. Applying the asymptotic equivalence between propensity score and balancing weights (\citealp{zhao2017entropy, ben2021balancing}) allows the extension of some propensity score weight-based approaches to balancing approaches (e.g., \citealp{hartman2024sensitivity}). However, this extension does not carry over to matching, stratification, or other approaches, as do our tools. Methods that follow \cite{robins1999association} and \cite{robins2000sensitivity}  (e.g., \citealp{brumback2004sensitivity}; \citealp{blackwell2014selection}; \citealp{li2011propensity}) are even more general, applying to any estimator that would be consistent under Assumption~\ref{asm:ci}  (\citealp{blackwell2014selection}). However, they require specifying a ``bias", or ``confounding", function of $X$, which is challenging. \cite{vanderweele2011unmeasured} is similarly general, but requires $Z$ to be binary. More broadly, our tools refrain from  distributional assumptions on $Z$, which are common in simulation-based methods (e.g., \citealp{ichino2008temporary}; \citealp{carnegie2016assessing}; \citealp{huang2020sensitivity}).

Further, we argue that the interpretability of the sensitivity parameters in our tools is a meaningful contribution, particularly that of $\hat{R}^2_w (D \sim Z | X)$, which describes the relationship between $Z$ and $D$, and is bounded between 0 and 1. One branch of the literature (e.g., \citealp{mccaffrey2004propensity, ridgeway2006assessing}) uses Rosenbaum Bounds (\citealp{rosenbaum1987sensitivity,rosenbaum2002sensitivity}) to specify this relationship, choosing a $\Lambda_{\mathrm{RB}}$ such that
    \begin{align}\label{eq:rbounds}
        \frac{1}{\Lambda_{\mathrm{RB}}} \leq \mathrm{OR} \biggr( p(D = 1 \ | \ X=x, Z = z_1), \ p(D = 1 \ | \ X=x, Z = z_2) \biggr) \leq \Lambda_{\mathrm{RB}}
    \end{align}
where $\mathrm{OR} (p_1, p_2) = \frac{p_1}{1-p_1} / \frac{p_2}{1-p_2}$. In words, $\Lambda_{\mathrm{RB}}$ bounds the odds ratio of the probability to be treated for two units that share the same value for $X$ but differ on $Z$. However, $\Lambda_{\mathrm{RB}}$ is unbounded, unlike $\hat{R}^2_w$. The modification of Rosenbaum Bounds introduced by \cite{tan2006distributional}, and explored by others (e.g., \citealp{zhao2017sensitivity}; \citealp{soriano2021interpretable}), shares this limitation, assuming a $\Lambda_{\mathrm{TB}}$ such that
    \begin{align}\label{eq:tbounds}
        \frac{1}{\Lambda_{\mathrm{TB}}} \leq \mathrm{OR} \biggr( p(D = 1 \ | \ X=x, Y(d) = y_1), \ p(D = 1 \ | \ X=x, Y(d) = y_2) \biggr) \leq \Lambda_{\mathrm{TB}}
    \end{align}
for any $d$. This modification replaces $Z$ in Rosenbaum's model in Expression~\ref{eq:rbounds} with $Y (d)$, and thus directly quantifies violations to Assumption~\ref{asm:ci} (i.e., $\Lambda_{\mathrm{TB}} = 1$ under Assumption~\ref{asm:ci}).
\cite{shen2011sensitivity} and \cite{hong2020did} take a different approach, defining a discrepancy between the $w_i$ and adjusted weights that properly account for $X$ and $Z$. The sensitivity parameter that describes the relationship between $Z$ and $D$ is then the variance of this discrepancy. The  $\hat{R}_w^2$ parameter we use provides an alternative scaling that may offer more intuitive traction for at least some users.

\subsection{Limitations and future directions}\label{sec:conclusion}

We note four limitations and/or directions for future research. First, we hope it is possible for future work to improve upon our benchmarking procedure for $\hat{R}^2_w (D \sim Z | X)$, the sensitivity parameter that quantifies the strength of the relationship between the treatment and the unobserved confounding. At present, when the weighted and semi-weighted distributions show stark differences, we can only recommend investigators  entertain strengths of the unobserved confounding, in terms of a factor of the strength of select dimensions of the observed covariates, higher than they might otherwise. This is clearly unsatisfying. A more formal approach to characterizing how different the weighted and semi-weighted distributions are, and what values of the translator (in Expression~\ref{eq:wsa_kappa_rewrite}) this implies, would be very useful. That said, we emphasize that this limitation is related only to the benchmarking exercise, and does not jeopardize the meaning and use of the two $\hat{R}^2_w$ sensitivity parameters, and related values such as the robustness value. 

Second, the adjusted inference procedures we describe are largely satisfying (albeit computationally intensive) for the random design, super-population case using Option 1 (pairs bootstrap with weight re-estimation). However, this approach may fail for matching estimators in some cases. The options we provide for fixed design inference are imperfect: Option 2 does not adjust for the possible variance reduction/inflation from the inclusion of the proposed $Z$, and is conservative due to (i) its partial reliance on a bootstrap that reflects variation that would not arise in a fixed design setting, and (ii) folding $Z$ into random noise, when it is truly non-random with a fixed design.  Option 3 attempts to address these shortcomings, but depends on a uniform shrinkage assumption on the corresponding residuals that likely does not hold.

Third, the augmented (weighted) estimator is another commonly used estimator in settings with a binary treatment and weights. This estimator is the usual form of doubly-robust estimators (e.g., \citealp{robins1994estimation}; \citealp{robins1995semiparametric}; \citealp{kang2007demystifying}; \citealp{van2006targeted}; \citealp{chernozhukov2018double}), which are consistent in an inverse propensity score weights setting when the investigator has correctly specified the propensity score or the conditional expectation function of the outcome. Augmented estimators have also been applied with covariate mean balancing weights (e.g., \citealp{athey2018approximate}; \citealp{hirshberg2017augmented}). The weighted least squares regression we consider here is in fact in the form of an augmented estimator, but 
an extension of our sensitivity tools to these estimators more generally would be a meaningful contribution.

Fourth, and finally, due to our tools' generality, it would be natural and valuable to consider their use for sensitivity analysis in synthetic control analysis (e.g. \citealp{abadie2003economic}; \citealp{abadie2010synthetic}).

\bibliographystyle{apalike}
\nocite{*}
\bibliography{WSAdraft}

\newpage
\appendix

\section{Appendix}\label{app:appendix}

\subsection{Standard error estimation: Further detail, discussion, and extensions}\label{app:variance_choices}

In this appendix, we further detail, discuss, and extend the three adjusted inference options introduced in Section~\ref{subsubsec:wsa_se}. To aid the reader, we also provide Table~\ref{tab:inference} below, which summarizes these options and their strengths and weaknesses. 

	\begin{table}[!h]
	\vspace{0.15in}
	\begin{center}
	\caption{Summary of adjusted inference options}\label{tab:inference}
    \scriptsize
     \begin{tabular}{ m{0.05cm} m{3.20cm} | m{0.95cm} | m{1.05cm} | m{3.52cm} | m{5.6cm} }
     \hline
     \hline
     & \textbf{Option} & \textbf{Design} & \textbf{Weights} & \textbf{Strengths} & \textbf{Weaknesses}  \\
     \hline
     1. & Pairs bootstrap: & Random & Smooth & No additional assumptions & Computationally intensive, \\
     & $\widehat{\se}_{\text{boot}} (\hat{\tau}_{\mathrm{target}})$& & & & May be incompatible with matching,  \\
     & & & & & Adjustment procedure variance \\  
     \hline
     2. & Fixed weights bootstrap: & Fixed & Honest & No additional assumptions, & Still computationally intensive,  \\
     & $\widehat{\se}_{\text{fix-boot}} (\hat{\tau}_{\mathrm{target}})$ & & & Allows matching, & Overestimates with high $\hat{R}_w^2 (Y \sim Z | D, X)$, \\
     & & & & Faster than Option 1, &  Adjustment procedure variance \\   
     & & & & Conservative &  \\     
     \hline
     3. & Closed form: & Fixed & Honest & Fast, & Relies on Assumption~\ref{asm:uniform_shrinkage} \\
     & $\widehat{\se}_{\text{adj-HC0}} (\hat{\tau}_{\mathrm{target}})$  & & & Allows matching, &  \\   
     & & & & Implied-coefficient variance &  \\    
     \hline
     \end{tabular}
     \normalsize
     \end{center}
     \end{table}

\subsubsection{Option 1: Pairs bootstrap with weight re-estimation}\label{subsec:se_random_bootstrap}

Our first option involves a bootstrap procedure, which proceeds as follows:
    \begin{enumerate}
        \item Draw $B$ (e.g., $B=1000$) bootstrap samples, each consisting of $n$ tuples $(D_i, X_i, Y_i)$ drawn with replacement from the data.
        \item Within each bootstrap sample, recalculate the weights (i.e., using the same process that formed the original $w_i$) and the corresponding $\hat{\tau}_{\mathrm{wls}}$, $\widehat{\sd}_w (Y^{\perp_w X, D})$, and $\widehat{\sd}_w (D^{\perp_w X})$.
        \item Choose values for the sensitivity parameters.
        \item Within each bootstrap sample, use Expression~\ref{eq:wsa_bias} to calculate $\hat{\tau}_{\mathrm{target}}$, where the necessary $\hat{\tau}_{\mathrm{wls}}$, $\widehat{\sd}_w (Y^{\perp_w X, D})$, and $\widehat{\sd}_w (D^{\perp_w X})$ come from Step 2, and the sensitivity parameters have been fixed at the values from Step 3.  
        \item Estimate a standard error as:
            \begin{align}
                \widehat{\se}_{\mathrm{boot}} (\hat{\tau}_{\mathrm{target}}) = \widehat{\sd} ( \{ \hat{\tau}_{\mathrm{target}}^{(1)}, \dots, \hat{\tau}_{\mathrm{target}}^{(B)} \} ) \nonumber
            \end{align}
        where $\hat{\tau}_{\mathrm{target}}^{(b)}$ is the estimator from the $b$th bootstrap sample, calculated in Step 4.
    \end{enumerate}
Confidence intervals can be obtained through the normal approximation method, using the adjusted standard error from Step 5, or the percentile method, choosing the the appropriate quantiles of the bootstrap distribution constructed in Step 4. 

This option has two main benefits. First, it does not require additional assumptions on the data beyond assumed values for the sensitivity parameters. Second, this option should be valid with smooth weights in random design settings, which we demonstrate in Appendix~\ref{app:bootstrap}. 

This option has three drawbacks, however. First, it is computationally intensive, particularly re-estimating the weights in each bootstrap sample in Step 2. Second, analytical work (e.g., \citealp{abadie2008failure}) has proven the inconsistency of the bootstrap that re-estimates the weights for matching \textit{with} replacement with a fixed number of matches.\footnote{Several simulation studies (e.g., \citealp{hill2006interval, bodory2020finite}) have demonstrated that a  bootstrap with re-estimated weights performs admirably with matching despite its theoretical inconsistency. We also find that a bootstrap that re-estimates the weights performs reasonably well for one-to-one matching with replacement in Appendix~\ref{app:bootstrap}, though the corresponding confidence intervals tend to show undercoverage as $n$ increases.}$^{,}$\footnote{One proposed solution by \cite{otsu2017bootstrap} for the bias-corrected matching estimator studied by \cite{abadie2011bias}, which takes a similar form as the weighted least squares estimator studied here when the weights come from matching, is a wild bootstrap. However, this wild bootstrap requires for each observation the predicted values from a model built on a set of confounders that satisfies the no unobserved confounding assumption. This is impossible in our setting without additional assumptions because that would require observing the omitted confounder $Z$, which is by definition unobserved.} Thus, we do not recommend this option with matching, or other non-smooth, weights. Although, recent work by \cite{lin2024consistency} also suggests that a bootstrap that re-estimates the weights becomes consistent when the number of matches is allowed to diverge, instead of staying fixed. 

Third, and finally, ideally an adjusted inference procedure would target the variance of the \textit{implied coefficient}, i.e., the standard error the regression would have produced had a $Z$ characterized by the postulated $\hat{R}^2_w$ values actually been included. This adjusted inference option, however, instead targets the variance of the \textit{adjustment procedure}: how $\hat{\tau}_{\mathrm{target}}$, at fixed sensitivity parameter values, would vary under resampling. For many, this variance estimand is less intuitive than the variance of the implied coefficient. For example, the variance of the implied coefficient can shrink when $\hat{R}^2_w (Y \sim Z \mid D, X)$ is large because $Z$ would have absorbed residual variation in $Y$, and can also inflate when $\hat{R}^2_w (D \sim Z \mid X)$ is large because including $Z$ leaves less variation in $D$ after partialing out $(X, Z)$. The variance of the adjustment procedure does not exhibit similarly predictable behavior.  



\subsubsection{Option 2: Fixed weights with bootstrap}\label{subsec:se_fixed_bootstrap}

If an adjusted standard error for matching is required, or re-estimating the weights for Option 1 is computational infeasible, we propose a second option that modifies the bootstrap from Option 1 slightly. For this option, we require that the weights be honest.
We then take a fixed design perspective for inference, assuming that variation only comes from $Y$, and treating $(\mathbf{X}, \mathbf{D}, \mathbf{Z})$ as fixed, and thus also the weights. This option's bootstrap then modifies that from Option 1 as follows:
    \begin{itemize}
        \item In Step 1 of Option 1, instead draw tuples $(D_i, X_i, Y_i, w_i)$ for each bootstrap sample.
        \item In Step 2 of Option 1, do not re-estimate the weights, and instead recalculate $\hat{\tau}_{\mathrm{wls}}$, $\widehat{\sd}_w (Y^{\perp_w X, D})$, and $\widehat{\sd}_w (D^{\perp_w X})$ in each bootstrap sample using the corresponding weights drawn in Step 1.
    \end{itemize}
To summarize, Option 2 resamples the original weights for each bootstrap sample rather than re-estimating them in each bootstrap sample for Option 1. We denote the resulting standard error from Option 2 as $\widehat{\se}_{\text{fix-boot}} (\hat{\tau}_{\mathrm{target}})$.

This option is conservative in fixed design settings, in that it overestimates the standard error of $\hat{\tau}_{\mathrm{target}}$. We demonstrate this through simulation in Appendix~\ref{app:fixed_bootstrap}. This over-estimation comes from two sources. First, Option 2 resamples $D$ and $X$ as if they were to vary from sample to sample, but $\textbf{X}$ and $\textbf{D}$ are fixed in truth. Thus, the option assumes more variability than exists in a fixed design setting. Second, and more importantly, because $\mathbf{Z}$ is unobserved it cannot be accounted for in this option, and is thus necessarily treated as a source of randomness in $Y$ that is fixed in truth. As a result, this option overestimates the true standard error of $\hat{\tau}_{\mathrm{target}}$ more as the true $\hat{R}^2_w (Y \sim Z \ | \ X, D)$ increases. For example, in Appendix~\ref{app:fixed_bootstrap}, we demonstrate a setting where the option over-estimates by 75-90\% when $\hat{R}^2_w (Y \sim Z \ | \ X, D) \approx 0.80$.

Nevertheless, this option has several appealing features. First, unlike Option 1, it applies with matching weights without theoretical, asymptotic concerns.  Second, it is also much faster than Option 1 because the weights are not re-estimated in each bootstrap sample. Third, like Option 1 it does not require additional assumptions beyond assumed values for the sensitivity parameters. While the result is conservative, sensitivity analysis typically is; e.g., it is wise to entertain higher levels of unobserved confounding than we might expect out of an abundance of caution.

The main worry is that this conservativeness may becomes so extreme that it damages the usefulness of the approach.  In our demonstrations below (see Appendix~\ref{app:fixed_bootstrap}), we find such concerning behavior when $\hat{R}^2_w (Y \sim Z \ | \ X, D) \gg 0.50$. Thus, when the user intends to choose a $\hat{R}^2_w (Y \sim Z \ | \ X, D)$ value that is meaningfully higher than 0.50, Options 1 or 3 (see below) may be preferable. Additionally, although this procedure is much faster than Option 1, it can still become computationally intensive in large samples. Finally, like Option 1, this inference option targets the variance of the adjustment procedure, rather than the variance of the implied coefficient, which is often preferred.


\subsubsection{Option 3: Closed-form implied-coefficient standard error for a fixed design}\label{subsec:se_closed}

Our final option offers an adjusted closed-form standard error at the cost of an additional assumption beyond assumed values for the sensitivity parameters. This is the uniform shrinkage assumption (Assumption~\ref{asm:uniform_shrinkage} in the main text), which states that, for all $i$,
\begin{enumerate}
    \item $(Y_i^{\perp_w X, D, Z})^2 \approx  (Y_i^{\perp_w X, D})^2 * \big( 1 - \hat{R}^2_w (Y \sim Z | X, D) \big)$, and
    \item $(D_i^{\perp_w X, Z})^2 \approx (D_i^{\perp_w X})^2 * \big( 1 - \hat{R}^2_w (D \sim Z | X) \big).$
\end{enumerate}
In words, the first statement of Assumption~\ref{asm:uniform_shrinkage} says that each squared residual from the $Y$ on $(X, D, Z)$ weighted regression can be approximated by shrinking the corresponding squared residual from the $Y$ on $(X, D)$ weighted regression by a function of $\hat{R}^2_w (Y \sim Z | X, D)$, the sensitivity parameter that pertains to $Y$. Similarly, the second statement of  Assumption~\ref{asm:uniform_shrinkage}  says that the squared residual from the $D$ on $(X, Z)$ weighted regression can approximated by shrinking the corresponding squared residual from the $D$ on $X$ weighted regression by a function of $\hat{R}^2_w (D \sim Z | X)$, the sensitivity parameter that pertains to $D$. Note that both statements of  Assumption~\ref{asm:uniform_shrinkage} hold on average in the weighted distribution by construction. For example, 
\begin{align}\label{eq:average_shrinkage}
    \underbrace{\frac{1}{n} \sum_{i=1}^n w_i (Y_i^{\perp_w X, D, Z})^2}_{\widehat{\var}_w (Y^{\perp_w X, D, Z})} =  \underbrace{ \frac{1}{n} \sum_{i=1}^n w_i (Y_i^{\perp_w X, D})^2 }_{\widehat{\var}_w (Y^{\perp_w X, D})} * \biggr( 1 - \hat{R}^2_w (Y \sim Z | X, D) \biggr)
\end{align}
However, Assumption~\ref{asm:uniform_shrinkage} demands a stronger condition in order to estimate a HC standard error: that the shrinkage in  Expression~\ref{eq:average_shrinkage} above occurs uniformly, albeit approximately, at the \textit{unit} level, not just on average. 

As for the resulting standard error, we again assume a fixed design $(\mathbf{X}, \mathbf{D}, \mathbf{Z})$ and honest weights. Then, let $\widehat{\se}_{\mathrm{HC0}} (\cdot)$ be the HC0 sandwich standard error (\citealp{white1980heteroskedasticity}) estimate of a regression coefficient. Under Assumption~\ref{asm:uniform_shrinkage},
\begin{align}\label{eq:adjusted_hc0_appendix}
    \widehat{\se}_{\mathrm{HC0}} ( \hat{\tau}_{\mathrm{target}}) \approx \underbrace{ \sqrt{ \frac{1 - \hat{R}^2_w (Y \sim Z | X, D)}{1 - \hat{R}^2_w (D \sim Z | X )} } \times \widehat{\se}_{\mathrm{HC0}} ( \hat{\tau}_{\mathrm{wls}}) }_{\widehat{\se}_{\text{adj-HC0}} ( \hat{\tau}_{\mathrm{target}})}
\end{align}
Proof is given in Appendix~\ref{app:proofs_wsa_se}. This option thus estimates the standard error of $\hat{\tau}_{\mathrm{target}}$ with the right-hand-side of Expression~\ref{eq:adjusted_hc0_appendix} above, which we denote $\widehat{\se}_{\text{adj-HC0}} ( \hat{\tau}_{\mathrm{target}})$. In Appendix~\ref{app:proofs_wsa_se} we also generalize this option to other closed-form standard errors (e.g., HC1 from \citealp{mackinnon1985some}). Naturally, this option is the fastest of the three, as it does not require a bootstrap. It also targets the variance of the implied coefficient, which may be conceptually preferred to the variance of the adjustment procedure, which Options 1 and 2 target. Relatedly, Expression~\ref{eq:adjusted_hc0_appendix} makes it clear how $\widehat{\se}_{\text{adj-HC0}} ( \hat{\tau}_{\mathrm{target}})$ can shrink and expand depending on the postulated values of the sensitivity parameters. We demonstrate in Appendix~\ref{app:fixed_bootstrap} that this option well-estimates the true standard error of $\hat{\tau}_{\mathrm{target}}$ when Assumption~\ref{asm:uniform_shrinkage} reasonably holds. 

The drawback with this option, of course, is that it relies on Assumption~\ref{asm:uniform_shrinkage}. In Appendix~\ref{app:proofs_wsa_se} we show that Assumption~\ref{asm:uniform_shrinkage} is used to make the approximation:
\begin{align}
    & \sum_{i=1}^n w_i^2 \big( D_i^{\perp_w X, Z}\big)^2 \big( Y_i^{\perp_w X, D, Z}\big)^2  \approx  \nonumber \\
    & \ \ \ \ \ \ \ \ \ \ \ \big( 1 - \hat{R}^2_w (Y \sim Z | X, D) \big) \big( 1 - \hat{R}^2_w (D \sim Z | X) \big) \sum_{i=1}^n w_i^2 \big( D_i^{\perp_w X}\big)^2 \big( Y_i^{\perp_w X, D}\big)^2
\end{align}
It is easily shown that this approximation amounts to:
\begin{align}\label{eq:uniform_shrinkage_sensemaking}
    & \underbrace{ \biggr( \widehat{\sd} \big( w ( D^{\perp_w X, Z} )^2 \big) \times \widehat{\sd} \big( w ( Y^{\perp_w X, D, Z} )^2 \big) \biggr) }_{(a)} \times \ \widehat{\cor}\big( w ( D^{\perp_w X, Z} )^2, \ w ( Y^{\perp_w X, D, Z} )^2 \big)   \approx \nonumber \\
    & \ \ \ \ \ \ \ \ \ \ \underbrace{  \biggr[ \widehat{\sd} \biggr( w \big(1 - \hat{R}^2_w (D \sim Z | X) \big) ( D^{\perp_w X} )^2 \biggr) \times  \widehat{\sd} \biggr( w \big( 1 - \hat{R}^2_w (Y \sim Z | X, D) \big) ( Y^{\perp_w X, D} )^2 \biggr) \biggr] }_{(b)} \times \nonumber \\
    & \ \ \ \ \ \ \ \ \ \    \widehat{\cor}\big( w ( D^{\perp_w X} )^2, \ w ( Y^{\perp_w X, D} )^2 \big)   
\end{align}
Thus, the extent to which Assumption~\ref{asm:uniform_shrinkage} is a valid approximation is partly driven by the difference in the correlation terms of (\ref{eq:uniform_shrinkage_sensemaking}): (i) the correlation between the weighted, squared \textit{full} regression residuals, $w_i (Y_i^{\perp_w X, D, Z})^2$ and $w_i (D_i^{\perp_w X, Z})^2$, and (ii) the correlation between the weighted, squared \textit{restricted} regression residuals, $w_i (Y_i^{\perp_w X, D})^2$ and $w_i (D_i^{\perp_w X})^2$. In other words, the adjusted standard error may underestimate or overestimate the true standard error of $\hat{\tau}_{\mathrm{target}}$ depending on if replacing the (weighted, squared) full residuals with the (weighted, squared, shrunken) restricted residuals masks or intensifies their correlation. Similarly, any difference between terms $(a)$ and $(b)$ in (\ref{eq:uniform_shrinkage_sensemaking}) also contributes to the performance of the adjusted standard error, and is driven by how replacing the (weighted, squared) full residuals with the (weighted, squared, shrunken) restricted residuals inflates or deflates their variance. Accordingly, we demonstrate in Appendix~\ref{app:fixed_bootstrap} that this proposed adjusted option can meaningfully overestimate or underestimate the true standard error of $\hat{\tau}_{\mathrm{target}}$ when Assumption~\ref{asm:uniform_shrinkage} fails.


\subsubsection{Extensions for clustered data}\label{subsec:se_extension} 




When data is clustered (e.g., students within schools), it may be necessary to account for intracluster dependence in the outcome in uncertainty estimates, or else risk downwardly biased standard error estimates for estimated regression coefficients (e.g., \citealp{hazlett2022understanding}). To do this here for Options 1 and 2, we suggest replacing the random sampling in the corresponding bootstraps with cluster-bootstrap sampling --- letting the data be partitioned by $G$ clusters, randomly sample $G$ \textit{clusters} with replacement to make up each bootstrap sample.\footnote{See \cite{cameron2015practitioner} for guidance on how large $G$ must be for cluster-robust inference.} We demonstrate in Appendix~\ref{app:cluster_bootstrap} that modifying Option 1 in this way shows strong results empirically. As for Option 3, we generalize this option to cluster-robust standard errors (\citealp{white1984asymptotic}) in Appendix~\ref{app:proofs_wsa_se_crse}. Relatedly, for matching weights \textit{without} replacement, we recommend clustering on the matched pairs (or sets) of observations, as do \cite{abadie2022robust} and \cite{austin2014use}. Our \href{https://github.com/lwainstein/sensewls}{\texttt{sensewls}} package allows for these cluster-robust modifications to all three inference options.

\subsection{Effective sample size}\label{subsec:w_distr_normalize}

While weighted regressions consider all $n$ units when the $w_i > 0$, units are nearly discarded when their $w_i$ are close to 0. One way to describe how many units a weighted regression meaningfully incorporates is the effective sample size,   
	\begin{align} 
		\mathrm{ESS} (w) = \frac{(\sum_{i=1}^n w_i)^2}{\sum_{i=1}^n w_i^2}
	\end{align}

When $D$ is binary, we also define $\mathrm{ESS}_d (w)$ to be the effective sample size of the weights within the group with treatment status $d$:
	\begin{align}\label{eq:ess_d}
		\mathrm{ESS}_{d} (w) = \frac{(\sum_{i: D_i = d} w_i)^2}{\sum_{i: D_i = d} w_i^2} 
	\end{align}
Note that it is not necessarily true that $\mathrm{ESS} (w) = \mathrm{ESS}_0 (w) + \mathrm{ESS}_1 (w)$, because the weights for the control group and those for the treated group may be on different scales.\footnote{
For example, if all $w_i = 1$, then $\mathrm{ESS}_d (w) = n_d$ and  $\mathrm{ESS} (w) = \mathrm{ESS}_0 (w) + \mathrm{ESS}_1 (w) = n$. However, if
    \begin{align*}
        w_i \propto \begin{cases} 
            1 & \text{if} \ D_i=0 \\
            \frac{1}{2} & \text{if} \ D_i=1 \\
        \end{cases}
    \end{align*}
then $\mathrm{ESS}_d (w) = n_d$, so $\mathrm{ESS}_0 (w) + \mathrm{ESS}_1 (w) = n$. However, $\mathrm{ESS} (w) = (n - \frac{1}{2} n_1)^2 / (n - \frac{3}{4} n_1) < n$.}
Though it is not required for the proposed tools, we thus suggest, for starting weights $\tilde{w}_i$, rescaled weights
	\begin{align}\label{eq:normalizedw}
		w_i = n \times \biggr( \frac{ \tilde{w_i} }{ \sum_{i:D_i=d} \tilde{w}_i} \biggr) \times \biggr( \frac{ \mathrm{ESS}_{d} (\tilde{w}) }{\mathrm{ESS}_0 (\tilde{w})  +  \mathrm{ESS}_1 (\tilde{w})  } \biggr) \ \ \ \text{if} \ \ \ D_i = d
	\end{align}  
With this rescaling, the effective sample size in the full sample reflects the effective sample size in the treated and control groups, i.e., $\mathrm{ESS} (w) = \mathrm{ESS}_0 (w) + \mathrm{ESS}_1 (w)$.

\newpage
\clearpage

\subsection{Demonstration: Inference Option 1 in a random design setting}\label{app:bootstrap}

This section demonstrates the merits our adjusted inference procedures detailed in Section~\ref{subsubsec:wsa_se} in a random design setting through simulation. The data generating process (DGP) here is as follows:
    \begin{align}\label{eq:heterodpg}
        Y &= X + Z + \delta D + \epsilon \ \ \text{and} \ \  p(D = 1 | X, Z)= \frac{\mathrm{exp} (X + Z - 1)}{1 + \mathrm{exp} (X + Z - 1)} \nonumber \\
        \text{where} \ \ [X \ Z]^{\top} &\overset{iid}{\sim} \mathcal{N}(0, I_2), \ \ \epsilon \overset{iid}{\sim} N(0, 2), \ \ \text{and} \ \ \delta \overset{iid}{\sim} N(0, \theta^2) \tag{DGP 1}
    \end{align}
where $(\delta D + \epsilon)$ makes up a combined error term, and $\theta^2 \in \{0, 4, 16\}$ determines the extent of the error's heteroscedasticity. Note that when $\theta^2 = 0$, the error is homoscedastic. Furthermore, there is no treatment effect (i.e., the ATE, ATT, and ATC are all 0). Finally, this DGP has a random design: $\mathbf{X}$, $\mathbf{Z}$, and $\mathbf{D}$ are  regenerated in each iteration of the simulation.

\begin{figure}[!h]
	\vspace{0.15in}
	\begin{center}
	\caption{Coverage rates in \ref{eq:heterodpg} for inverse propensity score weights for the ATE}\label{fig.bootstrap1}
    \vspace{-.6in}
    
    \includegraphics[scale=0.45]{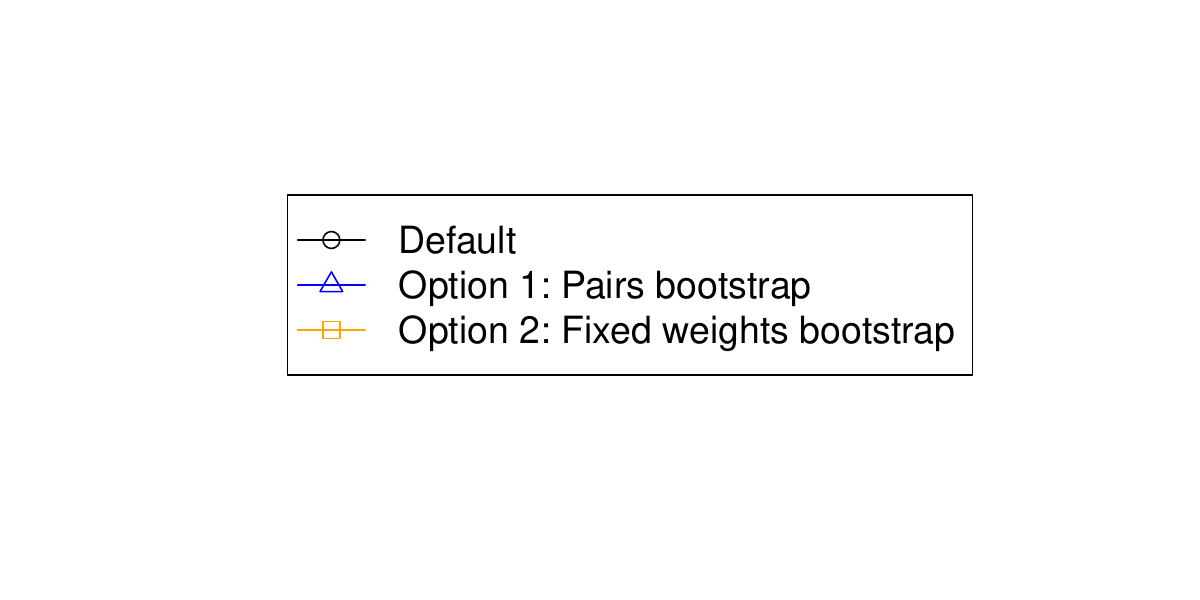} 

    \vspace{-0.9in}

    \begin{subfigure}{.30\textwidth}
    \begin{center}
    \includegraphics[scale=0.45]{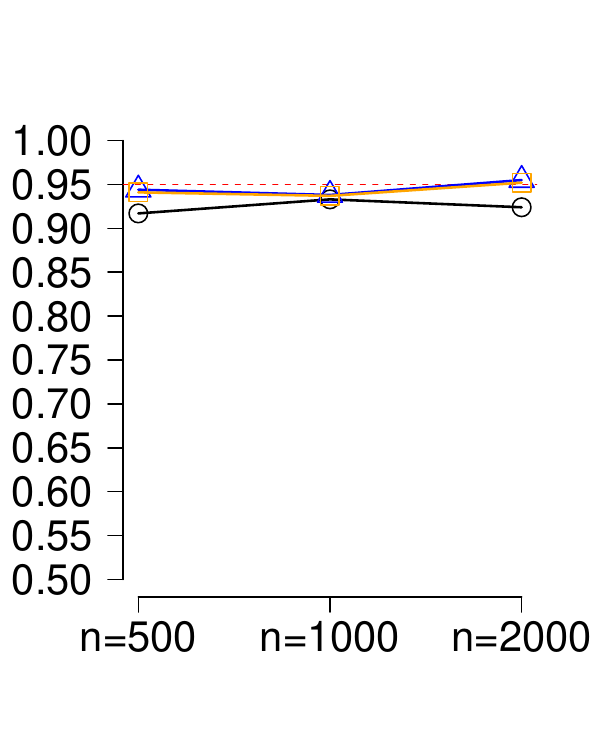} 
    \end{center}
    \vspace{-.4in}
    \subcaption{$\theta^2 = 0$}\label{fig.bootstrap1.y1}
    \end{subfigure}
    \begin{subfigure}{.30\textwidth}
    \begin{center}
    \includegraphics[scale=0.45]{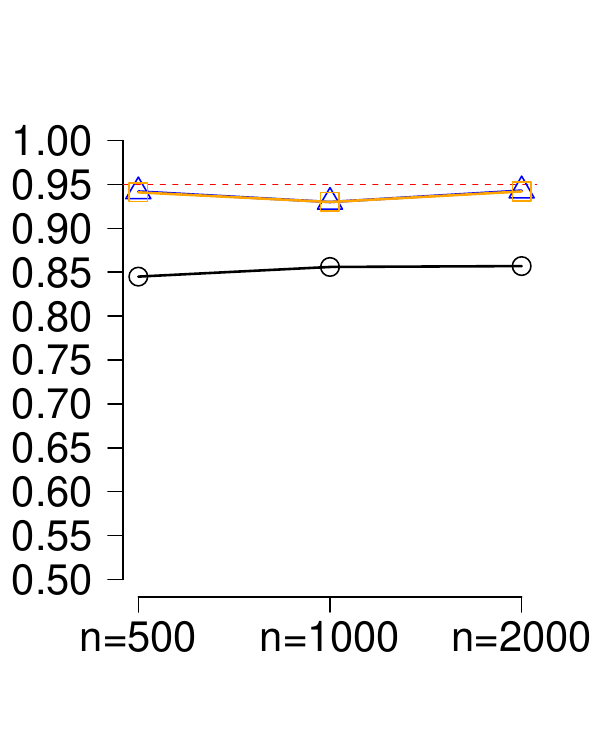} 
    \end{center}
    \vspace{-.4in}
    \subcaption{$\theta^2 = 4$}\label{fig.bootstrap1.y2}
    \end{subfigure}
    \begin{subfigure}{.30\textwidth}
    \begin{center}
    \includegraphics[scale=0.45]{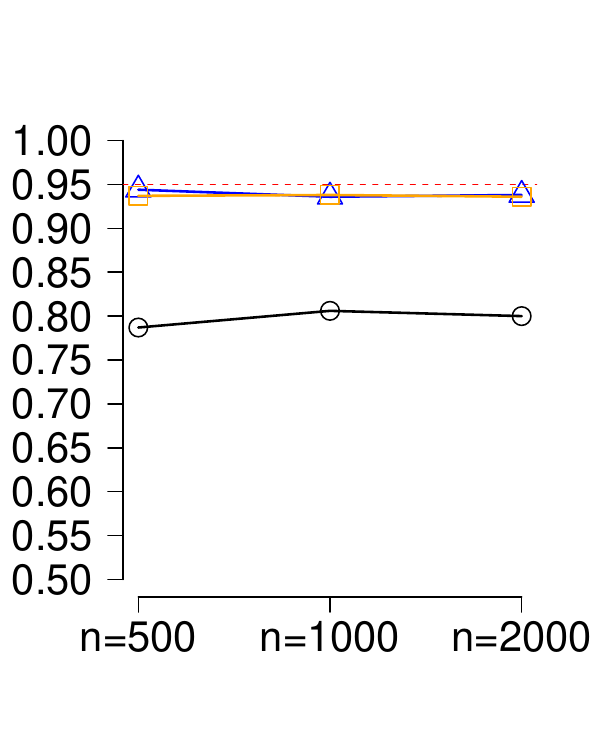} 
    \end{center}
    \vspace{-.4in}
    \subcaption{$\theta^2 = 16$}\label{fig.bootstrap1.y3}
    \end{subfigure}

    \subcaption*{\textit{Note:} Coverage rates of 95\% confidence intervals across 1000 iterations of \ref{eq:heterodpg}. The dashed line indicates the target coverage rate of 0.95. Weights are inverse propensity score weights for the ATE, and they employ the rescaling proposed in Section~\ref{subsec:w_distr_normalize}. For inference Options 1 and 2, we draw $B = 1000$ bootstrap samples at each iteration of \ref{eq:heterodpg}. }
	\end{center}
	\vspace{-0.4in}
	\end{figure}

\begin{figure}[!h]
	\vspace{0.15in}
	\begin{center}
	\caption{Coverage rates in \ref{eq:heterodpg} for entropy balancing weights for the ATT}\label{fig.bootstrap2}
    \vspace{-.6in}
    
    \includegraphics[scale=0.45]{Figures/BootstrapandSEDemonstration_Coverage_Legend.pdf} 

    \vspace{-0.9in}

    \begin{subfigure}{.30\textwidth}
    \begin{center}
    \includegraphics[scale=0.45]{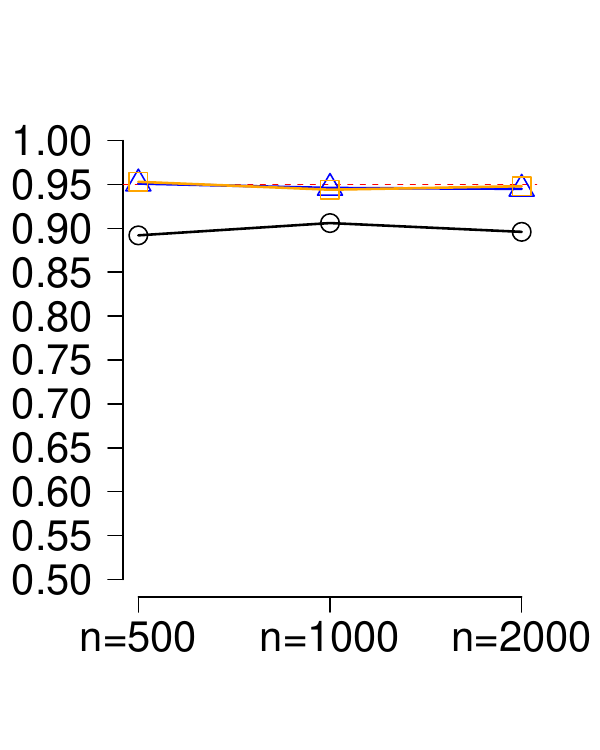} 
    \end{center}
    \vspace{-.4in}
    \subcaption{$\theta^2 = 0$}\label{fig.bootstrap2.y1}
    \end{subfigure}
    \begin{subfigure}{.30\textwidth}
    \begin{center}
    \includegraphics[scale=0.45]{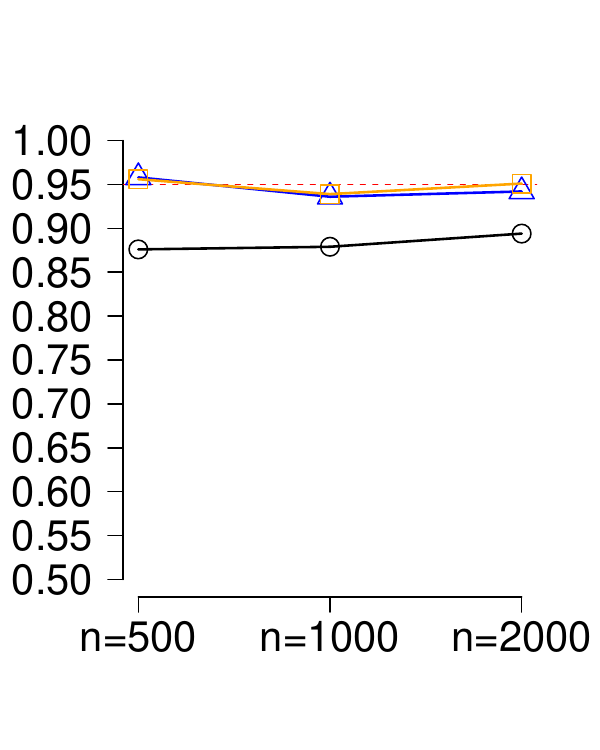} 
    \end{center}
    \vspace{-.4in}
    \subcaption{$\theta^2 = 4$}\label{fig.bootstrap2.y2}
    \end{subfigure}
    \begin{subfigure}{.30\textwidth}
    \begin{center}
    \includegraphics[scale=0.45]{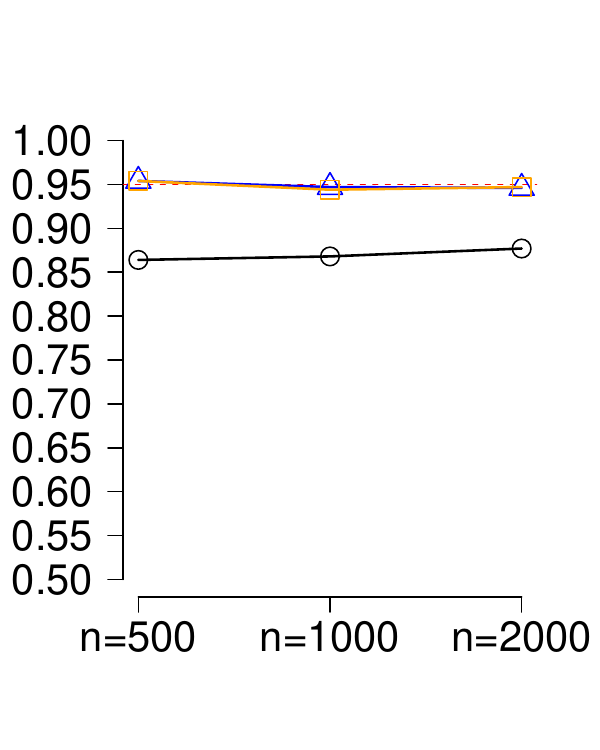} 
    \end{center}
    \vspace{-.4in}
    \subcaption{$\theta^2 = 16$}\label{fig.bootstrap2.y3}
    \end{subfigure}
    \subcaption*{\textit{Note:} Coverage rates of 95\% confidence intervals across 1000 iterations of \ref{eq:heterodpg}. The dashed line indicates the target coverage rate of 0.95. Weights are entropy balancing weights for the ATT, and they employ the rescaling proposed in Section~\ref{subsec:w_distr_normalize}. For inference Options 1 and 2, we draw $B = 1000$ bootstrap samples at each iteration of \ref{eq:heterodpg}. }
	\end{center}
	\vspace{-0.25in}
	\end{figure}

\begin{figure}[!h]
	\vspace{0.15in}
	\begin{center}
	\caption{Coverage rates in \ref{eq:heterodpg} for propensity score matching for the ATT}\label{fig.bootstrap3}
    \vspace{-.6in}
    
    \includegraphics[scale=0.45]{Figures/BootstrapandSEDemonstration_Coverage_Legend.pdf} 

    \vspace{-0.9in}

    \begin{subfigure}{.30\textwidth}
    \begin{center}
    \includegraphics[scale=0.45]{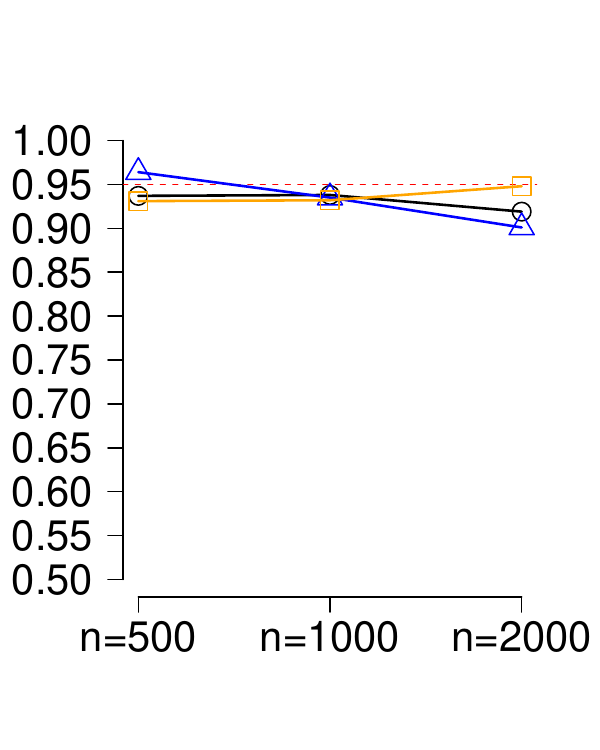} 
    \end{center}
    \vspace{-.4in}
    \subcaption{$\theta^2 = 0$}\label{fig.bootstrap3.y1}
    \end{subfigure}
    \begin{subfigure}{.30\textwidth}
    \begin{center}
    \includegraphics[scale=0.45]{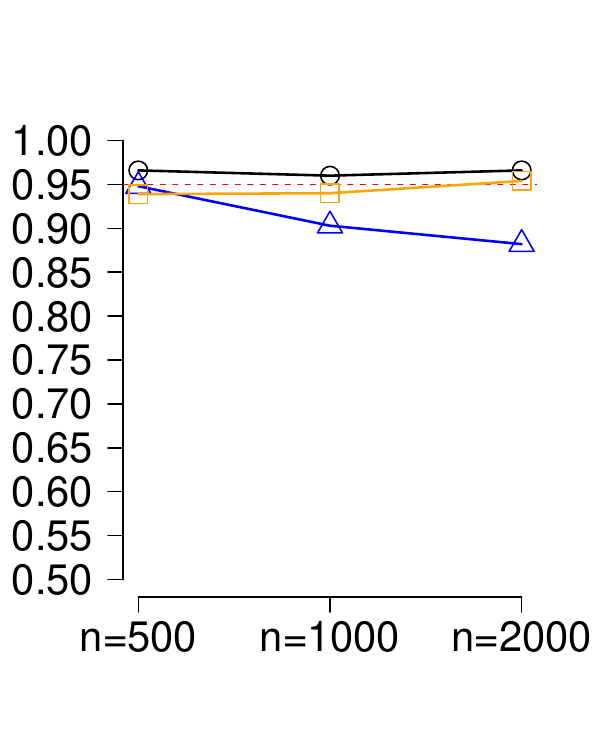} 
    \end{center}
    \vspace{-.4in}
    \subcaption{$\theta^2 = 4$}\label{fig.bootstrap3.y2}
    \end{subfigure}
    \begin{subfigure}{.30\textwidth}
    \begin{center}
    \includegraphics[scale=0.45]{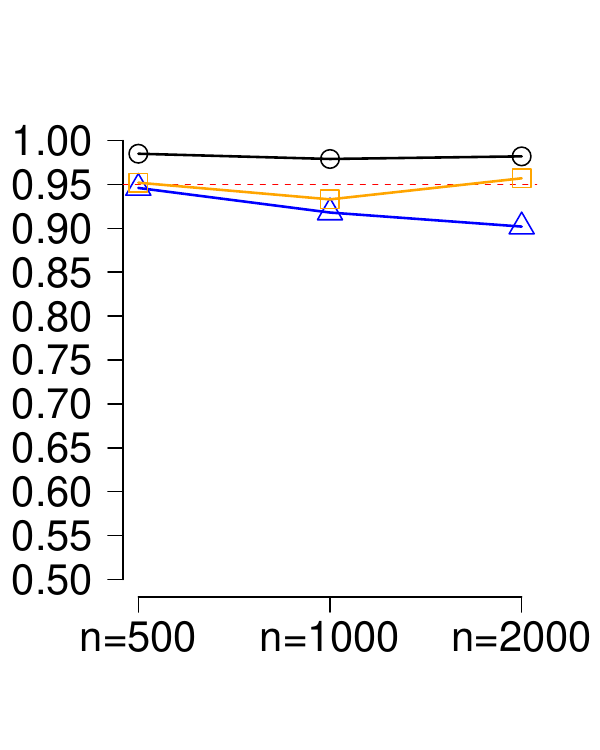} 
    \end{center}
    \vspace{-.4in}
    \subcaption{$\theta^2 = 16$}\label{fig.bootstrap3.y3}
    \end{subfigure}
    \subcaption*{\textit{Note:} Coverage rates of 95\% confidence intervals across 1000 iterations of \ref{eq:heterodpg}. The dashed line indicates the target coverage rate of 0.95. Weights are formed by one-to-one propensity score matching with replacement for the ATT, and they employ the rescaling proposed in Section~\ref{subsec:w_distr_normalize}. For inference Options 1 and 2, we draw $B = 1000$ bootstrap samples at each iteration of \ref{eq:heterodpg}.}
	\end{center}
	\vspace{-0.25in}
	\end{figure}

We apply our adjusted inference Option 1 (Section~\ref{subsec:se_random_bootstrap}), which is most appropriate in random design settings, and with smooth weights. We examine 95\% confidence intervals, obtained through the percentile method, for the desired estimand (all 0, in this DGP) with three types of weights: inverse propensity score weights for the ATE (Figure~\ref{fig.bootstrap1}), entropy balancing weights for the ATT (Figure~\ref{fig.bootstrap2}), and one-to-one propensity score matching with replacement for the ATT (Figure~\ref{fig.bootstrap3}). For each weighting method and each $\theta^2$ (see \ref{eq:heterodpg}), we fix the sensitivity parameters at their approximate probability limits.\footnote{We approximate the probability limits of the sensitivity parameters by taking their empirical averages across 1000 draws of \ref{eq:heterodpg} with $n = 10000$. For the inverse propensity score weights (Figure~\ref{fig.bootstrap1}), we set $\hat{R}_w^2 (D \sim Z | X) = 0.1442$. We then set $\hat{R}_w^2 (Y \sim Z | D, X) = 0.3006$ when $\theta^2 = 0$, $\hat{R}_w^2 (Y \sim Z | D, X) = 0.2172$ when $\theta^2 = 4$, and $\hat{R}_w^2 (Y \sim Z | D, X) = 0.1187$ when $\theta^2 = 16$. For entropy balancing weights (Figure~\ref{fig.bootstrap2}), we set $\hat{R}_w^2 (D \sim Z | X) = 0.1736$. We then set $\hat{R}_w^2 (Y \sim Z | D, X) = 0.2960$ when $\theta^2 = 0$, $\hat{R}_w^2 (Y \sim Z | D, X) = 0.1777$ when $\theta^2 = 4$, and $\hat{R}_w^2 (Y \sim Z | D, X) = 0.0809$ when $\theta^2 = 16$. For the matching weights (Figure~\ref{fig.bootstrap3}), we set $\hat{R}_w^2 (D \sim Z | X) = 0.1502$. We then set $\hat{R}_w^2 (Y \sim Z | D, X) = 0.2963$ when $\theta^2 = 0$, $\hat{R}_w^2 (Y \sim Z | D, X) = 0.1498$ when $\theta^2 = 4$, and $\hat{R}_w^2 (Y \sim Z | D, X) = 0.0606$ when $\theta^2 = 16$.} Note that one should expect the confidence intervals from Option 1 to struggle for the matching weights (Figure~\ref{fig.bootstrap3}), because these weights are non-smooth.

We compare confidence intervals from inference Option 1 with two other methods for inference. First, we employ the ``default", homoscedastic 95\% confidence intervals for $\hat{\tau}_{\mathrm{target}}$ directly pulled from the unadjusted \texttt{lm()} output (in \texttt{R}) from a weighted regression of $Y$ on $X$, $D$, and $Z$. Note that a researcher would not be able to implement these confidence intervals in practice, as they require that $Z$ be observed. However, they are included here to emphasize the fact that our inference Option 1 is valid under heteroscedasticity, while these ``default" confidence intervals should struggle in heteroscedastic settings of \ref{eq:heterodpg} (i.e., when $\theta^2 > 0$). We also calculate percentile method confidence intervals from our adjusted inference Option 2 (Section~\ref{subsec:se_fixed_bootstrap}), which fixes the weights in its bootstrap procedure. We recommend inference Option 2 more for fixed design settings, but include it here for sake of comparison with Option 1.

As expected, for the smooth weighting methods, inverse propensity score weights (Figure~\ref{fig.bootstrap1}) and balancing weights  (Figure~\ref{fig.bootstrap2}), confidence intervals from our inference Option 1 achieve, or come very close to achieving, nominal coverage rates for all $\theta^2$. Meanwhile, the default, homoscedastic 95\% confidence intervals for $\hat{\tau}_{\mathrm{target}}$ show clear undercoverage of the true treatment effect, with worsening undercoverage as the level of heteroscedasticity ($\theta^2$) increases. 

For the matching weights (Figure~\ref{fig.bootstrap3}), the coverage rates of Option 1's confidence intervals decrease as $n$ increases for all $\theta^2$. This aligns with the mathematical inconsistency of the standard bootstrap for matching with replacement proven by \cite{abadie2008failure}. However, Option 2's confidence intervals do achieve (near-)nominal coverage for all $\theta^2$ for matching weights (Figure~\ref{fig.bootstrap3}), as well as for the smooth weighting methods (Figures~\ref{fig.bootstrap1} and~\ref{fig.bootstrap2}). When Option 2 is appropriate in random design settings is left to further research. Though, treating weights as fixed, as in Option 2, does echo the advice of \cite{ho2007matching}, who argue that matching weights be treated as fixed ``nonparametric preprocessing". Fixing the weights also mimics the advice of \cite{residualizedSEs}, in which the weights are taken as fixed in a second stage weighted regression re-including the covariates, though that approach employs robust analytical standard errors.


\newpage
\clearpage

\subsection{Demonstration: Inference Options 2 and 3 in a fixed design setting}\label{app:fixed_bootstrap}

This section demonstrates the merits of our adjusted inference procedures detailed in Section~\ref{subsubsec:wsa_se} in a fixed design setting through simulation.  The data generating process (DGP) here is as follows:
    \begin{align}\label{eq:fixdgp}
        Y = X + Z + \epsilon, \ \ \text{where} \ \ \epsilon = \sigma * \biggr[ N \biggr(0, \frac{D}{2(0.57)^2} \biggr) + N \biggr(0, \frac{1}{2}\biggr) \biggr] \tag{DGP 2}
    \end{align}
where $\sigma$ is varied across values that correspond to $\hat{R}^2 (Y \sim Z | X, D) \in \{0.20, 0.35, 0.50, 0.65, 0.80\}$. Note that $\epsilon$ is heteroscedastic here, as its variance depends on $D$. Note also that there is no treatment effect here (i.e., the ATE, ATT, and ATC are all 0). This DGP then has a fixed design, where for each $n$ tried, $(\mathbf{X}, \mathbf{D}, \mathbf{Z})$ is \textit{fixed} across all iterations after simulating it \textit{once} according to: 
\begin{align}\label{eq:dgp1_xzd}
    p(D = 1 | X, Z)= \frac{\mathrm{exp} (X + Z - 1)}{1 + \mathrm{exp} (X + Z - 1)}, \ \ \text{where} \ \ [X \ Z]^{\top} &\overset{iid}{\sim} \mathcal{N}(0, I_2)
\end{align}

\begin{figure}[!h]
	\vspace{0.15in}
	\begin{center}
	\caption{Coverage rates in \ref{eq:fixdgp} for inverse propensity score weights for the ATE}\label{fig.fixbootstrap1}
    \vspace{-.6in}
    
    \includegraphics[scale=0.45]{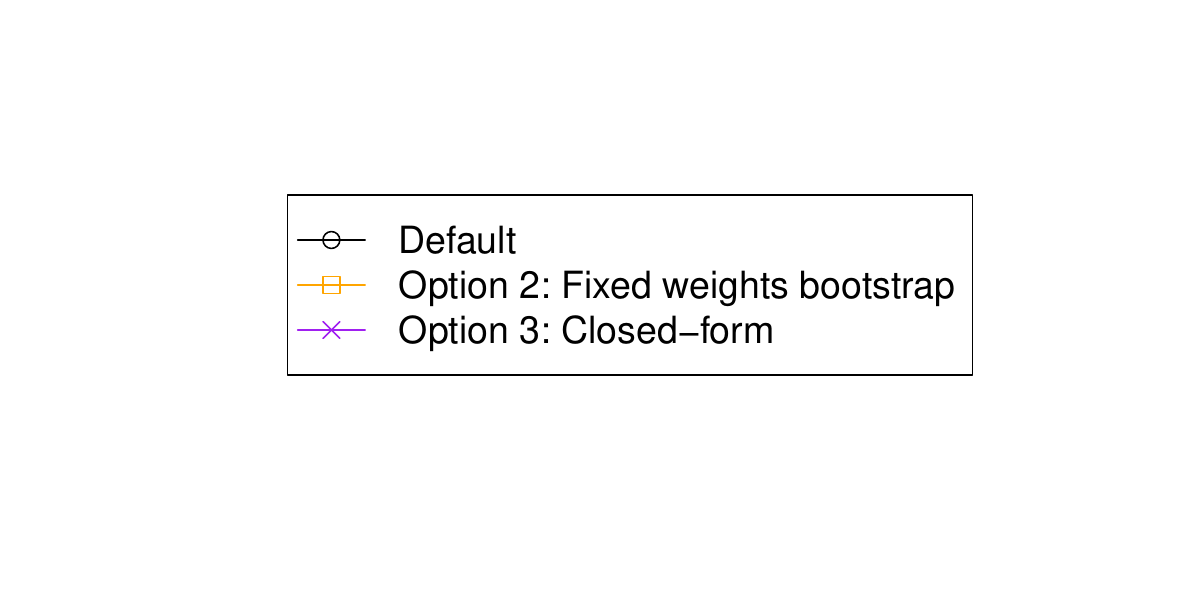} 

    \vspace{-0.9in}

    \begin{subfigure}{.30\textwidth}
    \begin{center}
    \includegraphics[scale=0.45]{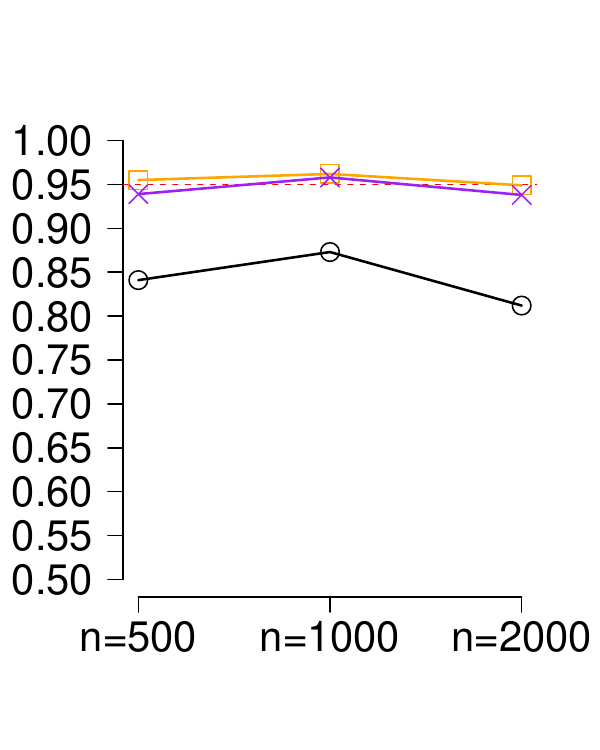} 
    \end{center}
    \vspace{-.4in}
    \subcaption{$\hat{R}^2 (Y \sim Z | X, D) \approx 0.20$}\label{fig.fixbootstrap1.y1}
    \end{subfigure}
    \begin{subfigure}{.30\textwidth}
    \begin{center}
    \includegraphics[scale=0.45]{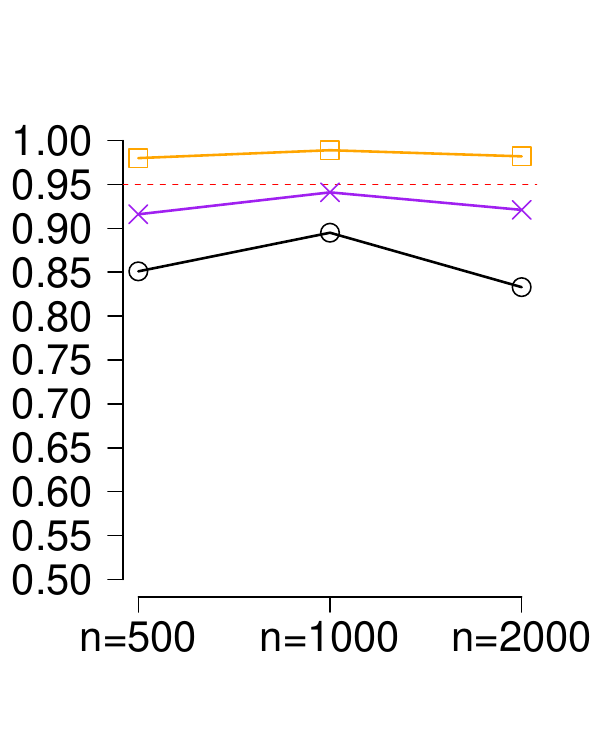} 
    \end{center}
    \vspace{-.4in}
    \subcaption{$\hat{R}^2 (Y \sim Z | X, D) \approx 0.50$}\label{fig.fixbootstrap1.y2}
    \end{subfigure}
    \begin{subfigure}{.30\textwidth}
    \begin{center}
    \includegraphics[scale=0.45]{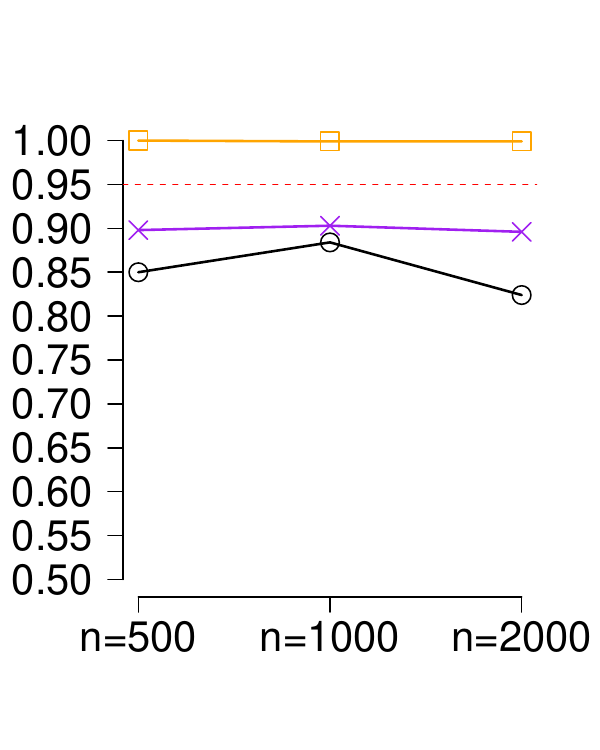} 
    \end{center}
    \vspace{-.4in}
    \subcaption{$\hat{R}^2 (Y \sim Z | X, D) \approx 0.80$}\label{fig.fixbootstrap1.y3}
    \end{subfigure}

    \subcaption*{\textit{Note:} Coverage rates of 95\% confidence intervals across 1000 iterations of \ref{eq:fixdgp}. The dashed line indicates the target coverage rate of 0.95. Weights are inverse propensity score weights for the ATE, and they employ the rescaling proposed in Section~\ref{subsec:w_distr_normalize}. For inference Option 2, we draw $B = 1000$ bootstrap samples at each iteration of \ref{eq:fixdgp}. }
	\end{center}
	\vspace{-0.4in}
	\end{figure}

\begin{figure}[!h]
	\vspace{0.15in}
	\begin{center}

	\caption{Coverage rates in \ref{eq:fixdgp} for entropy balancing weights for the ATT}\label{fig.fixbootstrap2}
    \vspace{-.6in}
    
    \includegraphics[scale=0.45]{Figures/InferenceDemonstration_FixedDesign_Coverage_Legend.pdf} 

    \vspace{-0.9in}

    \begin{subfigure}{.30\textwidth}
    \begin{center}
    \includegraphics[scale=0.45]{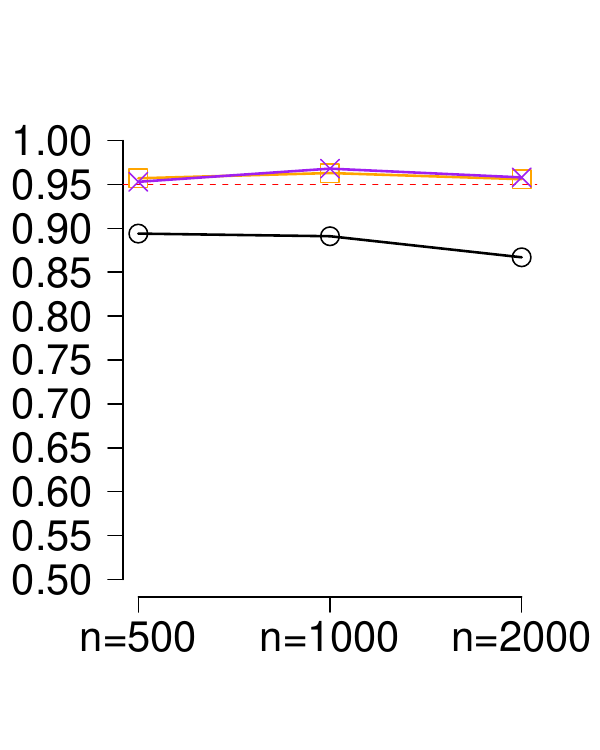} 
    \end{center}
    \vspace{-.4in}
    \subcaption{$\hat{R}^2 (Y \sim Z | X, D) \approx 0.20$}\label{fig.fixbootstrap2.y1}
    \end{subfigure}
    \begin{subfigure}{.30\textwidth}
    \begin{center}
    \includegraphics[scale=0.45]{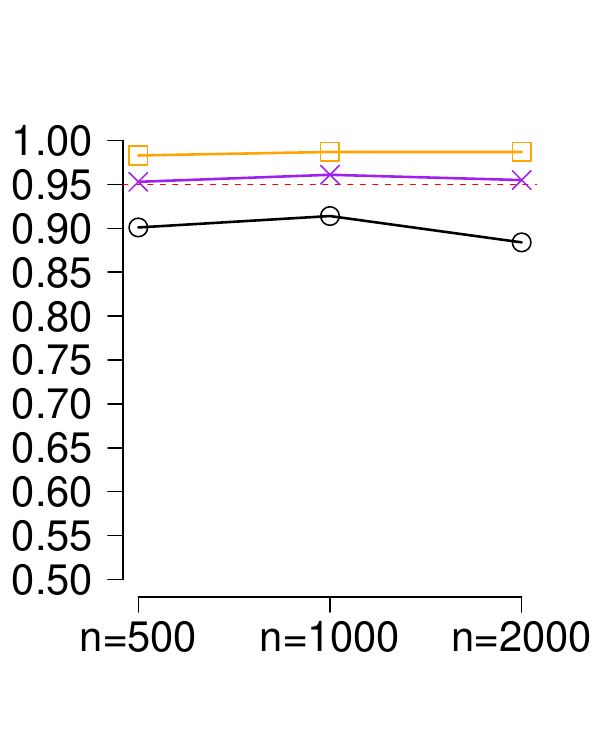} 
    \end{center}
    \vspace{-.4in}
    \subcaption{$\hat{R}^2 (Y \sim Z | X, D) \approx 0.50$}\label{fig.fixbootstrap2.y2}
    \end{subfigure}
    \begin{subfigure}{.30\textwidth}
    \begin{center}
    \includegraphics[scale=0.45]{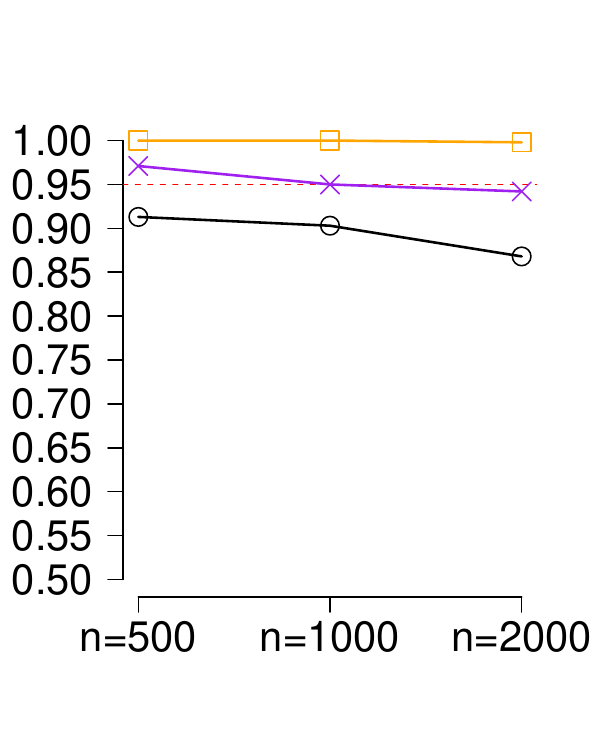} 
    \end{center}
    \vspace{-.4in}
    \subcaption{$\hat{R}^2 (Y \sim Z | X, D) \approx 0.80$}\label{fig.fixbootstrap2.y3}
    \end{subfigure}
    \subcaption*{\textit{Note:} Coverage rates of 95\% confidence intervals across 1000 iterations of \ref{eq:fixdgp}. The dashed line indicates the target coverage rate of 0.95. Weights are entropy balancing weights for the ATT, and they employ the rescaling proposed in Section~\ref{subsec:w_distr_normalize}. For inference Option 2, we draw $B = 1000$ bootstrap samples at each iteration of \ref{eq:fixdgp}. }
	\end{center}
	\vspace{-0.25in}
	\end{figure}

\begin{figure}[!h]
	\vspace{0.15in}
	\begin{center}
	\caption{Coverage rates in \ref{eq:fixdgp} for propensity score matching for the ATT}\label{fig.fixbootstrap3}
    \vspace{-.6in}
    
    \includegraphics[scale=0.45]{Figures/InferenceDemonstration_FixedDesign_Coverage_Legend.pdf} 

    \vspace{-0.9in}

    \begin{subfigure}{.30\textwidth}
    \begin{center}
    \includegraphics[scale=0.45]{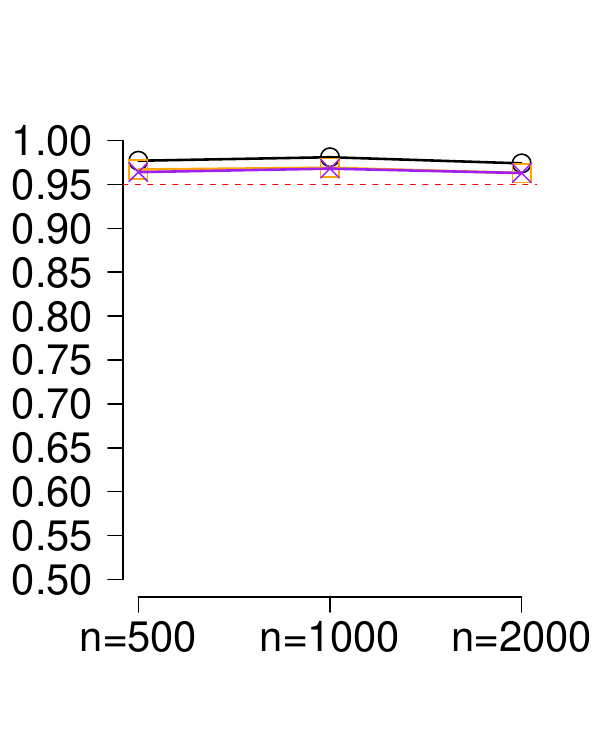} 
    \end{center}
    \vspace{-.4in}
    \subcaption{$\hat{R}^2 (Y \sim Z | X, D) \approx 0.20$}\label{fig.fixbootstrap3.y1}
    \end{subfigure}
    \begin{subfigure}{.30\textwidth}
    \begin{center}
    \includegraphics[scale=0.45]{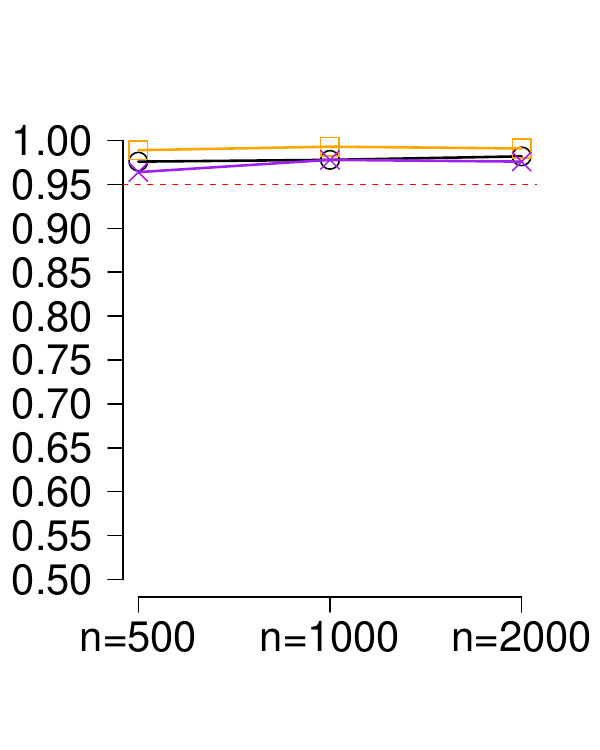} 
    \end{center}
    \vspace{-.4in}
    \subcaption{$\hat{R}^2 (Y \sim Z | X, D) \approx 0.50$}\label{fig.fixbootstrap3.y2}
    \end{subfigure}
    \begin{subfigure}{.30\textwidth}
    \begin{center}
    \includegraphics[scale=0.45]{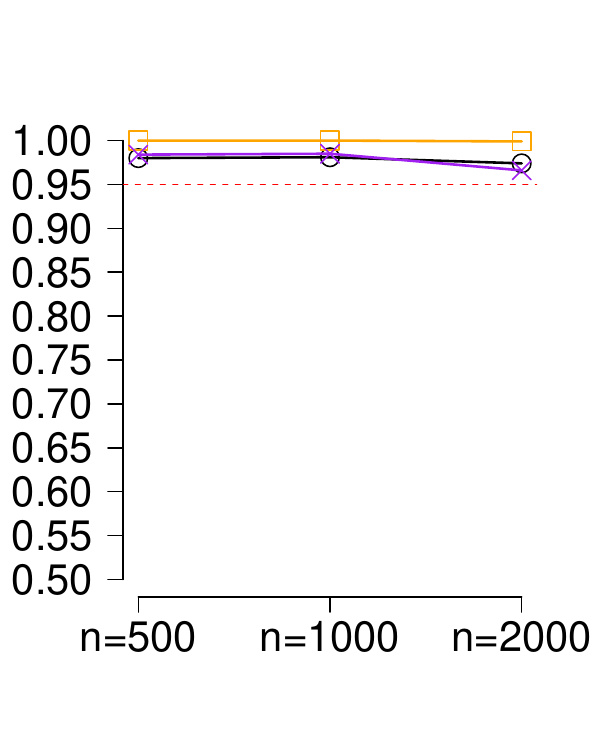} 
    \end{center}
    \vspace{-.4in}
    \subcaption{$\hat{R}^2 (Y \sim Z | X, D) \approx 0.80$}\label{fig.fixbootstrap3.y3}
    \end{subfigure}
    \subcaption*{\textit{Note:} Coverage rates of 95\% confidence intervals across 1000 iterations of \ref{eq:fixdgp}. The dashed line indicates the target coverage rate of 0.95. Weights are formed by one-to-one propensity score matching with replacement for the ATT, and they employ the rescaling proposed in Section~\ref{subsec:w_distr_normalize}. For inference Option 2, we draw $B = 1000$ bootstrap samples at each iteration of \ref{eq:fixdgp}.}
	\end{center}
	\vspace{-0.25in}
	\end{figure}

We apply our adjusted inference Options 2 (Section~\ref{subsec:se_fixed_bootstrap}) and 3 (Section~\ref{subsec:se_closed}), which are most appropriate in fixed design settings with honest weights. We examine 95\% confidence intervals, obtained through the standard error method, for the desired estimand (all 0, in this DGP) with three types of honest weights: inverse propensity score weights for the ATE (Figure~\ref{fig.fixbootstrap1}), entropy balancing weights for the ATT (Figure~\ref{fig.fixbootstrap2}), and one-to-one propensity score matching with replacement for the ATT (Figure~\ref{fig.fixbootstrap3}). For each weighting method, we fix $\hat{R}^2_w (D \sim Z | X)$ to be the true value in the fixed design. Then, for each weighting method and each $\sigma$ (see \ref{eq:fixdgp}), we fix $\hat{R}^2_w (Y \sim Z | X, D)$ to be its approximate probability limit.\footnote{We approximate the probability limit of $\hat{R}^2_w (Y \sim Z | X, D)$ with its empirical average across 1000 draws of \ref{eq:fixdgp} with $n = 10000$. 
For the inverse propensity score weights (Figure~\ref{fig.fixbootstrap1}), we set 
$\hat{R}_w^2 (Y \sim Z | D, X) = 0.1888$ when $\hat{R}^2 (Y \sim Z | D, X) \approx 0.20$; 
$\hat{R}_w^2 (Y \sim Z | D, X) = 0.3344$ when $\hat{R}^2 (Y \sim Z | D, X) \approx 0.35$; 
$\hat{R}_w^2 (Y \sim Z | D, X) = 0.4827$ when $\hat{R}^2 (Y \sim Z | D, X) \approx 0.50$; 
$\hat{R}_w^2 (Y \sim Z | D, X) = 0.6338$ when $\hat{R}^2 (Y \sim Z | D, X) \approx 0.65$; and 
$\hat{R}_w^2 (Y \sim Z | D, X) = 0.7888$ when $\hat{R}^2 (Y \sim Z | D, X) \approx 0.80$. 
For entropy balancing weights (Figure~\ref{fig.fixbootstrap2}), we set 
$\hat{R}_w^2 (Y \sim Z | D, X) = 0.1466$ when $\hat{R}^2 (Y \sim Z | D, X) \approx 0.20$; 
$\hat{R}_w^2 (Y \sim Z | D, X) = 0.2710$ when $\hat{R}^2 (Y \sim Z | D, X) \approx 0.35$; 
$\hat{R}_w^2 (Y \sim Z | D, X) = 0.4082$ when $\hat{R}^2 (Y \sim Z | D, X) \approx 0.50$; 
$\hat{R}_w^2 (Y \sim Z | D, X) = 0.5616$ when $\hat{R}^2 (Y \sim Z | D, X) \approx 0.65$; and 
$\hat{R}_w^2 (Y \sim Z | D, X) = 0.7343$ when $\hat{R}^2 (Y \sim Z | D, X) \approx 0.80$. 
For the matching weights (Figure~\ref{fig.fixbootstrap3}), we set 
$\hat{R}_w^2 (Y \sim Z | D, X) = 0.1171$ when $\hat{R}^2 (Y \sim Z | D, X) \approx 0.20$; 
$\hat{R}_w^2 (Y \sim Z | D, X) = 0.2235$ when $\hat{R}^2 (Y \sim Z | D, X) \approx 0.35$; 
$\hat{R}_w^2 (Y \sim Z | D, X) = 0.3481$ when $\hat{R}^2 (Y \sim Z | D, X) \approx 0.50$; 
$\hat{R}_w^2 (Y \sim Z | D, X) = 0.4978$ when $\hat{R}^2 (Y \sim Z | D, X) \approx 0.65$; and 
$\hat{R}_w^2 (Y \sim Z | D, X) = 0.6814$ when $\hat{R}^2 (Y \sim Z | D, X) \approx 0.80$.} 

As in Appendix~\ref{app:bootstrap}, we compare confidence intervals from inference Options 2 and 3 with the ``default", homoscedastic 95\% confidence intervals for $\hat{\tau}_{\mathrm{target}}$ directly pulled from the unadjusted \texttt{lm()} output (in \texttt{R}) from a weighted regression of $Y$ on $X$, $D$, and $Z$. These are included here to emphasize the fact that our inference Options 2 and 3 are valid under heteroscedasticity, while these ``default" confidence intervals should struggle in \ref{eq:fixdgp} here because the error $\epsilon$ is heteroscedastic. 

For inverse propensity score weights (Figure~\ref{fig.fixbootstrap1}) and entropy balancing weights (Figure~\ref{fig.fixbootstrap2}), Option 2 produces confidence intervals that achieve nominal coverage when $\hat{R}^2 (Y \sim Z \ | \ X, D)$ is low ($\approx 0.20$). Meanwhile, the ``default" 95\% confidence intervals show clear undercoverage due to the heteroscedasticity in \ref{eq:fixdgp} for all $\hat{R}^2 (Y \sim Z \ | \ X, D)$. As expected, however, Option 2 is increasingly conservative as $\hat{R}^2 (Y \sim Z \ | \ X, D)$ increases ($\approx 0.50$ or $\approx 0.80$). For the matching weights (Figure~\ref{fig.fixbootstrap3}), Option 2 is even noticeably conservative when $\hat{R}^2 (Y \sim Z \ | \ X, D) \approx 0.20$. We further investigate the relationship between in Option 2's conservativeness and $\hat{R}^2 (Y \sim Z \ | \ X, D)$ in Appendix~\ref{app:fixed_bootstrap_option2}.

Inference Option 3's performance varies widely depending on the choice of weights and the true $\hat{R}^2 (Y \sim Z \ | \ X, D)$. For the entropy balancing weights (Figure~\ref{fig.fixbootstrap2}), Option 3's 95\% confidence intervals achieve near nominal coverage for every $n$ and $\hat{R}^2 (Y \sim Z \ | \ X, D)$ shown. However, for matching weights (Figure~\ref{fig.fixbootstrap3}), Option 3 shows consistent, albeit slight, overcoverage. Option 3's performance with inverse propensity score weights for the ATE  (Figure~\ref{fig.fixbootstrap1}) is most concerning, showing undercoverage as $\hat{R}^2 (Y \sim Z \ | \ X, D)$ increases. Though, Option 3's confidence intervals are still superior to the ``default" confidence intervals for inverse propensity score weighting (Figure~\ref{fig.fixbootstrap1}) and matching (Figure~\ref{fig.fixbootstrap3}). 


\newpage

\clearpage

\subsubsection{Deeper dive: Option 2}\label{app:fixed_bootstrap_option2}
\begin{figure}[!h]
	\vspace{0.15in}
	\begin{center}
	\caption{Relationship between $\hat{R}^2 (Y \sim Z | X, D)$ and the overestimation of the standard error from inference Option 2 in \ref{eq:fixdgp}}\label{fig.se.overestimation}
    \vspace{-.8in}
    
    \includegraphics[scale=0.45]{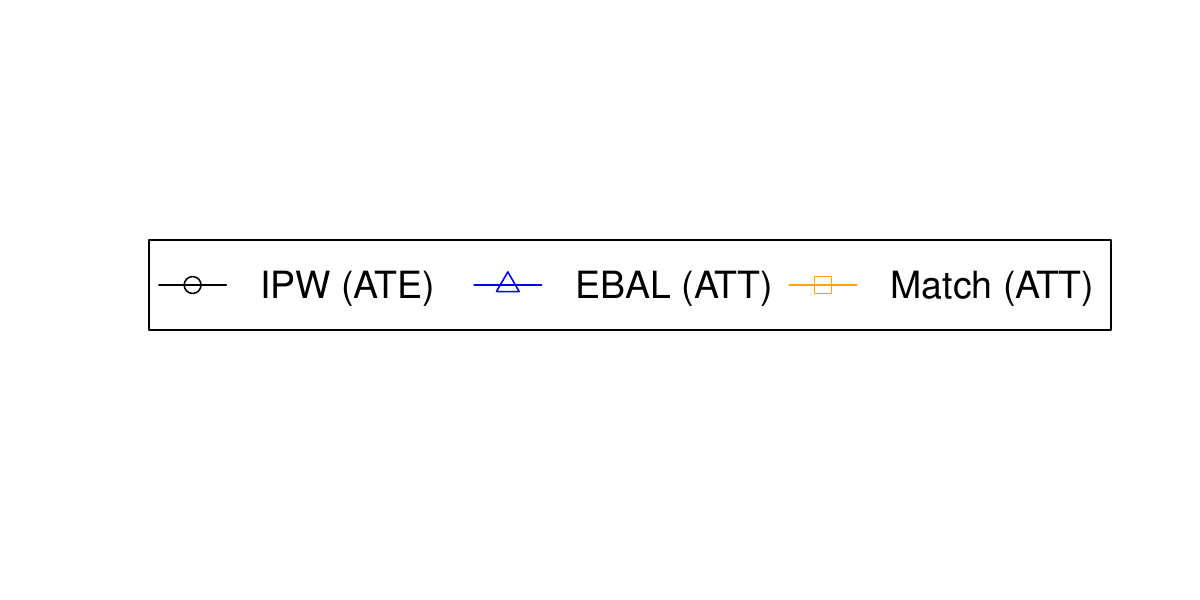} 

    \vspace{-0.9in}

    \begin{subfigure}{.48\textwidth}
    \begin{center}
    \includegraphics[scale=0.45]{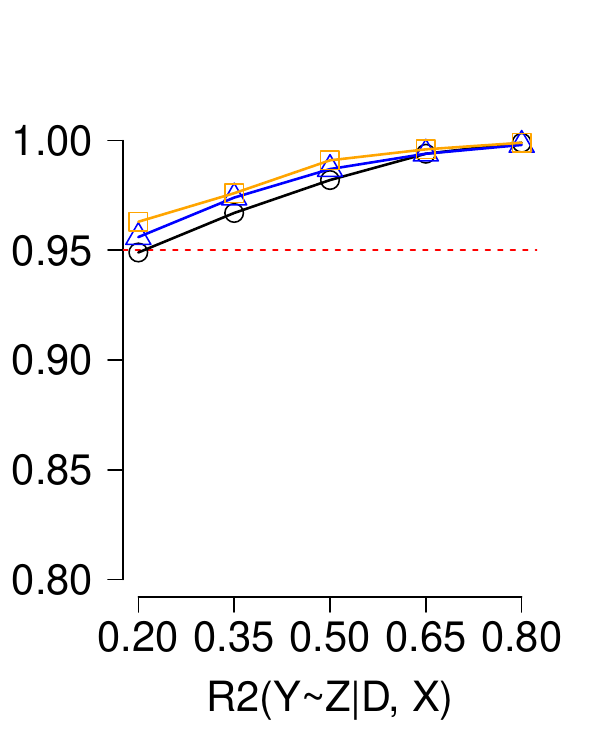} 
    \end{center}
    \vspace{-.2in}
    \subcaption{Coverage}\label{fig.se.overestimation.coverage}
    \end{subfigure}
    \begin{subfigure}{.48\textwidth}
    \begin{center}
    \includegraphics[scale=0.45]{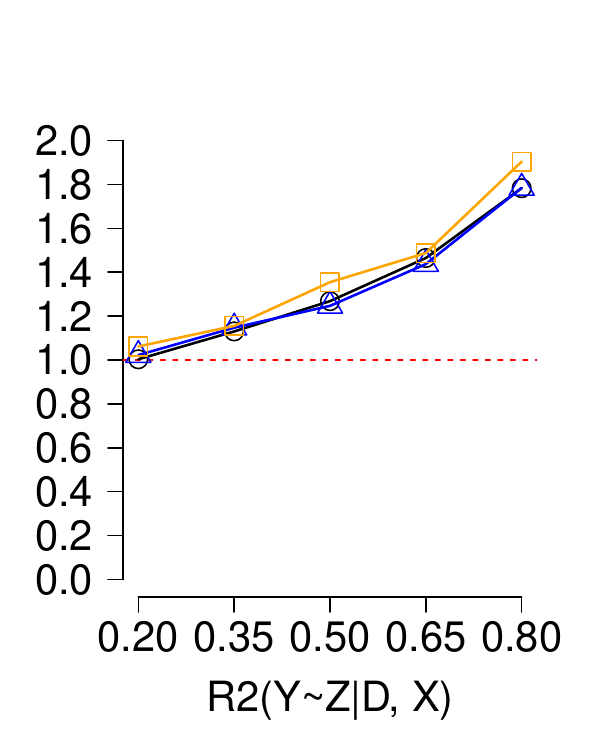} 
    \end{center}
    \vspace{-.2in}
    \subcaption{SE Ratio}\label{fig.se.overestimation.seratio}
    \end{subfigure}

    \subcaption*{\textit{Note:} Results across 1000 iterations of \ref{eq:fixdgp} for varying $\hat{R}^2 (Y \sim Z | X, D)$. Weights are inverse propensity score weights for the ATE (IPW), entropy balancing weights for the ATT (EBAL), and one-to-one propensity score matching for the ATT (Match). All weights employ the rescaling proposed in Section~\ref{subsec:w_distr_normalize}. For inference Option 2, we draw $B = 1000$ bootstrap samples at each iteration of \ref{eq:fixdgp}. \textit{(a)} Coverage rates of 95\% confidence intervals from inference Option 2. The dashed line indicates the target coverage rate of 0.95. \textit{(b)} Average SE Ratios from inference Option 2. The dashed line indicates the target ratio of 1.}
	\end{center}
	\vspace{-0.4in}
	\end{figure}

We inspect the relationship between $\hat{R}^2 (Y \sim Z \ | \ X, D)$ and Option 2's overestimation of the standard error more deeply in Figure~\ref{fig.se.overestimation}. Figure~\ref{fig.se.overestimation.coverage} reports coverage rates of 95\% confidence intervals from Option 2 over all values of  $\hat{R}^2 (Y \sim Z \ | \ X, D)$ tried. Figure~\ref{fig.se.overestimation.seratio} then calculates the ratio of the estimated standard errors at each iteration, and the true standard error: $\text{SE Ratio} = \frac{\widehat{\se}_{\text{fix-boot}} (\hat{\tau}_{\mathrm{target}})}{\se_{\mathrm{true}} (\hat{\tau}_{\mathrm{target}})}$. The target ratio is $\text{SE Ratio}=1$, which implies that $\widehat{\se}_{\text{fix-boot}} (\hat{\tau}_{\mathrm{target}})$ is accurately estimating $\se_{\mathrm{true}} (\hat{\tau}_{\mathrm{target}})$. If $\text{SE Ratio}>1$, then $\widehat{\se}_{\text{fix-boot}} (\hat{\tau}_{\mathrm{target}})$ from Option 2 overestimates the true standard error, and if $\text{SE Ratio}<1$, then $\widehat{\se}_{\text{fix-boot}} (\hat{\tau}_{\mathrm{target}})$ underestimates the true standard error. Figure~\ref{fig.se.overestimation.seratio} then reports the average SE Ratio across the iterations of each $\hat{R}^2 (Y \sim Z \ | \ X, D)$ tried. 

Figure~\ref{fig.se.overestimation} shows that Option 2's overestimation when  $\hat{R}^2 (Y \sim Z \ | \ X, D) \leq 0.50$ is acceptable in \ref{eq:fixdgp}. For each weighting method, Option 2 over-estimates the true standard error by roughly 30\% when $\hat{R}^2 (Y \sim Z \ | \ X, D) = 0.50$ (see Figure~\ref{fig.se.overestimation.seratio}). This produces coverage rates of 95\% confidence intervals around 98\%, which is acceptable (see Figure~\ref{fig.se.overestimation.coverage}). When $\hat{R}^2 (Y \sim Z \ | \ X, D) \leq 0.35$, Option 2 barely over-estimates the true standard error and produces only slightly higher-than-nominal coverages rates for 95\% confidence intervals. When $\hat{R}^2 (Y \sim Z \ | \ X, D) = 0.65$ or $0.80$, however, Option 2 becomes overly conservative: it overestimates the true standard error by over 50\% (see Figure~\ref{fig.se.overestimation.seratio}), and produces coverage rates that are at or near 100\% (see Figure~\ref{fig.se.overestimation.coverage}).

To summarize the lessons from this demonstration, Option 2 performs well in fixed design settings with honest weights. When the true $\hat{R}^2 (Y \sim Z \ | \ X, D)$ is lower (e.g., $\leq 0.35$), it well-estimates the true standard error and produces confidence intervals that achieve near-nominal rates. Option 2's conservativeness does produce over-coverage from 95\% confidence intervals for medium values of $\hat{R}^2 (Y \sim Z \ | \ X, D)$ (e.g., $\approx 0.50$), but this overcoverage is slight and acceptable. When the true $\hat{R}^2 (Y \sim Z \ | \ X, D)$ is higher (e.g., $>0.50$), Option 2 becomes overly conservative. Thus, in practice, when the researcher assumes $\hat{R}^2_w (Y \sim Z \ | \ X, D) \leq 0.50$, they can be reasonably certain that Option 2 will only be mildly conservative. When assuming $\hat{R}^2_w (Y \sim Z \ | \ X, D) \gg 0.50$, one can be fairly certain that Option 2 is overly-conservative.

\newpage
\clearpage

\subsection{Demonstration: Modifying inference Option 1 with a cluster-bootstrap}\label{app:cluster_bootstrap}

This section demonstrates the merits of modifying adjusted inference Option 1 (Section~\ref{subsec:se_random_bootstrap}) with a cluster-bootstrap, as described in Section~\ref{subsec:se_extension}. The data generating process (DGP) here expands on \ref{eq:heterodpg} (see Appendix~\ref{app:bootstrap}) to allow for clustering. First, we define new indices that can express clustered data. Let $g = 1, \dots, G$ index the group, and let $g[i]$ index unit $i$ in group $g$. Each group has size $n_g = 50$. The data is then generated as follows:
    \begin{align}\label{eq:clusterdgp}
        Y_{g[i]} &= X_{g[i]} + Z_{g[i]} + \delta_g D_{g[i]} + \epsilon_{g[i]} \nonumber \\
        \text{and} \ \  p(D_{g[i]} = 1 | X_{g[i]}, Z_{g[i]})& = \frac{\mathrm{exp} (X_{g[i]} + Z_{g[i]} - 1)}{1 + \mathrm{exp} (X_{g[i]} + Z_{g[i]} - 1)} \nonumber \\
        \text{where} \ \ [X_{g[i]} \ Z_{g[i]}]^{\top} &\overset{iid}{\sim} \mathcal{N}(0, I_2), \ \ \epsilon_{g[i]} \overset{iid}{\sim} N(0, 2), \ \ \text{and} \ \ \delta_g \overset{iid}{\sim} N(0, \theta^2) \tag{DGP 3}
    \end{align}
where $(\delta_g D_{g[i]} + \epsilon_{g[i]})$ makes up a combined error term that is clustered by groups, and $\theta^2 \in \{0, 4, 16\}$ determines the extent of the dependence within groups. Note that when $\theta^2 = 0$, the errors are mutually independent and homoscedastic. Furthermore, there is no treatment effect in \ref{eq:clusterdgp} (i.e., the ATE, ATT, and ATC are all 0). Finally, this DGP has a random design: $\mathbf{X}$, $\mathbf{Z}$, and $\mathbf{D}$ are  regenerated in each iteration of the simulation.

We apply our adjusted inference Option 1 (Section~\ref{subsec:se_random_bootstrap}), modified with a cluster bootstrap. We calculate 95\% confidence intervals, obtained through the percentile method, for the true treatment effect (of 0, in this DGP) with inverse propensity score weights for the ATE. For each $\theta^2$ (see \ref{eq:clusterdgp}), we fix the sensitivity parameters at their approximate probability limits.\footnote{We approximate the probability limits of the sensitivity parameters by taking their empirical averages across 1000 draws of \ref{eq:clusterdgp} with $n_g=200$. We set $\hat{R}_w^2 (D \sim Z | X) = 0.1441$, which is the average across 1000 draws when $G = 50$. For $\theta^2 = 0$, we set $\hat{R}_w^2 (Y \sim Z | D, X) = 0.3007$ when $G=15$, $\hat{R}_w^2 (Y \sim Z | D, X) = 0.3023$ when $G=25$, and $\hat{R}_w^2 (Y \sim Z | D, X) = 0.3008$ when $G=50$. For $\theta^2 = 4$, we set $\hat{R}_w^2 (Y \sim Z | D, X) = 0.2237$ when $G=15$, $\hat{R}_w^2 (Y \sim Z | D, X) = 0.2221$ when $G=25$, and $\hat{R}_w^2 (Y \sim Z | D, X) = 0.2199$ when $G=50$. For $\theta^2 = 16$, we set $\hat{R}_w^2 (Y \sim Z | D, X) = 0.1301$ when $G=15$, $\hat{R}_w^2 (Y \sim Z | D, X) = 0.1257$ when $G=25$, and $\hat{R}_w^2 (Y \sim Z | D, X) = 0.1227$ when $G=50$.} As in Appendix~\ref{app:bootstrap}, we compare our proposed confidence intervals from Option 1 with  the ``default", homoscedastic 95\% confidence intervals for $\hat{\tau}_{\mathrm{target}}$ directly pulled from the unadjusted \texttt{lm()} output (in \texttt{R}) from a weighted regression of $Y$ on $X$, $D$, and $Z$. These ``default" confidence intervals are expected to fail, as they assume no within-group dependence. 

\begin{figure}[!h]
	\vspace{0.15in}
	\begin{center}
	\caption{Coverage rates in \ref{eq:clusterdgp} for inverse propensity score weights for the ATE}\label{fig.clusterbootstrap1}
    \vspace{-.6in}
    
    \includegraphics[scale=0.45]{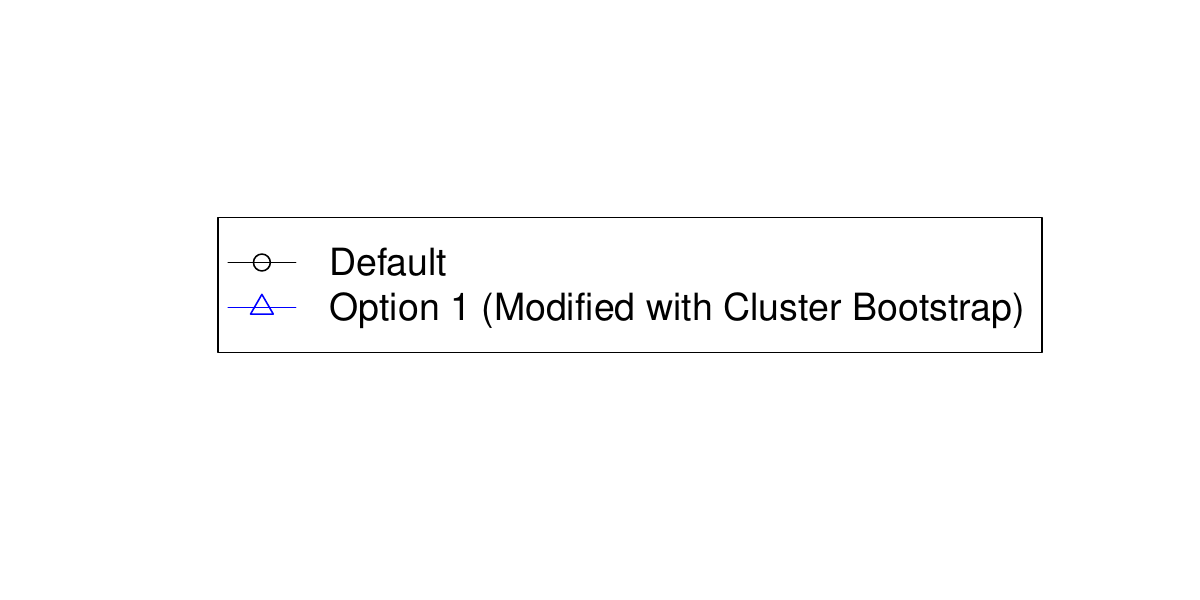} 

    \vspace{-0.9in}
    \begin{subfigure}{.30\textwidth}
    \begin{center}
    \includegraphics[scale=0.45]{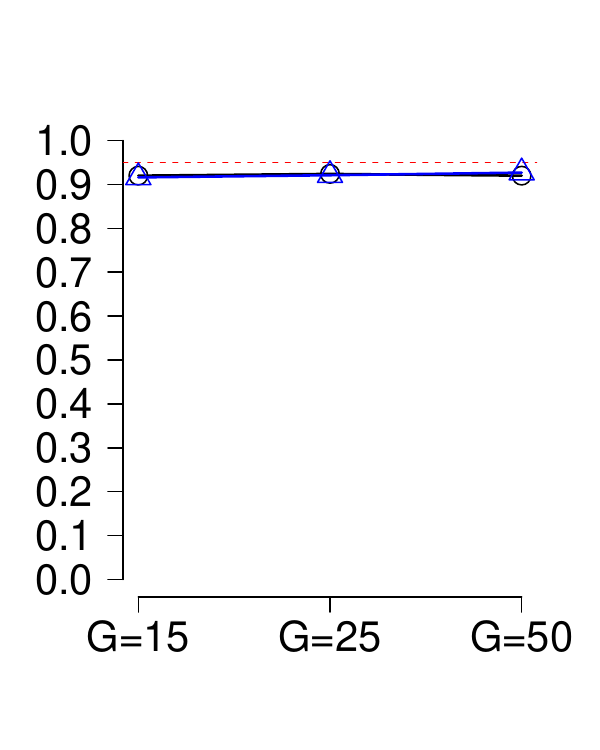} 
    \end{center}
    \vspace{-.25in}
    \subcaption{$\theta^2 = 0$}\label{fig.clusterbootstrap1.y1}
    \end{subfigure}
    \begin{subfigure}{.30\textwidth}
    \begin{center}
    \includegraphics[scale=0.45]{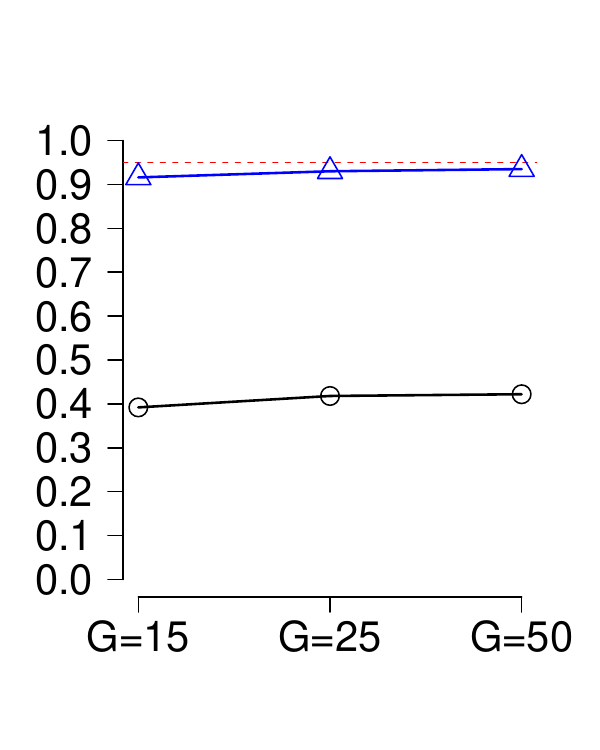} 
    \end{center}
    \vspace{-.25in}
    \subcaption{$\theta^2 = 4$}\label{fig.clusterbootstrap1.y2}
    \end{subfigure}
    \begin{subfigure}{.30\textwidth}
    \begin{center}
    \includegraphics[scale=0.45]{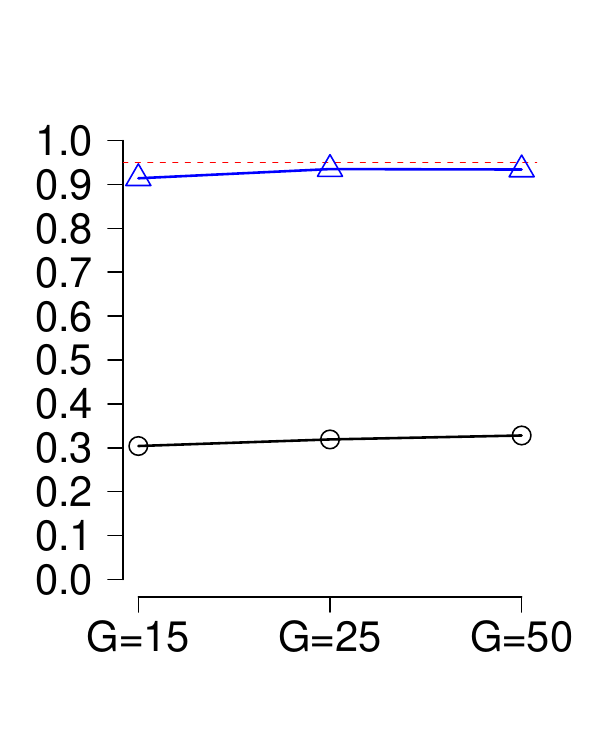} 
    \end{center}
    \vspace{-.25in}
    \subcaption{$\theta^2 = 16$}\label{fig.clusterbootstrap1.y3}
    \end{subfigure}
    \subcaption*{\textit{Note:} Coverage rates of 95\% confidence intervals across 1000 iterations of \ref{eq:clusterdgp}, with each $n_g=50$. The dashed line indicates the target coverage rate of 0.95. Weights are inverse propensity score weights for the ATE, and they employ the rescaling proposed in Section~\ref{subsec:w_distr_normalize}. For inference Option 1, we draw $B = 1000$ bootstrap samples at each iteration of \ref{eq:clusterdgp}. }
	\end{center}
	\vspace{-0.25in}
	\end{figure}

Figure~\ref{fig.clusterbootstrap1} shows that confidence intervals from our modified Option 1 achieve nominal coverage rates for all $\theta^2$  when $G$ is sufficiently large. Meanwhile, the ``default", homoscedastic 95\% confidence intervals achieve nominal coverage when $\theta^2 = 0$, but show worsening undercoverage as the dependence within groups increases (i.e., $\theta^2$ increases).

\subsection{Demonstration: Large ``translator" term in Expression~\ref{eq:wsa_kappa_rewrite} }\label{app:translator}

\begin{figure}[!h]
	\vspace{0.15in}
	\begin{center}
	\caption{Weighted distribution in \ref{eq:problemdgp}}\label{fig.problemdgp}
    \vspace{-.25in}
    \begin{subfigure}{.30\textwidth}
    \begin{center}
    \includegraphics[scale=0.35]{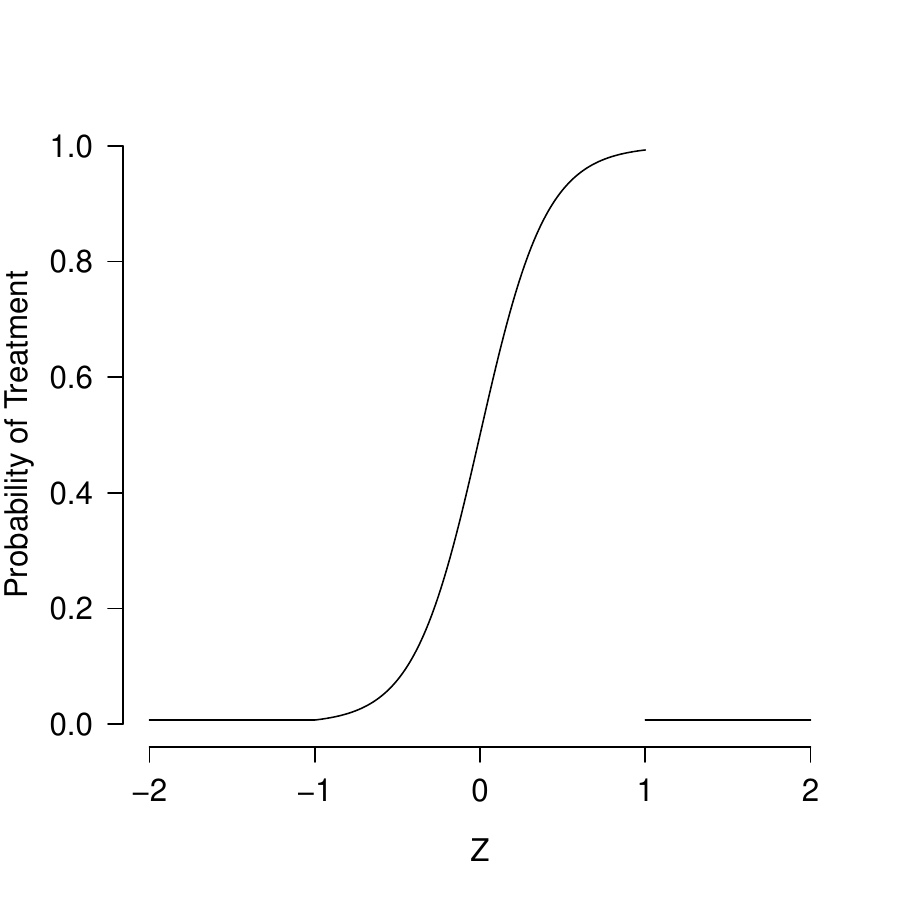} 
    \end{center}
    \vspace{-.25in}
    \subcaption{Probability of treatment}\label{fig.problemdgp.ps}
    \end{subfigure}
    \begin{subfigure}{.30\textwidth}
    \begin{center}
    \includegraphics[scale=0.35]{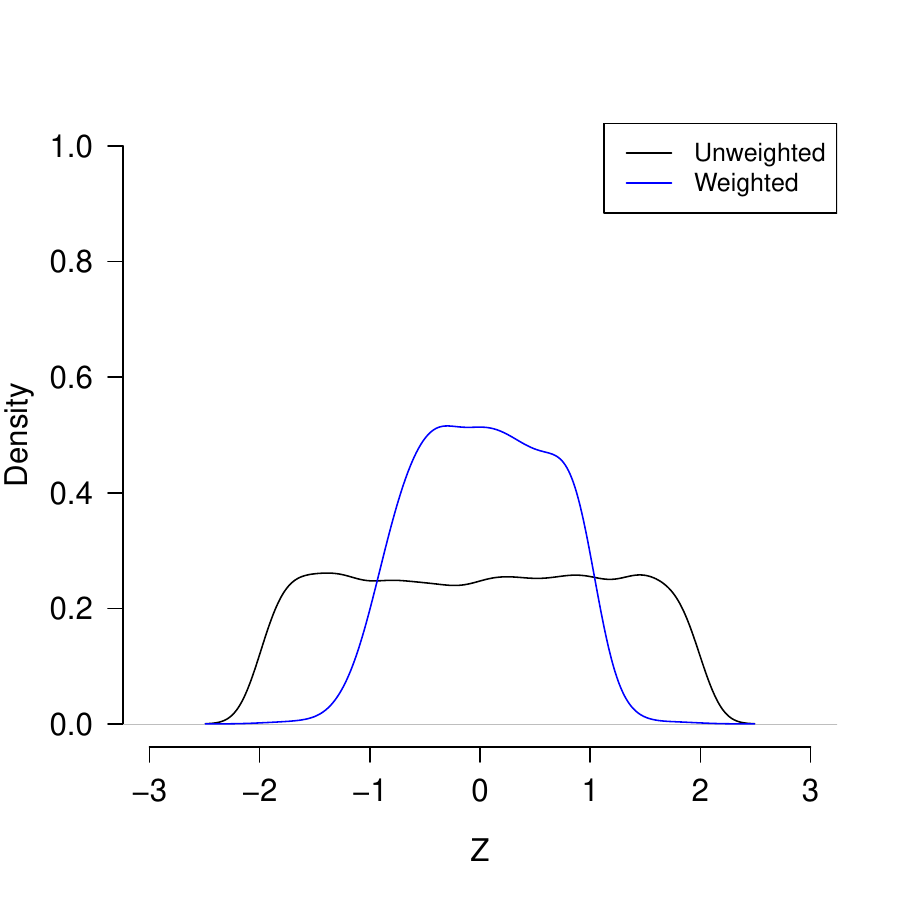} 
    \end{center}
    \vspace{-.25in}
    \subcaption{Density of $Z$}\label{fig.problemdgp.zdensity}
    \end{subfigure}
    \begin{subfigure}{.30\textwidth}
    \begin{center}
    \includegraphics[scale=0.35]{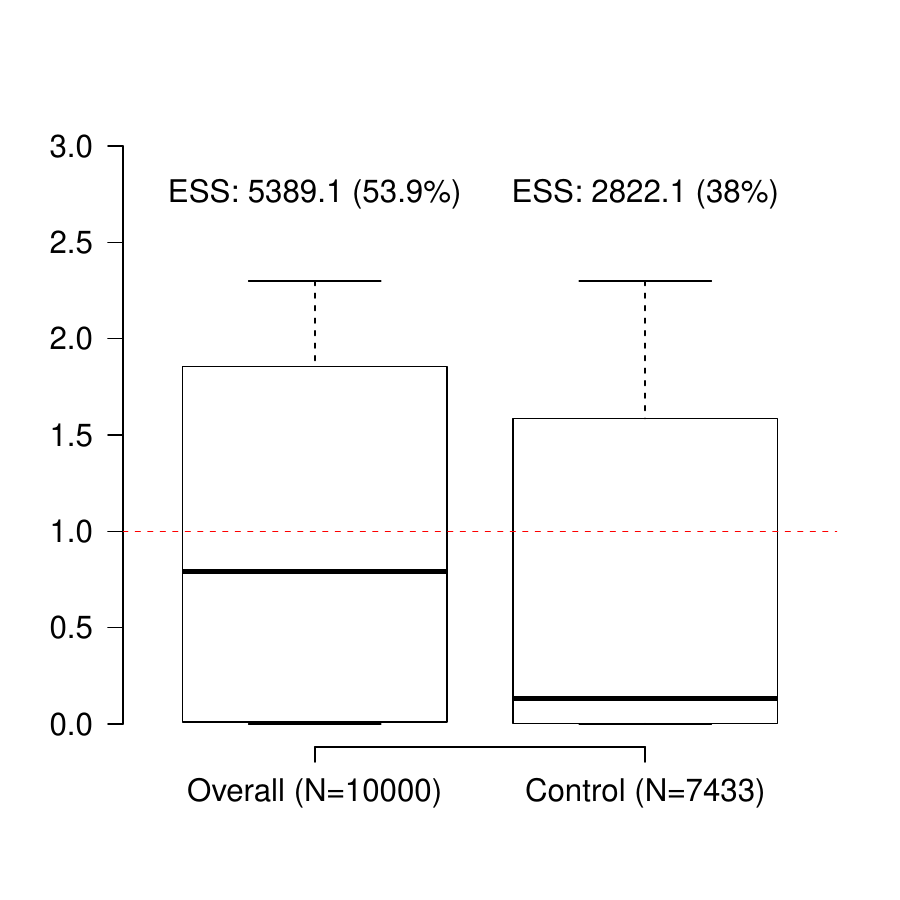} 
    \end{center}
    \vspace{-.25in}
    \subcaption{Distribution of weights}\label{fig.problemdgp.wdistribution}
    \end{subfigure}
    \subcaption*{\textit{Note:} Results across one iteration of \ref{eq:problemdgp} with $n = 10000$. Weights are found by Entropy Balancing in (\ref{eq:ebal}), with the rescaling in Section~\ref{subsec:w_distr_normalize}. \textit{(a)} Probability of treatment across $Z$. \textit{(b)} Weighted kernel density plot of $Z$ in the semi-weighted (here, unweighted) and weighted distributions. \textit{(c)} Distribution of the weights overall and within the control group. Percentages represent the effective sample size divided by the sample size within the group (i.e., $(100 \times \frac{\text{ESS}}{\mathrm{N}})\%$). }
	\end{center}
	\vspace{-0.25in}
	\end{figure}

To illustrate how the translator term in Expression~\ref{eq:wsa_kappa_rewrite} can be large, consider a data-generating process (DGP) where the probability of treatment is entirely determined by a $Z$: 
    \begin{align}\label{eq:problemdgp}
        p(D = 1 \ | \ Z) = \begin{cases}
            \ \ \ 0.007 & \text{if} \ |Z| > 1 \\
            \frac{\mathrm{exp} (5*Z)}{1 + \mathrm{exp} (5*Z)} & \text{if} \ |Z| \leq 1
            \end{cases} \ \ \text{where} \ \ Z \overset{iid}{\sim} \mathrm{Unif}(-2, 2) \tag{DGP 4}
    \end{align}
However, the researcher observes only $X = Z^4$. Because $X$ is one-dimensional, were it used to benchmark the strength of $Z$, the semi-weights would be uniform weights (i.e., $\hat{R}_{w^{(-j)}}^2 = \hat{R}^2$) and $X^{(-j)}$ would be an empty vector (i.e., $\hat{R}_w^2 (D \sim Z | X^{(-j)}) = \hat{R}_w^2 (D \sim Z)$ and $\hat{R}^2 (D \sim Z | X^{(-j)}) = \hat{R}^2 ( D \sim Z)$). Therefore, the translator in Expression~\ref{eq:wsa_kappa_rewrite} could be rewritten as $\frac{ \hat{R}_w^2 (D \sim Z | X^{(-j)})}{ \hat{R}_{w^{(-j)}}^2 (D \sim Z | X^{(-j)})} = \frac{\hat{R}_w^2 (D \sim Z)}{\hat{R}^2 (D \sim Z)}$, or the squared ratio of the weighted and unweighted correlations of $D$ and $Z$. From Figure~\ref{fig.problemdgp.ps}, it is apparent that $Z$ and $D$ are moderately correlated overall (at approximately 0.218), but are highly correlated for $|Z| \leq 1$ (at approximately 0.758). Thus, were $X$ used to benchmark the strength of $Z$, weights that neglect (i.e., set $w_i \approx 0$) units with $|Z_i|>1$ would yield a large translator term. Figure~\ref{fig.problemdgp.zdensity} shows that this occurs with balancing weights from Expression~\ref{eq:ebal} --- these weights focus on units with $|Z_i| \leq 1$, and thus the translator is $\frac{\hat{R}_w^2 (D \sim Z)}{\hat{R}^2 (D \sim Z)} \approx \frac{0.394}{0.051} = 7.770$.
This occurs because, while about half of all units have $|Z_i| > 1$ (or $X_i > 1$), these are essentially all control units because their probability of treatment is minuscule. Thus, Entropy Balancing gives these control units small weights because treated units almost all have $|Z_i| \leq 1$ (or $X_i \leq 1$). Figure~\ref{fig.problemdgp.wdistribution} shows the effect of this on the weighted distribution --- a large portion of the weights nears 0, and the effective sample size within the control group represents only 38.0\% of the group.

While extreme, this DGP is instructive: in settings where the weighted and semi-weighted distributions are very different,  one risks underestimating the strength of the relationship between $Z$ and $D$ by neglecting the translator in Expression~\ref{eq:wsa_kappa_rewrite}.

\newpage
\clearpage

\section{Proofs}\label{app:proofs}

\subsection{Derivation of $\widehat{\mathrm{bias}} (\hat{\tau}_{\mathrm{wls}} )$}\label{app:proofs_wsa_bias}

Without loss of generality, let $X$, $D$, $Y$, and $Z$ be centered by their weighted sample means (i.e., $\widehat{\E}_w (\cdot)$). 
$\hat{\tau}_{\mathrm{wls}}$ results from a weighted regression of $Y^{\perp_w X}$ on $D^{\perp_w X}$, so
	\begin{align}\label{eq:proofs_wsa_bias_1}
		\hat{\tau}_{\mathrm{wls}} &= \frac{\widehat{\cov}_w (D^{\perp_w X}, Y^{\perp_w X}) }{\widehat{\var}_w (D^{\perp_w X})}
	\end{align}
Additionally, $\hat{\tau}_{\mathrm{target}}$ and $\hat{\gamma}_{\mathrm{target}}$ result from a weighted regression of $Y^{\perp_w X}$ on $(D^{\perp_w X}, Z^{\perp_w X})$. Therefore,
	\begin{align}\label{eq:proofs_wsa_bias_2}
		 & \widehat{\cov}_w \biggr( D^{\perp_w X}, Y^{\perp_w X} - (\hat{\tau}_{\mathrm{target}} D^{\perp_w X} + \hat{\gamma}_{\mathrm{target}} Z^{\perp_w X}) \biggr)  = 0 \nonumber \\
		 \implies & \widehat{\cov}_w ( D^{\perp_w X}, Y^{\perp_w X}) = \widehat{\cov}_w ( D^{\perp_w X}, \hat{\tau}_{\mathrm{target}} D^{\perp_w X} + \hat{\gamma}_{\mathrm{target}} Z^{\perp_w X})
	\end{align}
which allows (\ref{eq:proofs_wsa_bias_1}) to continue as
	\begin{align}\label{eq:proofs_wsa_bias_3}
		\hat{\tau}_{\mathrm{wls}} &= \frac{\widehat{\cov}_w ( D^{\perp_w X}, \hat{\tau}_{\mathrm{target}} D^{\perp_w X} + \hat{\gamma}_{\mathrm{target}}  Z^{\perp_w X} ) }{\widehat{\var}_w (D^{\perp_w X})}  \nonumber \\
		 &=  \hat{\tau}_{\mathrm{target}} + \hat{\gamma}_{\mathrm{target}}  \frac{\widehat{\cov}_w ( D^{\perp_w X},  Z^{\perp_w X} ) }{\widehat{\var}_w (D^{\perp_w X})}
	\end{align}
Then, 
$\hat{\gamma}_{\mathrm{target}} $ results from a regression of $Y^{\perp_w X, D}$ on $Z^{\perp_w X, D}$, meaning that
	\begin{align}\label{eq:proofs_wsa_bias_4}
		\hat{\gamma}_{\mathrm{target}} &= \frac{\widehat{\cov}_w (Z^{\perp_w X, D}, Y^{\perp_w X, D}) }{\widehat{\var}_w (Z^{\perp_w X, D})} 
	\end{align}
Applying (\ref{eq:proofs_wsa_bias_4}) then allows (\ref{eq:proofs_wsa_bias_3}) to continue as
	\begin{align}\label{eq:proofs_wsa_bias_5}	
		\hat{\tau}_{\mathrm{wls}} &= \hat{\tau}_{\mathrm{target}} + \biggr[ \frac{\widehat{\cov}_w (Z^{\perp_w X, D}, Y^{\perp_w X, D}) }{\widehat{\var}_w (Z^{\perp_w X, D})}  \biggr] \biggr[ \frac{\widehat{\cov}_w ( D^{\perp_w X},  Z^{\perp_w X} ) }{\widehat{\var}_w (D^{\perp_w X})} \biggr] \nonumber \\
		&=  \hat{\tau}_{\mathrm{target}}  +  \biggr[ \hat{R}_w (Y \sim Z | X, D) \biggr( \frac{ \widehat{\sd}_w (Y^{\perp_w X, D})}{\widehat{\sd}_w (Z^{\perp_w X, D})} \biggr)\biggr]  \biggr[ \hat{R}_w (D \sim Z | X) \biggr( \frac{ \widehat{\sd}_w (Z^{\perp_w X})}{\widehat{\sd}_w (D^{\perp_w X})} \biggr) \biggr] \nonumber \\
		&=  \hat{\tau}_{\mathrm{target}} + \biggr( \frac{ \hat{R}_w (Y \sim Z | X, D) \times \hat{R}_w (D \sim Z | X)}{ \frac{\widehat{\sd}_w (Z^{\perp_w X, D})}{\widehat{\sd}_w (Z^{\perp_w X})}} \biggr) \biggr( \frac{ \widehat{\sd}_w (Y^{\perp_w X, D})}{\widehat{\sd}_w (D^{\perp_w X})}{} \biggr)
	\end{align}
Further noting that $\frac{\widehat{\sd}_w (Z^{\perp_w X, D})}{\widehat{\sd}_w (Z^{\perp_w X})} = \sqrt{ 1 - \hat{R}^2_w (Z \sim D | X) } = \sqrt{ 1 - \hat{R}_w^2 (D \sim Z | X) }$ allows (\ref{eq:proofs_wsa_bias_5}) to continue as
	\begin{align}\label{eq:proofs_wsa_bias_6}	
		\hat{\tau}_{\mathrm{wls}} = \hat{\tau}_{\mathrm{target}}  + \biggr( \frac{ \hat{R}_w (Y \sim Z | X, D) \times \hat{R}_w (D \sim Z | X)}{\sqrt{ 1 - \hat{R}_w^2 (D \sim Z | X) }} \biggr) \biggr( \frac{ \widehat{\sd}_w (Y^{\perp_w X, D})}{ \widehat{\sd}_w (D^{\perp_w X})}{} \biggr)
	\end{align}
Subtracting $\hat{\tau}_{\mathrm{target}}$ from both sides of (\ref{eq:proofs_wsa_bias_6}) completes the proof. 

\begin{flushright}
$\square$
\end{flushright}


\subsection{Derivation of closed-form standard errors}\label{app:proofs_wsa_se}

\subsubsection{Adjusted HC0}\label{app:proofs_wsa_se_hc0}

We first derive the adjusted HC0 standard error, $\widehat{\se}_{\text{adj-HC0}} ( \hat{\tau}_{\mathrm{target}})$, given in (\ref{eq:adjusted_hc0}). The HC0 standard error for $\hat{\tau}_{\mathrm{target}}$ is:
\begin{align}\label{eq:hc0_1}
    \widehat{\se}_{\mathrm{HC0}} ( \hat{\tau}_{\mathrm{target}}) = \sqrt{ \frac{\frac{1}{n^2} \sum_{i=1}^n w_i^2 \big( D_i^{\perp_w X, Z}\big)^2 \big( Y_i^{\perp_w X, D, Z}\big)^2}{ \big( \widehat{\var}_w (D^{\perp_w X, Z}) \big)^2}} 
\end{align}
Note that we can simplify the denominator of (\ref{eq:hc0_1}) with the identity:
\begin{align}\label{eq:hc0_2}
    \widehat{\var}_w (D^{\perp_w X, Z}) = \biggr( 1 - \hat{R}^2_w (D \sim Z | X) \biggr) \widehat{\var}_w (D^{\perp_w X})
\end{align}
Applying (\ref{eq:hc0_2}) to (\ref{eq:hc0_1}) yields:
\begin{align}\label{eq:hc0_3}
    \widehat{\se}_{\mathrm{HC0}} ( \hat{\tau}_{\mathrm{target}}) = \frac{1}{1 - \hat{R}^2_w (D \sim Z | X)} * \sqrt{ \frac{\frac{1}{n^2} \sum_{i=1}^n w_i^2 \big( D_i^{\perp_w X, Z}\big)^2 \big( Y_i^{\perp_w X, D, Z}\big)^2}{ \big( \widehat{\var}_w (D^{\perp_w X}) \big)^2}} 
\end{align}
Then, we can apply the uniform shrinkage approximation (Assumption~\ref{asm:uniform_shrinkage}) to simplify the numerator in (\ref{eq:hc0_3}) above, which completes the proof: 
\begin{align}\label{eq:hc0_4}
    \widehat{\se}_{\mathrm{HC0}} ( \hat{\tau}_{\mathrm{target}}) &\approx \sqrt{\frac{1 - \hat{R}^2_w (Y \sim Z | X, D)} {1 - \hat{R}^2_w (D \sim Z | X)} } * \underbrace{ \sqrt{ \frac{\frac{1}{n^2} \sum_{i=1}^n w_i^2 \big( D_i^{\perp_w X}\big)^2 \big( Y_i^{\perp_w X, D}\big)^2}{ \big( \widehat{\var}_w (D^{\perp_w X}) \big)^2}}}_{\widehat{\se}_{\mathrm{HC0}} ( \hat{\tau}_{\mathrm{wls}})} \nonumber \\
    &= \widehat{\se}_{\text{adj-HC0}} ( \hat{\tau}_{\mathrm{target}})
\end{align}

\begin{flushright}
$\square$
\end{flushright}

\subsubsection{Adjusted HC1}\label{app:proofs_wsa_se_hc1}

The HC1 standard error (\citealp{mackinnon1985some}) adjusts the HC0 standard error by multiplying it by a degrees of freedom constant: $\sqrt{ \frac{n}{n - \text{Number of Regressors}}}$. Because the number of regressors in the $Y \sim X + D + Z$ regression is $P + 2$, the HC1 standard error for $\hat{\tau}_{\mathrm{target}}$ is:
\begin{align}\label{eq:hc1_1}
    \widehat{\se}_{\mathrm{HC1}} (\hat{\tau}_{\mathrm{target}}) = \sqrt{ \frac{n}{n - P - 2}} \times \widehat{\se}_{\mathrm{HC0}} (\hat{\tau}_{\mathrm{target}})
\end{align}
Thus, we propose an adjusted HC1 standard error of:
\begin{align}
    \widehat{\se}_{\text{adj-HC1}} ( \hat{\tau}_{\mathrm{target}}) &= \sqrt{ \frac{n}{n - P - 2}} \times \widehat{\se}_{\text{adj-HC0}} ( \hat{\tau}_{\mathrm{target}}) \nonumber \\
    &= \sqrt{ \frac{n}{n - P - 2}} \times \sqrt{\frac{1 - \hat{R}^2_w (Y \sim Z | X, D)} {1 - \hat{R}^2_w (D \sim Z | X)} } \times \widehat{\se}_{\mathrm{HC0}} ( \hat{\tau}_{\mathrm{wls}})
\end{align}


\subsubsection{Adjusted cluster-robust standard error}\label{app:proofs_wsa_se_crse}

We first define notation that can express clustered data. Let $g = 1, \dots, G$ index the group, and let $g[i]$ index unit $i$ in group $g$. Each group $g$ then has size $n_g$. Then, let
\begin{align}
 \mathbf{X}_g = 
 \begin{bmatrix} 
 X_{g[1]}^{\top} \\
 \vdots \\
 X_{g[n_g]}^{\top} 
 \end{bmatrix} \in \mathbb{R}^{n_g \times P}, \ \ 
 \mathbf{D}_g = 
 \begin{bmatrix} 
 D_{g[1]} \\
 \vdots \\
 D_{g[n_g]} 
 \end{bmatrix} \in \mathbb{R}^{n_g}, \ \ 
 \mathbf{Z}_g = 
 \begin{bmatrix} 
 Z_{g[1]} \\
 \vdots \\
 Z_{g[n_g]} 
 \end{bmatrix} \in \mathbb{R}^{n_g}, \ \ 
 \mathbf{Y}_g = 
 \begin{bmatrix} 
 Y_{g[1]} \\
 \vdots \\
 Y_{g[n_g]} 
 \end{bmatrix} \in \mathbb{R}^{n_g}
\end{align}
be the matrices/vectors of $X$, $D$, $Z$, and $Y$ for group $g$, respectively. Finally, let \newline 
$\mathbf{W}_g = \textrm{diag}(w_{g[1]}, \dots, w_{g[n_g]})$ be the diagonal matrix of $w$ for group $g$.

Before any finite-sample corrections, the cluster-robust standard error (\citealp{white1984asymptotic}) for $\hat{\tau}_{\mathrm{target}}$ takes the form:
\begin{align}\label{eq:crse_1}
    \widetilde{\se}_{\mathrm{CRSE}} ( \hat{\tau}_{\mathrm{target}}) = \sqrt{ \frac{\frac{1}{n^2} \sum_{g=1}^G \big( \mathbf{D}_g^{\perp_w X, Z} \big)^{\top} \mathbf{W}_g \mathbf{Y}_g^{\perp_w X, D, Z} \big(\mathbf{Y}_g^{\perp_w X, D, Z}\big)^{\top} \mathbf{W}_g \mathbf{D}_g^{\perp_w X, Z}  }{ \big( \widehat{\var}_w (D^{\perp_w X, Z}) \big)^2}}
\end{align}
We simplify the denominator of (\ref{eq:crse_1}) with the identity:
\begin{align}\label{eq:crse_2}
    \widehat{\var}_w (D^{\perp_w X, Z}) = \biggr( 1 - \hat{R}^2_w (D \sim Z | X) \biggr) \widehat{\var}_w (D^{\perp_w X})
\end{align}
Applying (\ref{eq:crse_2}) to (\ref{eq:crse_1}) yields:
\begin{align}\label{eq:crse_3}
    & \widetilde{\se}_{\mathrm{CRSE}} ( \hat{\tau}_{\mathrm{target}}) = \nonumber \\
    & \ \ \ \ \ \ \ \ \ \frac{1}{1 - \hat{R}^2_w (D \sim Z | X)} \times \nonumber \\
    & \ \ \ \ \ \ \ \ \ \sqrt{ \frac{\frac{1}{n^2} \sum_{g=1}^G \big( \mathbf{D}_g^{\perp_w X, Z} \big)^{\top} \mathbf{W}_g \mathbf{Y}_g^{\perp_w X, D, Z} \big(\mathbf{Y}_g^{\perp_w X, D, Z}\big)^{\top} \mathbf{W}_g \mathbf{D}_g^{\perp_w X, Z}  }{ \big( \widehat{\var}_w (D^{\perp_w X}) \big)^2}}
\end{align}
We then apply a slightly stronger approximation than that in Assumption~\ref{asm:uniform_shrinkage}, which is that:
\begin{align}\label{eq:crse_4}
    Y_i^{\perp_w X, D, Z} &\approx  Y_i^{\perp_w X, D} * \sqrt{ 1 - \hat{R}^2_w (Y \sim Z | X, D) } \nonumber \\
    \text{and} \ \ D_i^{\perp_w X, Z} &\approx  D_i^{\perp_w X} * \sqrt{ 1 - \hat{R}^2_w (D \sim Z | X) }
\end{align}
Applying (\ref{eq:crse_4}) to (\ref{eq:crse_3}) yields:
\begin{align}\label{eq:crse_5}
    & \widetilde{\se}_{\mathrm{CRSE}} ( \hat{\tau}_{\mathrm{target}}) \approx \nonumber \\
    & \ \ \ \ \ \ \ \ \ \sqrt{\frac{1 - \hat{R}^2_w (Y \sim Z | X, D)} {1 - \hat{R}^2_w (D \sim Z | X)} } \times \nonumber \\
    & \ \ \ \ \ \ \ \ \ \underbrace{\sqrt{ \frac{\frac{1}{n^2} \sum_{g=1}^G \big( \mathbf{D}_g^{\perp_w X} \big)^{\top} \mathbf{W}_g \mathbf{Y}_g^{\perp_w X, D} \big(\mathbf{Y}_g^{\perp_w X, D}\big)^{\top} \mathbf{W}_g \mathbf{D}_g^{\perp_w X}  }{ \big( \widehat{\var}_w (D^{\perp_w X}) \big)^2}}}_{\widetilde{\se}_{\mathrm{CRSE}} ( \hat{\tau}_{\mathrm{wls}})}
\end{align}
The right-hand-side of (\ref{eq:crse_5}) is thus an adjusted $\widetilde{\se}_{\mathrm{CRSE}} ( \hat{\tau}_{\mathrm{target}})$. However, it is common to apply a finite sample correction to $\widetilde{\se}_{\mathrm{CRSE}} ( \cdot )$ (e.g., \citealp{cameron2015practitioner}) by multiplying it by:
\begin{align}\label{eq:crse_6}
    c = \sqrt{\frac{G}{G-1} \times \frac{n-1}{n - \text{Number of Regressors}}}
\end{align}
For $\hat{\tau}_{\mathrm{target}}$, the number of regressors is $P+2$. Thus, our proposed adjusted cluster-robust standard error is:
\begin{align}\label{eq:crse_7}
    & \widehat{\se}_{\mathrm{adj-CRSE}} ( \hat{\tau}_{\mathrm{target}}) = \nonumber \\
    & \ \ \ \ \ \ \ \ \  \sqrt{\frac{G}{G-1} \times \frac{n-1}{n - P - 2}}  \times \nonumber \\
    & \ \ \ \ \ \ \ \ \ \sqrt{\frac{1 - \hat{R}^2_w (Y \sim Z | X, D)} {1 - \hat{R}^2_w (D \sim Z | X)} } \times \nonumber \\
    & \ \ \ \ \ \ \ \ \ \underbrace{\sqrt{ \frac{\frac{1}{n^2} \sum_{g=1}^G \big( \mathbf{D}_g^{\perp_w X} \big)^{\top} \mathbf{W}_g \mathbf{Y}_g^{\perp_w X, D} \big(\mathbf{Y}_g^{\perp_w X, D}\big)^{\top} \mathbf{W}_g \mathbf{D}_g^{\perp_w X}  }{ \big( \widehat{\var}_w (D^{\perp_w X}) \big)^2}}}_{\widetilde{\se}_{\mathrm{CRSE}} ( \hat{\tau}_{\mathrm{wls}})}
\end{align}

\subsection{Derivation of $\mathrm{RV}_{q} (\hat{\tau}_{\mathrm{wls}})$ }\label{app:proofs_wsa_rv}
%
If $\hat{R}_w^2 (Y \sim Z | X, D) = \hat{R}_w^2 (D \sim Z | X) = x$ and $\hat{\tau}_{\mathrm{target}} = (1 - q) \hat{\tau}_{\mathrm{wls}}$, then from (\ref{eq:wsa_bias}),
	\begin{align}\label{eq:proofs_wsa_rv_q_1}
		q \hat{\tau}_{\mathrm{wls}}  = \frac{x}{\sqrt{1 - x}} \times \frac{ \widehat{\sd}_w ( Y^{\perp_w X, D})} {\widehat{\sd}_w ( D^{\perp_w X})} 
	\end{align}
(\ref{eq:proofs_wsa_rv_q_1}) can be rewritten as
	\begin{align}\label{eq:proofs_wsa_rv_q_2}
		x^2 + \biggr( q \hat{\tau}_{\mathrm{wls}}  \times \frac{ \widehat{\sd}_w ( D^{\perp_w X})} {\widehat{\sd}_w ( Y^{\perp_w X, D})}\biggr)^2 x -   \biggr( q \hat{\tau}_{\mathrm{wls}}  \times \frac{ \widehat{\sd}_w ( D^{\perp_w X})} {\widehat{\sd}_w ( Y^{\perp_w X, D})}\biggr)^2  = 0
	\end{align}
Noticing that
	\begin{align}\label{eq:proofs_wsa_rv_q_3}
		q \hat{\tau}_{\mathrm{wls}}  \times \frac{ \widehat{\sd}_w ( D^{\perp_w X})} {\widehat{\sd}_w ( Y^{\perp_w X, D})}
		&= q \times \frac{ \widehat{\cov}_w ( D^{\perp_w X}, Y^{\perp_w X})} {\widehat{\var}_w ( D^{\perp_w X})} \times  \frac{ \widehat{\sd}_w ( D^{\perp_w X})} {\widehat{\sd}_w ( Y^{\perp_w X, D})} \nonumber \\
		&= q \times \frac{ \widehat{\cov}_w ( D^{\perp_w X}, Y^{\perp_w X})} {\widehat{\sd}_w ( D^{\perp_w X}) \widehat{\sd}_w ( Y^{\perp_w X})} \times  \frac{ \widehat{\sd}_w ( Y^{\perp_w X})} {\widehat{\sd}_w ( Y^{\perp_w X, D})} \nonumber \\
		 &= q \times \frac{ \hat{R}_w (Y \sim D | X) } {\sqrt{ 1 - \hat{R}^2_w (Y \sim D | X)}} \nonumber \\
 		&= \omega_q
	\end{align}
allows (\ref{eq:proofs_wsa_rv_q_2}) to continue as
	\begin{align}\label{eq:proofs_wsa_rv_q_4}
		x^2 + \omega_q^2 x -   \omega_q^2  = 0
	\end{align}
Solving for $x$ using the quadratic formula then gives,
	\begin{align}\label{eq:proofs_wsa_rv_q_5}
		x = \frac{1}{2} (- \omega_q^2 \pm \sqrt{\omega_q^4 + 4\omega_q^2})
	\end{align}
Finally, noticing that $\hat{R}_w^2 (Y \sim Z | X, D) = \hat{R}_w^2 (D \sim Z | X) = x$ must be positive implies that only the upper bound of (\ref{eq:proofs_wsa_rv_q_5}) can hold, completing the proof.

\begin{flushright}
$\square$
\end{flushright} 

\subsection{Derivation of $\hat{R}^2_w (Y \sim D | X)$ as an extreme scenario }\label{app:proofs_wsa_extreme}

If $\hat{R}^2_w (Y \sim Z | D, X) = 1$, then additionally setting $\hat{\tau}_{\mathrm{target}} = 0$ in (\ref{eq:wsa_bias}) yields
	\begin{align}\label{eq:proofs_wsa_extreme_1}
		\hat{\tau}_{\mathrm{wls}} = \frac{\hat{R}_w (D \sim Z | X)}{\sqrt{1 - \hat{R}^2_w (D \sim Z | X)}} \times \frac{\widehat{\sd}_w (Y^{\perp_w X, D})}{ \widehat{\sd}_w (D^{\perp_w X}) }
	\end{align}
Using that
	\begin{align}\label{eq:proofs_wsa_extreme_2}
		\hat{\tau}_{\mathrm{wls}} = \frac{ \widehat{\cov}_w (D^{\perp_w X},  Y^{\perp_w X}) } {\widehat{\var}_w (D^{\perp_w X})} = \hat{R}_w (Y \sim D | X)  \times \frac{ \widehat{\sd}_w (Y^{\perp_w X}) } { \widehat{\sd}_w (D^{\perp_w X}) }
	\end{align}
allows (\ref{eq:proofs_wsa_extreme_1}) to continue as
	\begin{align}\label{eq:proofs_wsa_extreme_3}
		\hat{R}_w (Y \sim D | X)  \times \frac{ \widehat{\sd}_w (Y^{\perp_w X}) } { \widehat{\sd}_w (D^{\perp_w X}) } = \frac{\hat{R}_w (D \sim Z | X)}{\sqrt{1 - \hat{R}^2_w (D \sim Z | X)}} \times \frac{\widehat{\sd}_w (Y^{\perp_w X, D})}{ \widehat{\sd}_w (D^{\perp_w X}) }
	\end{align}
Rearranging terms in (\ref{eq:proofs_wsa_extreme_3}) then gives
	\begin{align}\label{eq:proofs_wsa_extreme_4}
		\hat{R}_w (Y \sim D | X)  \times \frac{ \widehat{\sd}_w (Y^{\perp_w X}) } { \widehat{\sd}_w (Y^{\perp_w X, D}) } = \frac{\hat{R}_w (D \sim Z | X)}{\sqrt{1 - \hat{R}^2_w (D \sim Z | X)}} 
	\end{align}	
Finally, using that
	\begin{align}\label{eq:proofs_wsa_extreme_5}
		\frac{ \widehat{\sd}_w (Y^{\perp_w X}) } { \widehat{\sd}_w (Y^{\perp_w X, D}) } = \frac{1}{ \sqrt{ 1 - \hat{R}_w^2 (Y \sim D | X)} }
	\end{align}
and squaring both sides of (\ref{eq:proofs_wsa_extreme_4}) gives
	\begin{align}\label{eq:proofs_wsa_extreme_6}
		\frac{\hat{R}^2_w (Y \sim D | X)}{1 - \hat{R}^2_w (Y \sim D | X)} = \frac{\hat{R}^2_w (D \sim Z | X)}{1 - \hat{R}^2_w (D \sim Z | X)} 
	\end{align}
which completes the proof.	
	
\begin{flushright}
$\square$
\end{flushright}

\subsection{Bounding $\hat{R}^2_w (D \sim Z | X)$ and $\hat{R}^2_w (Y \sim Z | D, X)$ using multiple observed covariates}\label{app:proofs_wsa_benchmark}

We consider here bounding the sensitivity parameters use \textit{multiple} covariates, where $X^{(1:j)}$ contains the first $j$ dimensions of $X$, and $X^{(-1:j)}$ contains the remaining (i.e., the final $P-j$) dimensions of $X$. Then, let $w_i^{(-1:j)}$ be semi-weights, which are formed by the exact same process as are $w_i$, but after removing $X^{(1:j)}$ from $X$. We redefine the bounding constants (i.e., $\kappa$) accordingly as
	\begin{align}\label{eq:wsa_kappa_multiple}
		\kappa_{w/w^{(-1:j)}} (D) := \frac{ \hat{R}_w^2 (D \sim Z | X^{(-1:j)})}{ \hat{R}_{w^{(-1:j)}}^2 (D \sim X^{(1:j)} | X^{(-1:j)})} \ \ \text{and} \ \ \kappa_w (Y) := \frac{ \hat{R}_w^2 (Y \sim Z | D, X^{(-1:j)})}{ \hat{R}_w^2 (Y \sim X^{(1:j)} | D, X^{(-1:j)})}
	\end{align}
Here, $\kappa_{w/w^{(-1:j)}} (D)$ and $\kappa_w (Y)$
describe the strength of $Z$ in relation to that of (a multiple of) the \textit{combined} strength of $X^{(1:j)}$. These constants then define bounds on the sensitivity parameters:
	\begin{align}\label{eq:wsa_benchmarkd_multiple}
		\ \ \ \ \ \hat{R}_w^2 (D \sim Z | X) &= \kappa_{w/w^{(-1:j)}} (D) \times \frac{\hat{R}_{w^{(-1:j)}}^2 (D \sim X^{(1:j)} | X^{(-1:j)})}{1 - \hat{R}_{w}^2 (D \sim X^{(1:j)} | X^{(-1:j)})}
	\end{align}
	\vspace{-0.3in}
	\begin{align}\label{eq:wsa_benchmarky_multiple}
		\hat{R}_w^2 (Y \sim Z | D, X) &\leq \eta_{w/w^{(-1:j)}}^2 \times \frac{\hat{R}_w^2 (Y \sim X^{(1:j)} | D, X^{(-1:j)})}{1 - \hat{R}_w^2 (Y \sim X^{(1:j)} | D, X^{(-1:j)})}
	\end{align}
where
    \begin{align}\label{eq:wsa_eta_multiple}
        & \eta^2_{w/w^{(-1:j)}} = \biggr( \frac{ \sqrt{ \kappa_w (Y) } + |\hat{R}_w (Z \sim X^{(1:j)} | D, X^{(-1:j)}) | } {\sqrt{1 - \hat{R}_w^2 (Z \sim X^{(1:j)} | D, X^{(-1:j)})} } \biggr)^2 \nonumber \\
        \text{with} \ \ \ & \hat{R}_w^2 (Z \sim X^{(1:j)} | D, X^{(-1:j)}) = \biggr( \frac{ \kappa_{w/w^{(-1:j)}} ( D) \times \hat{R}^2_{w^{(-1:j)}} (D \sim X^{(1:j)} | X^{(-1:j)}) } {1 - \kappa_{w/w^{(-1:j)}} ( D) \times \hat{R}^2_{w^{(-1:j)} } (D \sim X^{(1:j)} | X^{(-1:j)}) } \biggr) \nonumber \\
        & \ \ \ \ \ \ \ \ \ \ \ \ \ \ \ \ \ \ \ \ \ \ \ \ \ \ \ \ \ \ \ \ \ \ \ \times \biggr( \frac{ \hat{R}^2_{w} (D \sim X^{(1:j)} | X^{(-1:j)}) } {1 - \hat{R}^2_{w} (D \sim X^{(1:j)} | X^{(-1:j)}) } \biggr)
    \end{align}
Note that the original bounds in (\ref{eq:wsa_benchmarkd}) and (\ref{eq:wsa_benchmarky}) are special cases of (\ref{eq:wsa_benchmarkd_multiple}) and (\ref{eq:wsa_benchmarky_multiple}) above, respectively, where $X^{(1:j)}$ is a single covariate, $X^{(j)}$.

\subsubsection{Proof of bound on $\hat{R}^2_w (D \sim Z | X)$}\label{app:proofs_wsa_benchmark_d}

Starting with identity,
	\begin{align}\label{eq:proofs_wsa_benchmark_d_1}
		& \hat{R}^2_w (D \sim X^{(1:j)} + Z | X^{(-1:j)} )  \nonumber 
		\\
		& \ \ \ \ \ = \hat{R}^2_w (D \sim X^{(1:j)}  | X^{(-1:j)} ) + \biggr(1 - \hat{R}^2_w (D \sim X^{(1:j)} | X^{(-1:j)} ) \biggr) \hat{R}^2_w (D \sim  Z | X)
	\end{align}
yields, after rearranging,
	\begin{align}\label{eq:proofs_wsa_benchmark_d_2}
		\hat{R}^2_w (D \sim  Z | X) = \frac { \hat{R}^2_w (D \sim X^{(1:j)} + Z | X^{(-1:j)} ) - \hat{R}^2_w (D \sim X^{(1:j)}  | X^{(-1:j)} )} {1 - \hat{R}^2_w (D \sim X^{(1:j)} | X^{(-1:j)} )}
	\end{align}
Without loss of generality, $Z$ can be chosen such that $\hat{R}^2_w (Z \sim  X ) = 0$. Thus, the numerator in (\ref{eq:proofs_wsa_benchmark_d_2}) simplifies to
	\begin{align}\label{eq:proofs_wsa_benchmark_d_3}
		& \hat{R}^2_w (D \sim X^{(1:j)} + Z | X^{(-1:j)} ) - \hat{R}^2_w (D \sim X^{(1:j)}  | X^{(-1:j)} ) \nonumber \\
		& \ \ \ \ \ = \hat{R}^2_w (D \sim X^{(1:j)} | X^{(-1:j)} ) + \hat{R}^2_w (D \sim Z | X^{(-1:j)} ) - \hat{R}^2_w (D \sim X^{(1:j)}  | X^{(-1:j)} ) \nonumber \\
		& \ \ \ \ \ = \hat{R}^2_w (D \sim Z | X^{(-1:j)} )
	\end{align}
Therefore, (\ref{eq:proofs_wsa_benchmark_d_2}) continues as
		\begin{align}\label{eq:proofs_wsa_benchmark_d_4}
		\hat{R}^2_w (D \sim  Z | X) &= \frac { \hat{R}^2_w (D \sim Z | X^{(-1:j)} ) } {1 - \hat{R}^2_w (D \sim X^{(1:j)} | X^{(-1:j)} )} \nonumber \\
		&= \kappa_{w/w^{(-1:j)}} (D) \times \frac { \hat{R}^2_{w^{(-1:j)}} (D \sim X^{(1:j)} | X^{(-1:j)} ) } {1 - \hat{R}^2_w (D \sim X^{(1:j)} | X^{(-1:j)} )}
	\end{align}
where the second line of (\ref{eq:proofs_wsa_benchmark_d_4}) above uses the definition of $\kappa_{w/w^{(-1:j)}} (D)$ in (\ref{eq:wsa_kappa_multiple}), completing the proof.

\begin{flushright}
$\square$
\end{flushright}

\subsubsection{Proof of bound on $\hat{R}^2_w (Y \sim Z | D, X)$}\label{app:proofs_wsa_benchmark_y}

First, let 
    \begin{align}\label{eq:proofs_wsa_benchmark_y_1}
        A_i = [X_i^{(1:j)}]^{\perp_w D, X^{(-1:j)}}
    \end{align}
In other words, $A$ is the result of partialing out $(D, X^{(-1:j)})$ from $X^{(1:j)}$. Then, let
    \begin{align}\label{eq:proofs_wsa_benchmark_y_2}
        \hat{\alpha}_w = [\widehat{\var}_w (A)]^{-1} \widehat{\cov}_w (A, Z^{\perp_w D, X^{(-1:j)}}) 
    \end{align}
be the coefficients from the weighted regression of $Z^{\perp_w D, X^{(-1:j)}}$ on $A$. This allows the partialing out of $A$ from $Z^{\perp_w D, X^{(-1:j)}}$ to be written as
    \begin{align}\label{eq:proofs_wsa_benchmark_y_3}
        [Z_i^{\perp_w D, X^{(-1:j)}}]^{\perp_w A} = Z_i^{\perp_w D, X^{(-1:j)}} - A_i^{\top} \hat{\alpha}_w
    \end{align}
Although messy, defining $A$ and $\hat{\alpha}_w$ as above greatly simplifies notation in the rest of the proof.

Now, using the definition of partial $\hat{R}_w^2$, the sensitivity parameter of interest can be rewritten as
    \begin{align}\label{eq:proofs_wsa_benchmark_y_4}
        \hat{R}_w^2 (Y \sim Z | D, X) = \hat{R}_w^2 (Y^{\perp_w D, X} \sim Z^{\perp_w D, X})
    \end{align}
Notice then that partialing out $X$ and $D$ is the same as first partialing out $X^{(-1:j)}$ and $D$, and then partialing out $A$. In other words,
    \begin{align}\label{eq:proofs_wsa_benchmark_y_5}
        Z_i^{\perp_w D, X} = [Z_i^{\perp_w D, X^{(-1:j)}}]^{\perp_w A} \ \ \text{and} \ \ Y_i^{\perp_w D, X} = [Y_i^{\perp_w D, X^{(-1:j)}}]^{\perp_w A}
    \end{align}
Therefore,
    \begin{align}\label{eq:proofs_wsa_benchmark_y_6}
        \hat{R}_w (Y \sim Z | D, X) &= \frac{ \widehat{\cov}_w ( Y^{\perp_w D, X}, Z^{\perp_w D, X}) }{ \widehat{\sd}_w (Y^{\perp_w D, X}) \widehat{\sd}_w (Z^{\perp_w D, X}) } \nonumber \\
        &= \frac{ \widehat{\cov}_w \biggr( [Y^{\perp_w D, X^{(-1:j)}}]^{\perp_w A}, [Z^{\perp_w D, X^{(-1:j)}}]^{\perp_w A} \biggr) }{ \widehat{\sd}_w (Y^{\perp_w D, X}) \widehat{\sd}_w ( Z^{\perp_w D, X}) } \nonumber \\
        &= \frac{ \widehat{\cov}_w \biggr( Y^{\perp_w D, X^{(-1:j)}}, [Z^{\perp_w D, X^{(-1:j)}}]^{\perp_w A} \biggr) }{ \widehat{\sd}_w (Y^{\perp_w D, X}) \widehat{\sd}_w ( Z^{\perp_w D, X}) }
    \end{align}
where the last line of (\ref{eq:proofs_wsa_benchmark_y_6}) comes from the fact that $\widehat{\cov}_w (C^{\perp_w B}, E^{\perp_w B}) = \widehat{\cov}_w (C, E^{\perp_w B})$ for arbitrary $B$, $C$, and $E$. Applying (\ref{eq:proofs_wsa_benchmark_y_3}) then allows (\ref{eq:proofs_wsa_benchmark_y_6}) to continue as
    \begin{align}\label{eq:proofs_wsa_benchmark_y_7}
        & \hat{R}_w (Y \sim Z | D, X) = \frac{ \widehat{\cov}_w ( Y^{\perp_w D, X^{(-1:j)}}, Z^{\perp_w D, X^{(-1:j)}} ) }{ \widehat{\sd}_w (Y^{\perp_w D, X}) \widehat{\sd}_w ( Z^{\perp_w D, X}) } -  \frac{ \widehat{\cov}_w ( Y^{\perp_w D, X^{(-1:j)}}, A \hat{\alpha}_w ) }{ \widehat{\sd}_w (Y^{\perp_w D, X}) \widehat{\sd}_w ( Z^{\perp_w D, X}) }
    \end{align}
Notice then that the terms in the denominators of (\ref{eq:proofs_wsa_benchmark_y_7}) can be rewritten as
    \begin{align}\label{eq:proofs_wsa_benchmark_y_8}
        \widehat{\sd}_w (Y^{\perp_w D, X}) &= \frac{\widehat{\sd}_w (Y^{\perp_w D, X})}{\widehat{\sd}_w (Y^{\perp_w D, X^{(-1:j)}})} \times \widehat{\sd}_w (Y^{\perp_w D, X^{(-1:j)}}) \nonumber \\
        &= \sqrt{1 - \hat{R}^2_w (Y \sim X^{(1:j)} | D, X^{(-1:j)})} \times \widehat{\sd}_w (Y^{\perp_w D, X^{(-1:j)}}) 
    \end{align}
and, similarly,
    \begin{align}\label{eq:proofs_wsa_benchmark_y_9}
        \widehat{\sd}_w (Z^{\perp_w D, X}) &= \frac{\widehat{\sd}_w (Z^{\perp_w D, X})}{\widehat{\sd}_w (Z^{\perp_w D, X^{(-1:j)}})} \times \widehat{\sd}_w (Z^{\perp_w D, X^{(-1:j)}}) \nonumber \\
        &= \sqrt{1 - \hat{R}^2_w (Z \sim X^{(1:j)} | D, X^{(-1:j)})} \times \widehat{\sd}_w (Z^{\perp_w D, X^{(-1:j)}}) 
    \end{align}
Thus, applying (\ref{eq:proofs_wsa_benchmark_y_8}) and (\ref{eq:proofs_wsa_benchmark_y_9}) allows the expression for $\hat{R}_w (Y \sim Z | D, X)$ in  (\ref{eq:proofs_wsa_benchmark_y_7}) to continue as
    \begin{align}\label{eq:proofs_wsa_benchmark_y_10}
        & \hat{R}_w (Y \sim Z | D, X) = \nonumber \\
        & \ \ \ \ \ \ \frac{1}{\sqrt{1 - \hat{R}^2_w (Y \sim X^{(1:j)} | D, X^{(-1:j)})} \sqrt{1 - \hat{R}^2_w (Z \sim X^{(1:j)} | D, X^{(-1:j)})}} \times \nonumber \\
        & \ \ \ \ \ \ \biggr( \frac{ \widehat{\cov}_w ( Y^{\perp_w D, X^{(-1:j)}}, Z^{\perp_w D, X^{(-1:j)}} ) }{ \widehat{\sd}_w (Y^{\perp_w D, X^{(-1:j)}}) \widehat{\sd}_w ( Z^{\perp_w D, X^{(-1:j)}}) } -  \frac{ \widehat{\cov}_w ( Y^{\perp_w D, X^{(-1:j)}}, A \hat{\alpha}_w ) }{ \widehat{\sd}_w (Y^{\perp_w D, X^{(-1:j)}}) \widehat{\sd}_w ( Z^{\perp_w D, X^{(-1:j)}}) } \biggr) 
    \end{align}
Using then that
    \begin{align}\label{eq:proofs_wsa_benchmark_y_11}
        \frac{ \widehat{\cov}_w ( Y^{\perp_w D, X^{(-1:j)}}, Z^{\perp_w D, X^{(-1:j)}} ) }{ \widehat{\sd}_w (Y^{\perp_w D, X^{(-1:j)}}) \widehat{\sd}_w ( Z^{\perp_w D, X^{(-1:j)}}) } = \hat{R}_w (Y \sim Z | D, X^{(-1:j)})
    \end{align}
allows (\ref{eq:proofs_wsa_benchmark_y_10}) to continue as
    \begin{align}\label{eq:proofs_wsa_benchmark_y_12}
        & \hat{R}_w (Y \sim Z | D, X) = \nonumber \\
        & \ \ \ \ \ \ \frac{1}{\sqrt{1 - \hat{R}^2_w (Y \sim X^{(1:j)} | D, X^{(-1:j)})} \sqrt{1 - \hat{R}^2_w (Z \sim X^{(1:j)} | D, X^{(-1:j)})}} \times \nonumber \\
        & \ \ \ \ \ \ \biggr( \hat{R}_w (Y \sim Z | D, X^{(-1:j)}) -  \frac{ \widehat{\cov}_w ( Y^{\perp_w D, X^{(-1:j)}}, A \hat{\alpha}_w ) }{ \widehat{\sd}_w (Y^{\perp_w D, X^{(-1:j)}}) \widehat{\sd}_w ( Z^{\perp_w D, X^{(-1:j)}}) } \biggr) 
    \end{align}
Meaning that
    \begin{align}\label{eq:proofs_wsa_benchmark_y_13}
        & |\hat{R}_w (Y \sim Z | D, X)| \leq \nonumber \\
        & \ \ \ \ \ \ \frac{1}{\sqrt{1 - \hat{R}^2_w (Y \sim X^{(1:j)} | D, X^{(-1:j)})} \sqrt{1 - \hat{R}^2_w (Z \sim X^{(1:j)} | D, X^{(-1:j)})}} \times \nonumber \\
        & \ \ \ \ \ \ \biggr( |\hat{R}_w (Y \sim Z | D, X^{(-1:j)})| + \underbrace{ \biggr|  \frac{ \widehat{\cov}_w ( Y^{\perp_w D, X^{(-1:j)}}, A \hat{\alpha}_w ) }{ \widehat{\sd}_w (Y^{\perp_w D, X^{(-1:j)}}) \widehat{\sd}_w ( Z^{\perp_w D, X^{(-1:j)}}) } \biggr| }_{(a)} \biggr) 
    \end{align}
We now proceed by bounding (a) in (\ref{eq:proofs_wsa_benchmark_y_13}) above with $\hat{R}_w^2$ values. We find
    \begin{align}\label{eq:proofs_wsa_benchmark_y_14}
        (a) = \biggr|  \frac{ \widehat{\cov}_w ( Y^{\perp_w D, X^{(-1:j)}}, A \hat{\alpha}_w ) }{ \widehat{\sd}_w (Y^{\perp_w D, X^{(-1:j)}}) \widehat{\sd}_w ( A \hat{\alpha}_w) } \biggr| \times \frac{ \widehat{\sd}_w (A \hat{\alpha}_w) }{\widehat{\sd}_w ( Z^{\perp_w D, X^{(-1:j)}})}
    \end{align}
The definition for the $A_i$ (in (\ref{eq:proofs_wsa_benchmark_y_1})) then implies that
    \begin{align}\label{eq:proofs_wsa_benchmark_y_15}
        \biggr|  \frac{ \widehat{\cov}_w ( Y^{\perp_w D, X^{(-1:j)}}, A \hat{\alpha}_w ) }{ \widehat{\sd}_w (Y^{\perp_w D, X^{(-1:j)}}) \widehat{\sd}_w ( A \hat{\alpha}_w) } \biggr| &= | \hat{R}_w (Y \sim X^{(1:j)} \hat{\alpha}_w | D, X^{-(1:j)} ) | \nonumber \\
        &\leq | \hat{R}_w (Y \sim X^{(1:j)} | D, X^{-(1:j)} ) |
    \end{align}
Additionally, the definition for $\hat{\alpha}_w$ (in (\ref{eq:proofs_wsa_benchmark_y_2})) and (\ref{eq:proofs_wsa_benchmark_y_3}) imply that 
    \begin{align}\label{eq:proofs_wsa_benchmark_y_16}
    \frac{\widehat{\sd}_w (A \hat{\alpha}_w) }{\widehat{\sd}_w ( Z^{\perp_w D, X^{(-1:j)}})} = | \hat{R}_w(Z \sim X^{(1:j)} | D, X^{(-1:j)}) |
    \end{align}
Applying (\ref{eq:proofs_wsa_benchmark_y_15}) and (\ref{eq:proofs_wsa_benchmark_y_16}) to (\ref{eq:proofs_wsa_benchmark_y_14}) then gives a bound for (a) in (\ref{eq:proofs_wsa_benchmark_y_13}):
    \begin{align}\label{eq:proofs_wsa_benchmark_y_17}
        (a) \leq | \hat{R}_w (Y \sim X^{(1:j)} | D, X^{-(1:j)} ) \times \hat{R}_w (Z \sim X^{(1:j)} | D, X^{-(1:j)} ) |
    \end{align}
Thus, the expression for $\hat{R}_w (Y \sim Z | D, X)$ in (\ref{eq:proofs_wsa_benchmark_y_13}) continues as
    \begin{align}\label{eq:proofs_wsa_benchmark_y_18}
        & |\hat{R}_w (Y \sim Z | D, X)| \leq \nonumber \\
        & \ \ \ \ \ \ \frac{ |\hat{R}_w (Y \sim Z | D, X^{(-1:j)})| + | \hat{R}_w (Y \sim X^{(1:j)} | D, X^{-(1:j)} ) \times \hat{R}_w (Z \sim X^{(1:j)} | D, X^{-(1:j)}) | }{\sqrt{1 - \hat{R}^2_w (Y \sim X^{(1:j)} | D, X^{(-1:j)})} \sqrt{1 - \hat{R}^2_w (Z \sim X^{(1:j)} | D, X^{(-1:j)})}} 
    \end{align}        
Then using the definition of $\kappa_w (Y)$ in  (\ref{eq:wsa_kappa_multiple}) allows (\ref{eq:proofs_wsa_benchmark_y_18}) to continue as
    \begin{align}\label{eq:proofs_wsa_benchmark_y_19}
        & |\hat{R}_w (Y \sim Z | D, X)| \leq \nonumber \\
        & \ \ \ \ \ \ \frac{ \sqrt{ \kappa_w (Y)} | \hat{R}_w (Y \sim X^{(1:j)} | D, X^{(-1:j)})| + | \hat{R}_w (Y \sim X^{(1:j)} | D, X^{-(1:j)} ) \times \hat{R}_w (Z \sim X^{(1:j)} | D, X^{-(1:j)} ) | }{\sqrt{1 - \hat{R}^2_w (Y \sim X^{(1:j)} | D, X^{(-1:j)})} \sqrt{1 - \hat{R}^2_w (Z \sim X^{(1:j)} | D, X^{(-1:j)})}}
    \end{align}  
Rearranging $(\ref{eq:proofs_wsa_benchmark_y_19})$ and squaring both sides then gives the desired bound in (\ref{eq:wsa_benchmarky_multiple}), restated below:
	\begin{align}\label{eq:proofs_wsa_benchmark_y_20}
		\hat{R}_w^2 (Y \sim Z | D, X) &\leq \eta_{w/w^{(-1:j)}}^2 \times \frac{\hat{R}_w^2 (Y \sim X^{(1:j)} | D, X^{(-1:j)})}{1 - \hat{R}_w^2 (Y \sim X^{(1:j)} | D, X^{(-1:j)})}
	\end{align}
where $\eta^2_{w/w^{(-1:j)}}$, defined in (\ref{eq:wsa_eta_multiple}), is also restated below:
    \begin{align}\label{eq:proofs_wsa_benchmark_y_21}
        \eta^2_{w/w^{(-1:j)}} = \biggr( \frac{ \sqrt{ \kappa_w (Y) } + |\hat{R}_w (Z \sim X^{(1:j)} | D, X^{(-1:j)}) | } {\sqrt{1 - \hat{R}_w^2 (Z \sim X^{(1:j)} | D, X^{(-1:j)})} } \biggr)^2 
    \end{align}
What remains to prove is that $\hat{R}_w^2 (Z \sim X^{(1:j)} | D, X^{(-1:j)})$ in the equation for $\eta^2_{w/w^{(-1:j)}}$ is equal to the expression in (\ref{eq:wsa_eta_multiple}). First, note that
    \begin{align}\label{eq:proofs_wsa_benchmark_y_22}
        & \hat{R}_w^2 (Z \sim X^{(1:j)} + D | X^{(-1:j)} ) = \nonumber \\
        & \ \ \ \ \ \hat{R}_w^2 (Z \sim X^{(1:j)} | X^{(-1:j)}) + \biggr( 1 -  \hat{R}_w^2 (Z \sim X^{(1:j)}  | X^{(-1:j)} ) \biggr)  \hat{R}_w^2 (Z \sim D | X)
    \end{align}
We may assume that $\hat{R}_w^2 (Z \sim X) = 0$. Thus, additionally using that $\hat{R}_w^2 (Z \sim D | X) = \hat{R}_w^2 (D \sim Z | X)$ allows (\ref{eq:proofs_wsa_benchmark_y_22}) to continue as
    \begin{align}\label{eq:proofs_wsa_benchmark_y_23}
        & \hat{R}_w^2 (Z \sim X^{(1:j)} + D | X^{(-1:j)} ) = \hat{R}_w^2 (D \sim Z | X)
    \end{align}
Notice then that $\hat{R}_w^2 (Z \sim X^{(1:j)} + D | X^{(-1:j)} )$ may also be written as
    \begin{align}\label{eq:proofs_wsa_benchmark_y_24}
        & \hat{R}_w^2 (Z \sim X^{(1:j)} + D | X^{(-1:j)} )  \nonumber \\
        & \ \ \ \ \ = \hat{R}_w^2 (Z \sim D | X^{(-1:j)}) + \biggr( 1 -  \hat{R}_w^2 (Z \sim D  | X^{(-1:j)} ) \biggr)  \hat{R}_w^2 (Z \sim X^{(1:j)} | D, X^{(-1:j)}) \nonumber \\
        & \ \ \ \ \ = \hat{R}_w^2 (D \sim Z | X^{(-1:j)}) + \biggr( 1 -  \hat{R}_w^2 (D \sim Z  | X^{(-1:j)} ) \biggr)  \hat{R}_w^2 (Z \sim X^{(1:j)} | D, X^{(-1:j)})  
    \end{align}
Equating the expressions for $\hat{R}_w^2 (Z \sim X^{(1:j)} + D | X^{(-1:j)} )$ in  (\ref{eq:proofs_wsa_benchmark_y_23}) and (\ref{eq:proofs_wsa_benchmark_y_24}) and rearranging then yields
    \begin{align}\label{eq:proofs_wsa_benchmark_y_25}
        \hat{R}_w^2 (Z \sim X^{(1:j)} | D, X^{(-1:j)}) = \frac{ \hat{R}_w^2 (D \sim Z | X) - \hat{R}_w^2 (D \sim Z | X^{(-1:j)}) }{ 1 -  \hat{R}_w^2 (D \sim Z  | X^{(-1:j)} ) }
    \end{align}
Finally, using definition of $\kappa_{w/w^{(-1:j)}} (D)$ in $(\ref{eq:wsa_kappa_multiple})$, as well as the resulting bound for $\hat{R}_w^2 (D \sim Z | X)$ in  (\ref{eq:wsa_benchmarkd_multiple})
allows (\ref{eq:proofs_wsa_benchmark_y_25}) to continue as
    \begin{align}\label{eq:proofs_wsa_benchmark_y_26}
        & \hat{R}_w^2 (Z \sim X^{(1:j)} | D, X^{(-1:j)}) = \nonumber \\
        & \ \ \ \frac{ \kappa_{w/w^{(-1:j)}} (D) \times \frac{\hat{R}_{w^{(-1:j)}}^2 (D \sim X^{(1:j)} | X^{(-1:j)})}{1 - \hat{R}_{w}^2 (D \sim X^{(1:j)} | X^{(-1:j)})} - \kappa_{w/w^{(-1:j)}} (D) \hat{R}^2_{w^{(-1:j)}} (D \sim X^{(1:j)} | X^{(-1:j)}) }{ 1 -  \kappa_{w/w^{(-1:j)}} (D) \hat{R}^2_{w^{(-1:j)}} (D \sim X^{(1:j)} | X^{(-1:j)}) }
    \end{align}
Rearranging terms in (\ref{eq:proofs_wsa_benchmark_y_26}) then gives the desired expression for $\hat{R}_w^2 (Z \sim X^{(1:j)} | D, X^{(-1:j)})$ in (\ref{eq:wsa_eta_multiple}), completing the proof.

\begin{flushright}
$\square$
\end{flushright}

\end{document}